\documentclass[12pt,twoside]{article}   

\usepackage[super,sort,comma]{natbib}
\usepackage{soul}
\usepackage{lscape}  
\usepackage{boxedminipage}
\usepackage{rotating,booktabs}
\usepackage{multirow}
\usepackage{multicol}
\usepackage{longtable}
\usepackage{caption}
\usepackage{todonotes}
\usepackage{tabularx}
\usepackage{colortbl}
\usepackage{enumitem}
\usepackage{rotating}
\usepackage{adjustbox}
\usepackage{graphicx}
\usepackage{multirow}
\usepackage{bm}
\usepackage{lscape}
\usepackage{comment}
\usepackage{amssymb}
\usepackage{calrsfs}

\usepackage{fancyhdr}		

\newcommand{\captionv}[3]{\begin{center}\parbox{#1cm}{\caption[#2]{{\sf #3}}}
        \end{center}}
        
\DeclareMathAlphabet{\pazocal}{OMS}{zplm}{m}{n}

\usepackage[section]{placeins}   %

\usepackage{graphicx}

\makeatletter \renewcommand\@biblabel[1]{$^{#1}$} \makeatother
\newcommand{\note}[1]{\mbox{}\\ \noindent \rule{16cm}{0.5mm} \\
{\em #1} \\ \noindent \rule{16cm}{0.5mm}
\typeout{    }
\typeout{***********note active on this page *************************}
\typeout{Note: #1  }
\typeout{****************************************end Note}
}

\newcommand{\cen}[1]{\begin{center} #1 \end{center}}

\usepackage[mathlines]{lineno}

\usepackage{hyperref}
\hypersetup{ colorlinks,
    citecolor=blue,
    filecolor=blue,
    linkcolor=blue,
    urlcolor=blue
}

\usepackage{xcolor}

\definecolor{gray}{rgb}{0.6,0.6,0.6}
\definecolor{red}{rgb}{0.85,0,0}
\definecolor{green}{rgb}{0,0.85,0}
\definecolor{blue}{rgb}{0,0,0.85}
\definecolor{beige}{rgb}{0.92,0.87,0.78}
\usepackage[all]{hypcap}    

\begin{document}

\cen{\sf {\Large {\bfseries Deep learning-based synthetic-CT generation in radiotherapy and PET: a review} \\  
\vspace*{10mm}
Maria Francesca Spadea$^{1,*}$, Matteo Maspero$^{2,3,*}$, Paolo Zaffino$^1$,  and Joao Seco$^{4,5}$ \\} 
$^1$ Department of Clinical and Experimental Medicine, University “Magna Graecia” of Catanzaro, 88100 Catanzaro, Italy\\
$^2$ Department of Radiotherapy, Division of Imaging \& Oncology, University Medical Center Utrecht, Heidelberglaan 100, 3508 GA Utrecht, The Netherlands \\
$^3$ Computational Imaging Group for MR diagnostics \& therapy, Center for
Image Sciences, University Medical Center Utrecht, Heidelberglaan 100, 3508
GA Utrecht, The Netherlands \\
$^4$ DKFZ German Cancer Research Center, Division of Biomedical Physics in Radiation Oncology, 69120 Heidelberg, Germany \\
$^5$ Department of Physics and Astronomy, Heidelberg University, 69120 Heidelberg, Germany\\
$*$ These authors equally contributed.
\vspace{5mm}\\
Version typeset: \today\\
}
\pagenumbering{arabic}
\setcounter{page}{1}

\pagestyle{plain}

\begin{abstract}
\noindent 

Recently, deep learning (DL)-based methods for the generation of synthetic computed tomography (sCT) have received significant research attention as an alternative to classical ones.
We present here a systematic review of these methods by grouping them into three categories, according to their clinical applications: \\
I) To replace CT in magnetic resonance (MR)-based treatment planning.\\
II) Facilitate cone-beam computed tomography (CBCT)-based image-guided adaptive radiotherapy.\\
III) Derive attenuation maps for the correction of positron emission tomography (PET).\\
Appropriate database searching was performed on journal articles published between January 2014 and December 2020.\\
The DL methods' key characteristics were extracted from each eligible study, and a comprehensive comparison among network architectures and metrics was reported. A detailed review of each category was given, highlighting essential contributions, identifying specific challenges, and summarising the achievements.
Lastly, the statistics of all the cited works from various aspects were analysed, revealing the popularity and future trends and the potential of DL-based sCT generation. The current status of DL-based sCT generation was evaluated, assessing the clinical readiness of the presented methods.
\end{abstract}
\note{\small Authors to whom correspondence should be addressed. Email: 
j.seco@dkfz.de}

\newpage     



\setlength{\baselineskip}{0.7cm}      

\pagestyle{fancy}
\section{Introduction}
Medical imaging's impact on oncological patients' diagnosis and therapy has grown significantly over the last decades\cite{Husband2016}. Especially in radiotherapy (RT)\cite{Beaton2019}, imaging plays a crucial role in the entire workflow, from treatment simulation to patient positioning and monitoring \cite{Verellen2007,jaffray2012image,seco2015imaging,IAEA2017}.\\ 
Traditionally, computed tomography (CT) is considered the primary imaging modality in RT. It provides accurate and high-resolution patient's geometry, enabling direct electron density conversion needed for dose calculations \cite{Seco2006}. 	
X-ray based imaging, including planar imaging and cone-beam computed tomography (CBCT), are widely adopted for patient positioning and monitoring before, during or after the dose delivery \cite{jaffray2012image}. 
Along with CT, positron emission tomography (PET) is commonly acquired to provide functional and metabolic information allowing tumour staging and improving tumour contouring \cite{Unterrainer2020pet}.
Magnetic resonance imaging (MRI) has also proved its added value for tumours and organs-at-risk (OARs) delineation, thanks to its superb soft tissue contrast~\cite{Dirix2014,Schmidt2015}.\\
To benefit from the complementary advantages offered by different imaging modalities, MRI is generally registered to CT \cite{Devic2012}.
However, residual misregistration and differences in patient set-up may introduce systematic errors that would affect the accuracy of the whole treatment~\cite{Nyholm2009,Ulin2010}.\\ 		
Recently, MR-only based RT has been proposed \cite{Fraas1987,Lee2003,Nyholm_counter} to eliminate residual registration errors.
Furthermore, it can simplify and speed up the workflow, decreasing patient's exposure to ionising radiation, which is particularly relevant for repeated simulations \cite{Kapanen2013} or fragile populations, e.g. children. Also, MR-only RT may reduce overall treatment costs~\cite{Owrangi2018} and workload \cite{Karlsson2009}.
Additionally, the development of MR-only techniques can be beneficial for MR-guided RT \cite{Lagendijk2014}.

The main obstacle regarding the introduction of MR-only radiotherapy is the lack of tissue attenuation information
required for accurate dose calculations \cite{Nyholm2009,Jonsson2010}.
Many methods have been proposed to convert MR to CT-equivalent representations, often known as synthetic CT (sCT), for treatment planning and dose calculation. These approaches are summarised in two specific reviews on this topic \cite{Edmund2017,Johnstone2018,Wafa2018}, in site-specific reviews~\cite{Owrangi2018,Kerkmeijer2018,Bird2019} or broader review on MR-guided~\cite{Thorwarth2021} or proton therapy~\cite{Hoffmann2020}. 

Additionally, similar techniques to derive sCT from a different imaging modality have been envisioned to improve the quality of CBCT \cite{taasti2020developments}. Cone-beam computed tomography plays a vital role in image-guided adaptive radiation therapy (IGART) for photon and proton therapy. However, due to the severe scatter noise and truncated projections, image reconstruction is affected by several artefacts, such as shading, streaking and cupping \cite{ZhuCTBCTnoise,ZhuCTBCTscatter}. For this reason, daily CBCT has not commonly been used for online plan adaptation. The conversion of CBCT-to-CT would allow accurate dose computation and improve the quality of IGART provided to the patients. 

Finally, sCT estimation is also crucial for PET attenuation correction. Accurate PET quantification requires a reliable photon attenuation correction (AC) map, usually derived from CT. In the new PET/MRI hybrid scanners, this step is not immediate, and MRI to sCT translation has been proposed to solve the MR attenuation correction (MRAC) issue. Besides, standalone PET scanners can benefit from the derivation of sCT from uncorrected PET \cite{mehranian2016vision,mecheter2020mr,Catana2020}.

In the last years, the derivation of sCT from MRI, PET or CBCT has raised increasing interest based on artificial intelligence algorithms such as machine learning or deep learning (DL) \cite{lecun2015deep}.
This paper aims to systematically review and summarise the latest developments, challenges and trends in DL-based sCT generation methods.
Deep learning is a branch of machine learning, a field of artificial intelligence that involves using neural networks to generate hierarchical representations of the input data to learn a specific task without hand-engineered features\cite{Goodfellow2016}. 
Recent reviews have discussed the application of deep learning in radiotherapy \cite{Meyer2018, Sahiner2018,Boon2018,Wang2019rev, Boldrini2019, Jarrett2019, Kiser2019}, and in PET attenuation correction \cite{Catana2020}. Convolutional neural networks (CNNs), which are the most successful models for image processing~\cite{Krizhevsky2012,Litjens2017}, have been proposed 
for sCT generation since 2016 \cite{nie2016estimating}, with a rapidly increasing number of published papers on the topic.  However, DL-based sCT generation has not been reviewed in details, except for applications in PET \cite{lee2020review}. 
With this survey, we aim at summarising the latest developments in DL-based sCT generation, highlighting the contributions based on the applications and providing detailed statistics discussing trends in terms of imaging protocols, DL architectures, and performance achieved. Finally, the clinical readiness of the reviewed methods will be discussed.

\section{Material and Methods}

A systematic review of techniques was carried out using the ~\href{https://www.sciencedirect.com/science/article/pii/S0360301617338403?via\%3Dihub\#bib14}{PRISMA
guidelines}.
PubMed, Scopus and Web of Science databases were searched from January 2014 to December 2020 using defined criteria (for more details, see Appendix~\ref{sec:appA}). Studies related to radiation therapy, either with photons or protons and attenuation correction for PET, were included when dealing with sCT generation from MRI, CBCT or PET. This review considered external beam radiation therapy, excluding, therefore, investigations that are focusing on brachytherapy. 
Conversion methods based on fundamental machine learning techniques were not considered in this review, preferring only deep learning-based approaches.
Also, the generation of dual-energy CT was not considered along with the direct estimation of corrected attenuation maps from PET. 
Finally, conference proceedings were excluded: proceedings can contain valid methodologies; however, the large number of relevant abstracts and incomplete report of information was considered not suitable for this review.  After the database search, duplicated articles were removed and records screened for eligibility. A citation search of the identified articles was performed. 

\begin{figure}[h]
  \centering
  \includegraphics[width=0.7\textwidth]{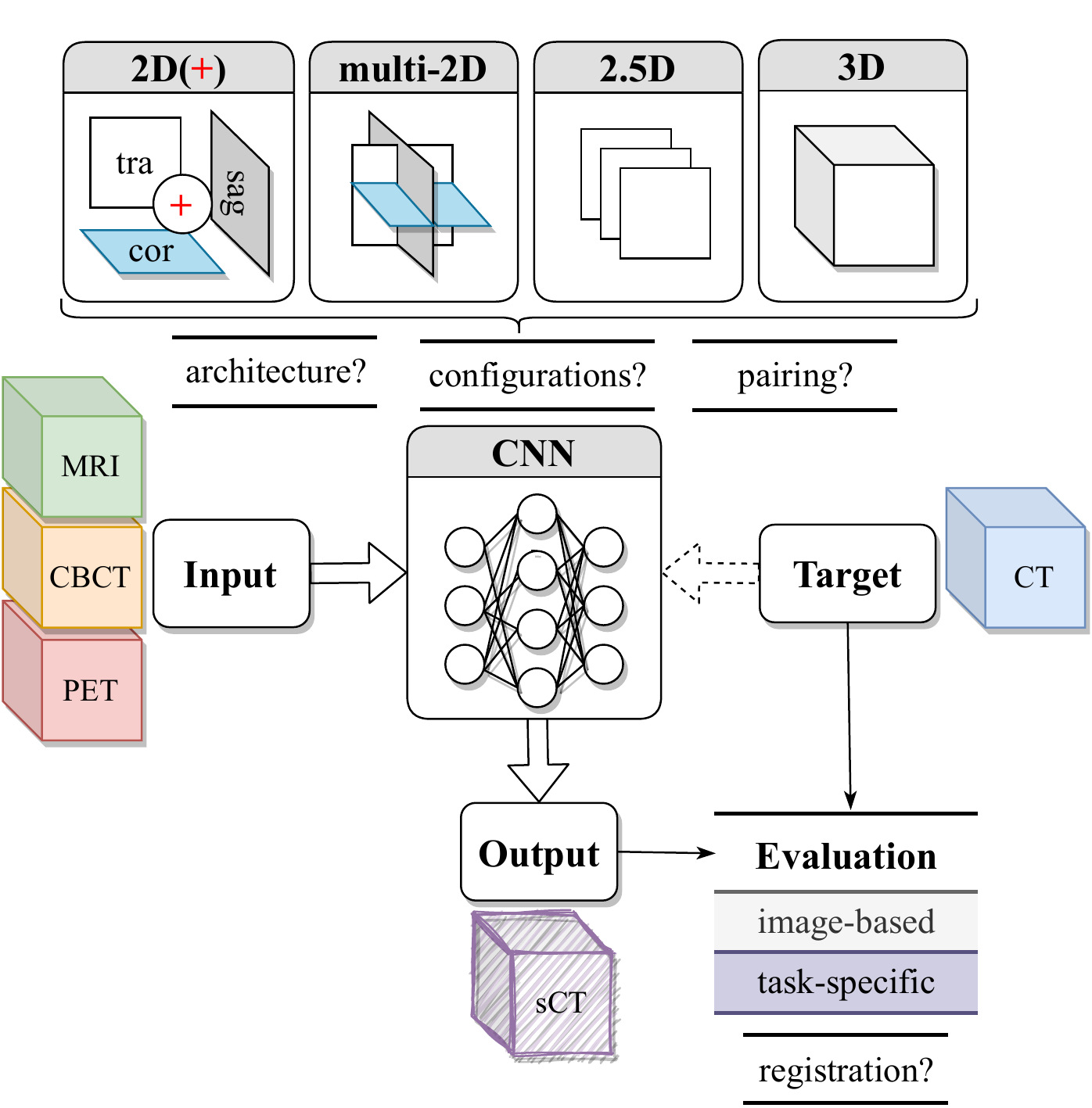}
  \caption{\textbf{Schematic of deep learning-based sCT generation study.} The input images/volumes, either being MRI (green), CBCT (yellow) or PET (red), are converted by a Convolutional Neural Network (CNN) into sCT. The CNN is trained to generate sCT similar to the target CT (blue). Several choices can be made in terms of network architecture, configuration, data pairing. After the sCT generation, the output image/volume is evaluated with image- and task-specific metrics.}
   \label{fig:schematic}\vspace{10pt}
\end{figure}

Each included study was assigned to a clinical application category. The selected categories were:
\begin{enumerate}[label=\textbf{\Roman*}]
\item \textbf{MR-only RT};
\item \textbf{CBCT-to-CT for image-guided (adaptive) radiotherapy};
\item \textbf{PET attenuation correction}.
\end{enumerate}
For each category, an overview of the methods was constructed in the form of tables\footnote{The tables presented in this review have been made publicly accessible at \url{https://matteomaspero.github.io/overview_sct}.}.
The tables were composed by capturing salient information of DL-based sCT generation approaches, which has been schematically depicted in Figure~\ref{fig:schematic}. 
\\
Independent of the input image, i.e. MRI, CBCT or PET, the chosen architecture (CNN) can be trained with paired or unpaired input data and different configurations. In this review, we define the following configurations: 2D (single slice, 2D, or patch, 2Dp) when training was performed considering transverse (tra), sagittal (sag) or coronal (cor) images; 2D+ when independently trained 2D networks in different views were combined during of after inference; multi-2D (m2D, also known as multi-plane) when slices from different views, e.g. transverse, sagittal and coronal, were provided to the same network; 2.5D when training was performed with neighbouring slices which were provided to multiple input channels of one network;  3D when volumes were considered as input (the whole volume, 3D, or patches, 3Dp). 
The architectures generally considered are introduced in the next section (\ref{DL_arch}).
The sCTs are generated inferring on an independent test set the trained network or combining an ensemble (ens) of trained networks. Finally, the quality of the sCT can be evaluated with image-based or task-specific metrics (\ref{metric}).

For each of the sCT generation category, we compiled tables providing a summary of the published techniques, including the key findings of each study and other pertinent factors, here indicated: the anatomic site investigated; the number of patients included; relevant information about the imaging protocol; DL architecture, the configuration chosen to sample the patient volume (2D or 2D+ or m2D, 2.5D or 3D); using paired/unpaired data during the network training;  
the radiation treatment adopted, where appropriate, along with the most popular metrics used to evaluate the quality of sCT (see \ref{metric}).

The year of publication for each category was noted according to the date of the first online appearance. 
Statistics in terms of popularity of the mentioned fields were calculated with pie charts for each category. Specifically, we subdivided the papers according to the anatomical region they dealt with: abdomen, brain, head \& neck (H\&N), thorax, pelvis and whole body; where available, tumour site was also reported.
A discussion of the clinical feasibility of each methodology and observed trends follows. 

The most common network architectures and metrics will be introduced in the following sections to facilitate the tables' interpretation. 

\subsection{Deep learning for image synthesis}\label{DL_arch}
Medical image synthesis can be formulated as an image-to-image translation problem, where a model that maps input image (A) to a target image (B) has to be found \cite{Yu2020}. Among all the possible strategies, DL methods have dramatically improved state of the art \cite{Wang2020review}.
\begin{figure}[b!]
  \centering
  \includegraphics[width=0.8\textwidth]{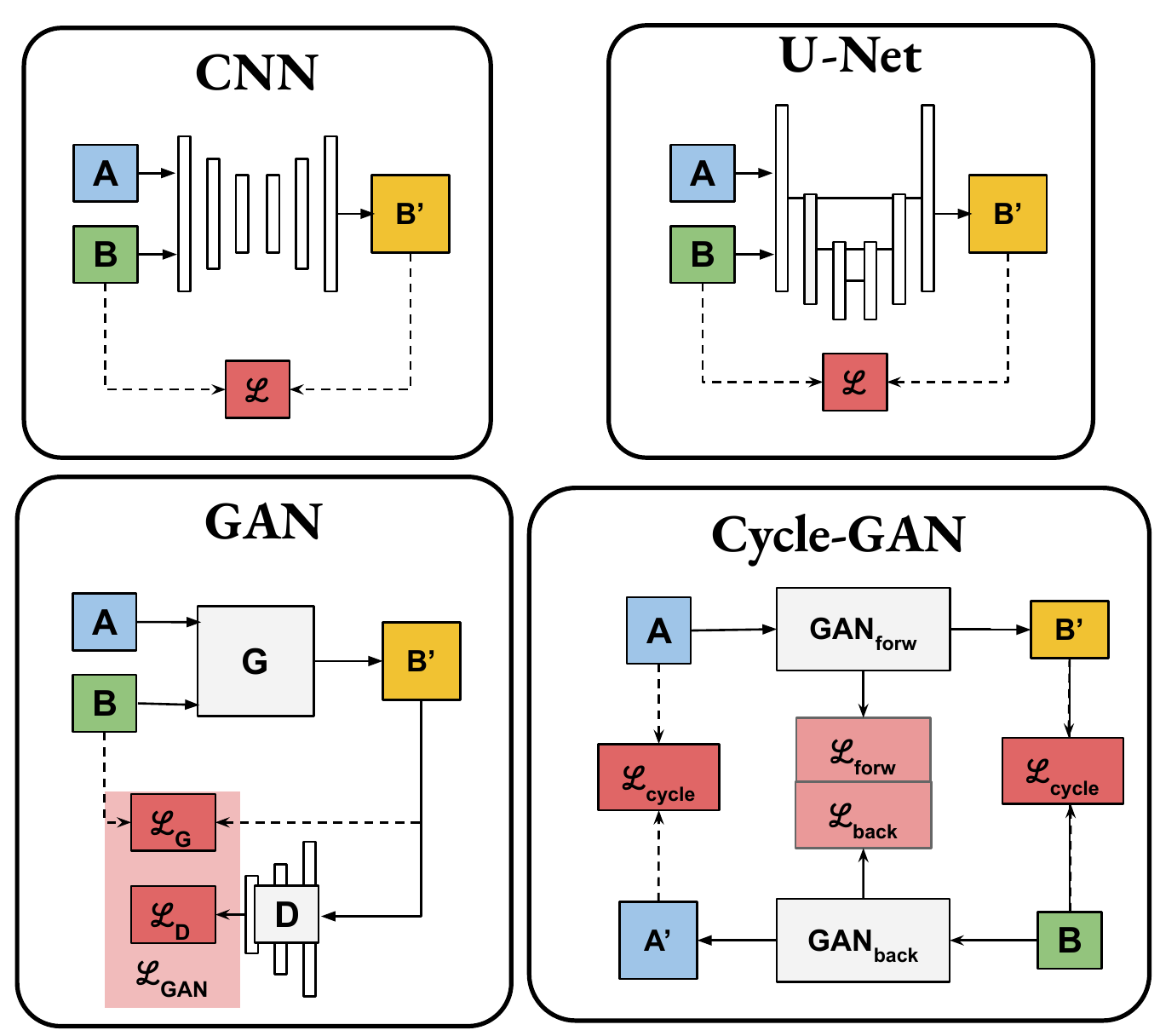}
  \caption{\textbf{Deep learning architectures used for image-to-image translation}. In the most straightforward configurations (CNN and U-Net, top left and right, respectively), a single loss function between input and output images is computed. GANs (bottom) use more than one CNN and loss to train the generator's performance (G). Cycle-GANs enable unsupervised learning by employing multiple GANs and cycle-consistency losses ($\mathcal{L}_{\textrm{\tiny cycle}}$).}
   \label{fig:DL_archs}
\end{figure}
DL approaches mainly used to synthesise sCT belong to the class of CNNs, where convolutional filters are combined through weights (also called parameters) learned during training. The depth is provided by using multiple layers of filters~\cite{Lecun2015}. The training is regulated by finding the "optimal" model parameters according to the search criterion defined by a loss function ($\mathcal{L}$).
Many CNN-based architectures have been proposed for image synthesis, with the most popular being the U-nets\cite{ronneberger2015u} and generative adversarial networks (GANs)\cite{goodfellow2014generative} (see Figure~\ref{fig:DL_archs}). 
U-net presents an encoding and a decoding path with additional skip connections to extract and reconstruct image features, thus learning to go from domain A to B. In the most simple GAN architecture, two networks are competing. A generator (G) that is trained to obtain synthetic images (B$'$) similar to the input set ($\mathcal{L}_{\textrm{\tiny G}}$), and a discriminator (D) that is trained to classify whether B$'$ is real or fake ($\mathcal{L}_{\textrm{\tiny D}}$), improving G's performances.\\
GANs learn a loss that combines both the tasks resulting in realistic images \cite{Isola2017}.
Given these premises, many variants of GANs can be arranged, with U-nets being employed as a possible generator in the GAN framework. We will not detail all possible configurations since it is not the scope of this review, and we address the interested reader to~\cite{wu2017survey,creswell2018generative,yi2019generative}. 
A particular derivation of GAN, called cycle-consistent GAN (cycle-GAN), is worth mentioning. Cycle-GANs opened the era of unpaired image-to-image translation~\cite{zhu2017unpaired}. Here, two GANs are trained, one going from A to B$'$, called forward pass (forw), and the second going from B$'$ to A, called backwards pass (back), are adopted with their related loss terms (Figure~\ref{fig:DL_archs} bottom right). Two consistency losses $\mathcal{L}_{c}$ are introduced, aiming at minimising differences between A and A$'$ and B and B$'$, enabling unpaired training.

\subsection{Metrics}
\label{metric}

An overview of the metrics used to assess and compare the reviewed publications' performances is summarised in Table~\ref{tab:metric_table}, subdivided in image similarity, geometric accuracy and task-specific as suggested in~\cite{Paganelli2018pati}.
\begin{table}[!hb]
\vspace{10pt}
\captionv{22}{Metrics reported}{\textbf{Overview of the most popular metrics} reported in the literature subdivi-\\ded into image similarity, geometric accuracy, task-specific metrics, and category.
\label{tab:metric_table}}
\vspace{-20pt}
\begin{adjustbox}{width=\columnwidth,center}
\begin{tabular}{|cc|c|}
\hline
         \multicolumn{2}{|c|}{\textbf{Category}}                         & \textbf{Metric} \\ \hline
\multicolumn{2}{|c|}{\multirow{11}{*}{\textbf{\begin{tabular}[c]{@{}c@{}}Image\\ similarity\end{tabular}}}} & \multirow{2}{*}{$\textrm{M\textcolor{red}{(A)}E} = \frac{\sum_{1}^{n} \textcolor{red}{|} \textrm{CT}_i - \textrm{sCT}_i  \textcolor{red}{|}}{n}$\small{, with $n$=voxel number in ROI;}  }     \\    &  &            \\ \cline{3-3} 
                        \multicolumn{2}{|c|}{\ }      & \multirow{2}{*}{$\textrm{\textcolor{red}{(R)}MSE} = {\color{red}(\sqrt{})} \frac{\sum_{1}^{n}(\textrm{CT}_i - \textrm{sCT}_i)^2}{n}$}              \\   & &       \\ \cline{3-3} 
                          \multicolumn{2}{|c|}{\ }      & \multirow{2}{*}{$\textrm{PSNR} = 10 log(\frac{MAX_{CT}^2}{\textrm{MSE}})$}                \\   & &     \\ \cline{3-3} 
                            \multicolumn{2}{|c|}{\ }      & \begin{tabular}[c]{@{}c@{}}$\textrm{SSIM} = \frac{(2\mu_{sCT}\mu_{CT}+c_1)(2\sigma_{sCT,rCT}+c_2)}{(\mu_{sCT}^{2}+\mu_{CT}^{2}+c_1)(\sigma_{sCT}^{2}+\sigma_{CT}^{2}+c_2)}$\\ with\\ $c_1=(k_1L)^2, c_2=(k_2L)^2$ \\ $\mu=$ mean, $\sigma=$ variance/covariance \\$L=$ dynamic range, $k_1=0.01$ and $k_2=0.03$ \end{tabular}   \\       \hline
\multicolumn{2}{|c|}{\textbf{\begin{tabular}[c]{@{}c@{}}Geometry\\ accuracy\end{tabular}}}              & $\textrm{DSC}(\textrm{Seg}_{CT}, \textrm{Seg}_{sCT}) = 2 \frac{\textrm{Seg}_{sCT} \cap \textrm{Seg}_{CT}}{\textrm{Seg}_{sCT}+\textrm{Seg}_{CT}}$      \\   \hline
\multirow{9}{*}{\textbf{\begin{tabular}[c]{@{}c@{}}Task \\specific\end{tabular}}}& \multirow{3}{*}{\textbf{MR-only $\&$}}    & \multirow{2}{*}{$\textrm{DD}=100 \cdot \frac{\textrm{D}_{sCT}-\textrm{D}_{CT}}{\textrm{D}_{CT}}\%$\small{, with $D$=dose;}} \\   & &      \\ \cline{3-3}
& \multirow{3}{*}{\textbf{CBCT-to-CT}}    &  DPR = \% of voxel with $DD<x$\% in an ROI \\ \cline{3-3}
 & & GPR$=$\% of voxel with $\gamma<1$ in an ROI \\ \cline{3-3}
  & & \begin{tabular}[c]{@{}c@{}}DVH$=$difference of specific points in\\ dose-volume histogram plot\end{tabular} \\ \cline{2-3}

& \multirow{2}{*}{\textbf{\begin{tabular}[c]{@{}c@{}}PET \\ reconstruction\end{tabular}}}    & \multirow{2}{*}{$\textrm{PET}_{\textcolor{red}{|}err\textcolor{red}{|}} = 100 \cdot \frac{\textcolor{red}{|}\textrm{PET}_{sCT}-\textrm{PET}_{CT}\textcolor{red}{|}}{\textrm{PET}_{CT} }\%$}        \\   & &     \\ 
             \hline
\end{tabular}
\end{adjustbox}
\end{table}

\textbf{Image similarity}
The most straightforward way to evaluate the quality of the sCT is to calculate the similarity of the sCT to the ground truth/target CT on a voxel-wise basis.
The calculation of voxel-based image similarity metrics implies that sCT and CT are aligned by translation, rigid (rig), affine (aff) or deformable (def) registrations.
Widespread similarity metrics for this task are reported in Table~\ref{tab:metric_table} and include: mean (absolute) error (M(A)E), sometimes referred to as mean absolute prediction error (MAPE), peak signal-to-noise ratio (PSNR) and structural similarity index measure (SSIM). 
Other less common metrics are cross-correlation (CC) and normalised cross-correlation (NCC), along with the (root) mean squared error ((R)MSE). 

M(A)E and (R)MSE are relatively easy to compute as the average of the (absolute) difference and difference in quadrature over a defined region of interest. For both the metrics, lower values indicate better prediction accuracy for sCT. MAE and ME are often computed together to represent the random and systematic error, respectively. MSE and RMSE are used to give more weight to higher errors, thus understanding the impact of possible outliers. PSNR is the ratio between the maximum in an image and the intensity of the corrupting noise affecting the fidelity of its representation, calculated as MSE. PSNR evaluates the noise introduced in the CT synthesis relatively to the ground truth CT.  
SSIM is a more sophisticated metric developed to take advantage of the known characteristics of the human visual system~\cite{Wang2004} perceiving the loss of image structure due to variations in lighting.

\textbf{Geometric accuracy}
Along with voxel-based metrics, the geometric accuracy of the generated sCT can also be assessed by comparing corresponding segmented structures on CT and sCT, e.g. bones, fat, muscle, air and body.
The segmentation can be performed manually but can also be automatic. In this context, the delineations are found after applying a threshold to CT and sCT and, if necessary, morphological operations on the obtained binary masks.
The metrics for geometric accuracy are, therefore, generally the same used for a segmentation task.
For example, the Dice similarity coefficient (DSC)~\cite{Dice1945} is a common metric that assesses the accuracy of depicting specific tissue classes/structures. DSC is twice the ratio between the correctly classified voxel and all the voxels in the mask from CT and sCT ($\textrm{Seg}_{CT}$ and $\textrm{Seg}_{sCT}$).
Additionally, metrics generally used to estimate the distance among segmentations can also be adopted as the Hausdorff distance (HD)~\cite{Huttenlocher1993} or mean absolute surface distance, which measures two sets of contours' maximum and average distance, respectively.
Even if segmentation-based metrics are common, choosing the right metric for the specific task is a non-trivial task, as recently highlighted by Reinke et al.~\cite{Reinke2021} and should be assessed on an application basis.

Other image-based metrics can be subdivided according to the application and presented in the following sections' appropriate sub-category.

\textbf{Task-specific metrics}
In MR-only RT and CBCT-to-CT for adaptive RT, dose calculation accuracy on sCT is generally compared to CT-based in specific ROIs for dose calculations performed either for photon ($x$) and proton ($p$) RT.

The most common voxelwise-based metric is the dose difference (DD), calculated as the average dose (D$_{\textrm{CT}}$ D$_{\textrm{sCT}}$) in ROIs as the whole body, target or other structures of interest. The dose difference can be expressed as an absolute value (Gy) or relative (\%), either to the prescribed dose, the maximum dose or the voxel-wise reference dose.
The dose pass rate (DPR) is directly correlated to DD, and it is calculated as the percentage of voxels with DD$<$ than a set threshold.

Gamma ($\gamma$) analysis allows combining dose and spatial criteria~\cite{Low2010}, and it can be performed either in 2D or 3D. Several parameters need to be set to perform $\gamma$-analysis, including dose criteria, distance-to-agreement criteria, local or global
analysis, and dose threshold. Interpretation and comparison between studies of gamma index results are challenging since they depend on the chosen parameters, dose grid size, and voxel resolution~\cite{clasie2012numerical, Hussein2017}. Results of $\gamma$-analysis are generally expressed as gamma pass rate (GPR), counting the percentage of voxels with $\gamma<1$ or the mean $\gamma$ in an ROI generally defined based on a threshold of the reference dose distribution.
  
Dose-volume histograms (DVHs) are one of the most diffused tools in the clinical routine~\cite{Drzymala1991}. DVH summarises 3D dose distributions in a graphical 2D format offering no spatial information.

For the evaluation of sCT, generally, the differences among clinically relevant DVH points is reported.

In proton RT, range shift (RS) analysis is also performed. Here, the ideal range (known as the prescribed range) is defined as the depth at which the dose has decreased to 80\% of the maximum dose, on the distal dose fall-off ($R_{\textrm{80}}$) \cite{paganetti2012range}. RS error (RSe) can be defined both as the absolute difference between the prescribed and the actual range ($RSe=R_{80CT}-R_{80sCT}$) and as relative RS (\%RS) error, expressed as the shift in \% relative to the prescribed range, along the beam direction \cite{pileggi2018proton}

\begin{equation}
\%RS = \left | \frac{R_{80CT}-R_{80sCT}}{R_{80CT}} \right | \times 100
\label{eq:RS}
\end{equation}

For sCT for PET attenuation correction, the relative error (signed $\textrm{PET}_{err}$ and unsigned $\textrm{PET}_{|err|}$) of  PET reconstruction is usually reported along with the difference in standard uptake values (SUV).

Please note that even if two papers calculate the same metric, differences could occur in the ROI where the metrics are calculated, making challenging performance comparisons. For example, MAE can be computed on the whole predicted volume, in a volume of interest or a cropped volume. In addition to that, the implementation of the metric computation can change.
In gamma analysis, for example, different dose difference and distance to agreement criteria can be stated ($\gamma_{3\%,3\textrm{mm}}$ ($\gamma_3$), $\gamma_{2\%,2\textrm{mm}}$ ($\gamma_2$) and $\gamma_{1\%,1\textrm{mm}}$ ($\gamma_1$)). Moreover, it can be calculated on ROI obtained from different dose thresholds and 2D or 3D algorithms.
In the following sections, we will highlight the possible differences speculating on the impact.

\section{Results}
\begin{figure}[h!]
  \centering
  \includegraphics[width=0.7\textwidth]{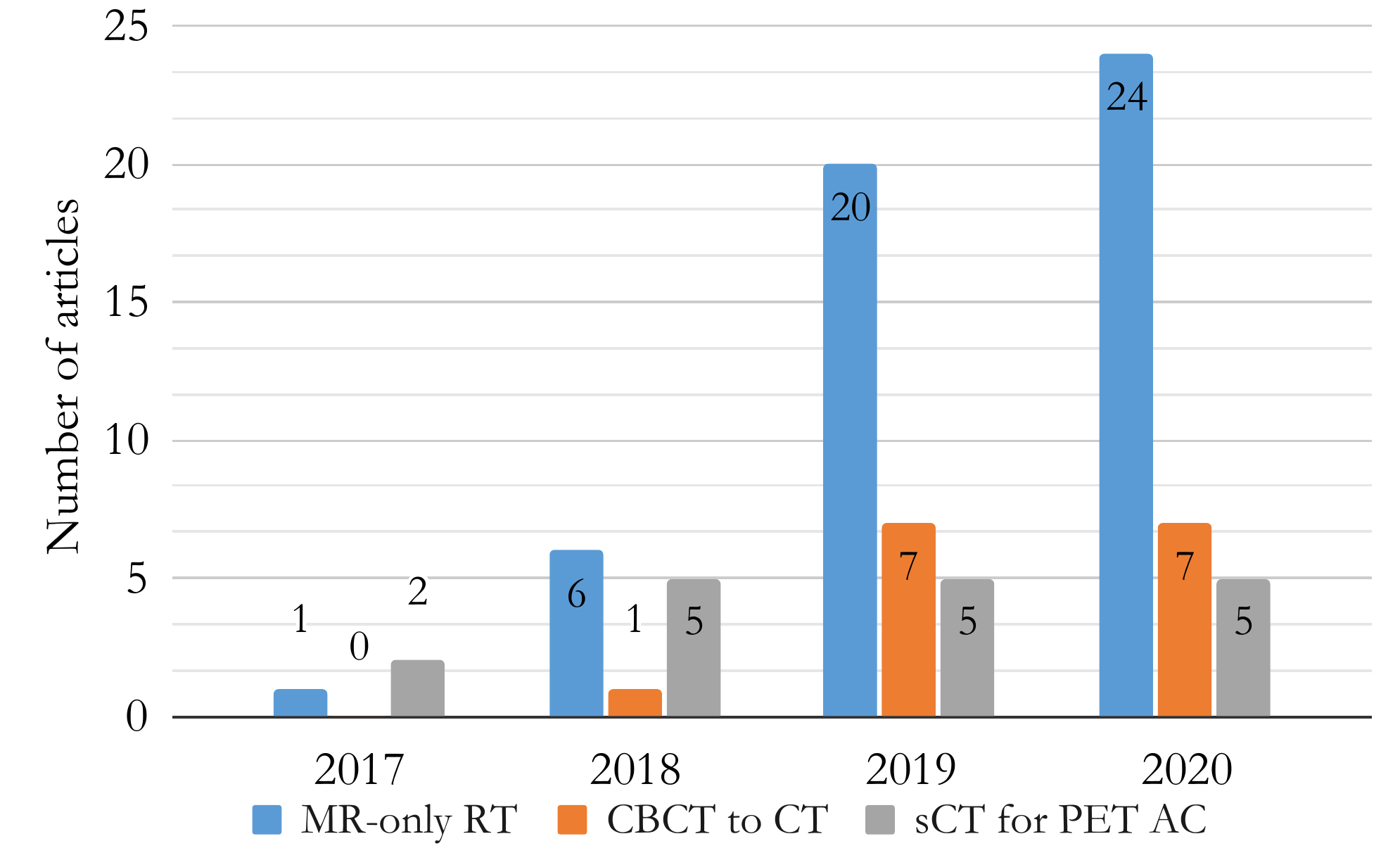}
  \includegraphics[width=0.9\textwidth]{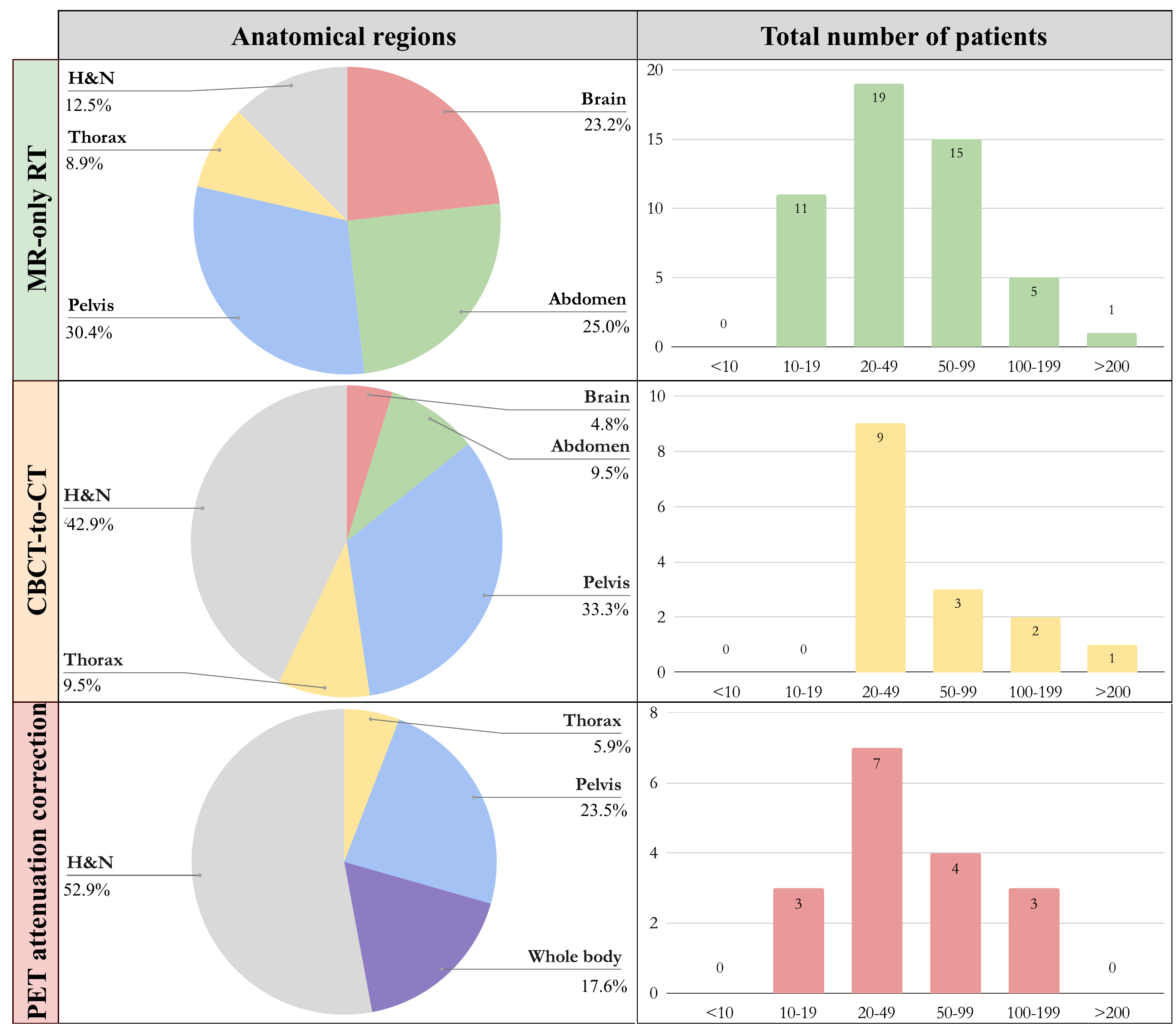}
  \caption{(Top) Number of published articles grouped by application and year; (middle) pie charts of the anatomical regions investigated for each application; (bottom) bar plot of the publications binned per the total number of patients included in the study.}
   \label{fig:chart}
\end{figure}

Database searching led to 91 records on PubMed, 98 on Scopus and 218 on Web of Science. After duplicates removal and content check, 83 eligible papers were found.\\
Figure~\ref{fig:chart} summarises the number of articles published by year, grouped in $51$ ($61.4\%$), 15 ($18.1\%$) and 17 ($20.5\%$) for MR-only RT (category I), CBCT-to-CT for adaptive RT (category II), and sCT for PET attenuation correction (category III), respectively. The first conference paper appeared in 2016~\cite{nie2016estimating}.
Given that we excluded conference papers from our search, we found that the first work was published in 2017. In general, the number of articles increased over the years, except for CBCT-to-CT and sCT for PET attenuation correction, which was stable in the last years. 
Figure~\ref{fig:chart} shows that the brain, pelvis and H$\&$N were the most popular anatomical regions investigated in DL-based sCT for MR-only RT, covering $\sim$80\% of the studies. For CBCT-to-CT, H$\&$N and pelvic regions were the most explored sites, being present in $>$75\% of the works. Finally, for PET AC, H$\&$N was investigated in the majority of the studies, followed by the pelvic region. Together, they covered $>$75\% of the publications.

The total number of patients included in the analysis was variable, but most studies dealt with  less than 50 patients for all three categories. The largest patient cohorts included 402~\cite{Andres2020} (I), 328~\cite{eckl2020evaluation} (II) and 193 patients~\cite{Peng2020} (I), while the smallest studies included 10 patients~\cite{Qian2020} and another 10 volunteers~\cite{Xu2019multi}(I).

Most papers enrolled adult patients. Paediatric (paed) patients represent a more heterogeneous dataset for network training, and its feasibility has been investigated first for  attenuation correction in PET \cite{ladefoged2019deep} (79 patients) and more recently for photon and proton RT\cite{Maspero2020,Florkow2020dose}.

All the models were trained to perform a regression task from the input to sCT, except for two studies where networks were trained to segment the input image into a pre-defined number of classes, thus performing a segmentation task \cite{Jeon2019,bradshaw2018feasibility}.

In most of the works, training was implemented in a paired manner, with unpaired training investigated in 13/83 articles.
Four studies compared paired against unpaired~\cite{Fu2020,Peng2020,Li2020comp,Xu2020}.
Over all the three categories, 2D networks were the most common adopted. Specifically, 2D networks were used about 61\% of the times, 2D+ 6\%, 2.5D 10\%, and 3D configuration 24\%. In some studies, multiple configurations were investigated, for example \cite{Fu2019,Neppl2019,Fu2020}. 
GANs were the most popular architectures (45-times), followed by U-nets (36) and other CNNs. 
Note that U-nest may be employed as generator of GANs, and that in this case, the architecture was categoraised as GAN. 

All the investigations employed registration between sCT and CT to evaluate the quality of the sCT, except for Xu et al. \cite{Xu2020} and Fetty et al. \cite{Fetty2020}, where metrics were defined to assess the quality of the sCT in an unpaired manner, e.g. Frechet inception distance (FID).

Main findings are reported in Table \ref{tab:MRonly} for studies on sCT for MR-only RT without dosimetric evaluations, in Table~\ref{tab:MRonlyDose1}, \ref{tab:MRonlyDose2} for studies on sCT for MR-only RT with dosimetric evaluations, in Table~\ref{tab:CBCT} for studies on CBCT-to-CT for IGART, and in Table~\ref{tab:PET} for studies on PET attenuation correction. Tables are organised by anatomical site and tumour location where available. Studies investigating the independent training and testing of several anatomical regions are reported for each specific site \cite{Xu2020,Xiang2018,Cusumano2020,eckl2020evaluation, harms2019paired}. Works using the same network to train or test data from different scanners and anatomy are reported at the bottom of the table \cite{maspero2020single,zhang2020improving}.
Detailed results based on these tables are presented in the following sections subdivided for each category.

\subsection{MR-only radiotherapy}

The first work ever published in this category, and in among all the categories, was by Han in 2017, where he proposed to use a paired U-net for brain sCT generation. After one year, the first work published with a dosimetric evaluation was presented by Maspero et al.~\cite{Maspero2018}, investigating a 2D paired GAN trained on prostate patients and evaluated on prostate, rectal and cervical cancer patients.

\begin{sidewaystable}[]
\captionv{22}{MR-only RT only image evaluation}{Overview sCT methods for MR-only radiotherapy with sole image-based evaluation.
\label{tab:MRonly}
}
\vspace{-20pt}
\begin{adjustbox}{width=\columnwidth,center}
\begin{tabular}{|cc|c|c|c|c|c|c|c|c|c|c|c|c|c|c|}
\hline
 \multicolumn{2}{|c|}{\multirow{2}{*}{\textbf{Tumor}}} & \multicolumn{4}{c|}{\textbf{Patients}}            & \multicolumn{2}{c|}{\textbf{MRI}}       & \multicolumn{2}{c|}{\textbf{DL method}}         & \multirow{3}{*}{\textbf{Reg}} & \multicolumn{4}{c|}{\textbf{Image-similarity}}                 & \multirow{3}{*}{\textbf{Reference}}     \\ \cline{3-10} \cline{12-15} 

  \multicolumn{2}{|c|}{\multirow{2}{*}{\textbf{site}} }  & \multirow{2}{*}{\textbf{train}} & \multirow{2}{*}{\textbf{val}} & \multirow{2}{*}{\textbf{test}} & \multirow{2}{*}{\textbf{x-fold}} & \textbf{field} & \multirow{2}{*}{\textbf{sequence}} & \multirow{2}{*}{\textbf{conf}} & \multirow{2}{*}{\textbf{arch}} &  & \textbf{MAE} & \textbf{PSNR} & \multirow{2}{*}{\textbf{SSIM}} & \multirow{2}{*}{\textbf{others}} &            \\ 
 & & & & &  & [T] & & & & & [HU] & [dB] & & &   \\ \hline

\parbox[t]{2mm}{\multirow{2}{*}{\rotatebox[origin=c]{90}{\textbf{Abd}}}}& \cellcolor{green!25} Abdomen &	10$^{v}$	& 	&	10	&	LoO	&	n.a.	&	mDixon &	2D pair	&	GAN$^*$ &	def	&	61$\pm$3	&		&		&	CC &  Xu2019\cite{Xu2019multi} \\ \cline{2-16}
& \cellcolor{green!25}Abdomen&  160 & & & LoO & n.a. & n.a. & 2D pair & GAN$^*$ & rig & 5.1$\pm$0.5& & .90$\pm$.43 & (F/M)SIM IS ... & Xu2020\cite{Xu2020}  \\ \hline
\hline 

\parbox[t]{2mm}{\multirow{13}{*}{\rotatebox[origin=c]{90}{\textbf{Brain}}}}&\cellcolor{red!25} Brain	&	18	&		&		&	6x	&	1.5 &	3D T1 GRE		&	2D pair	&	U-net		&	rig	&	85$\pm$17	&		&		&	MSE, ME	& Han2017\cite{Han2017}	\\ \cline{2-16}

& \cellcolor{red!25}Brain	&	16	&		&		&	LoO &	n.a.	&	T1	&	2.5Dp pair 	&		CNN+ 	&	rig	&	85$\pm$9	&	27.3$\pm$1.1	&		& & Xiang2018\cite{Xiang2018}	\\    \cline{2-16}

& \cellcolor{red!25}Brain	&	15	&		&		&	5x	&	1.0	&	T1 Gd	& 2D pair & \begin{tabular}[c]{@{}c@{}}CNN\\	GAN\end{tabular}	&	def	&	\begin{tabular}[c]{@{}c@{}}102$\pm$11\\89$\pm$10\end{tabular}	&\begin{tabular}[c]{@{}c@{}}25.4$\pm$1.1\\	26.6$\pm$1.2\end{tabular}	&	\begin{tabular}[c]{@{}c@{}}.79$\pm$.03\\.83$\pm$.03\end{tabular}	&	tissues	& 	Emami2018\cite{Emami2018} \\    \cline{2-16}

&\cellcolor{red!25}Brain	&	\begin{tabular}[c]{@{}c@{}}98CT \\84MR\end{tabular}	&		&	10	&		&	3	&	3D T2	&	\begin{tabular}[c]{@{}c@{}}2D\\ pair/unp$^*$\end{tabular} &		GAN	&	aff	&	19$\pm$3	&	65.4$\pm$0.9	&	.25$\pm$.01	&		& Jin2019\cite{Jin2019} \\ \cline{2-16}
  
&\cellcolor{red!25}Brain	&	24	&		&		&	LoO	&	n.a.	&	T1  	&	3Dp pair	&	GAN$^*$	&	rig	&	56$\pm$9	&	26.6$\pm$2.3	&		&	NCC, HD body	& Lei2019\cite{Lei2019mri} \\ \cline{2-16}

&\cellcolor{red!25}Brain	&	33	&		&		&	LoO	&	n.a.	&	T1$^b$	&	2D unp	&	GAN$^*$ &	No	&	9.0$\pm$0.8	&		&	.75$\pm$0.77	& (F/M)SIM IS ...	& Xu2020\cite{Xu2020} \\ \cline{2-16}

&\cellcolor{red!25}Brain	&	28$^t$	&	2	&	15 &		&	1.5	&	n.a.	&	2D pair$^*$	&	GAN$^*$	&	aff	&	134$\pm$12 & 24.0$\pm$0.9	& .76$\pm$.02	&		&  Yang2020\cite{Yang2020} \\ \cline{2-16}

&\multirow{4}{*}{\cellcolor{red!25}}	&	\multirow{4}{*}{81}	&		&	\multirow{4}{*}{11}	&	\multirow{4}{*}{8x}	&	\multirow{4}{*}{1.5}	&	3D T1 GRE 	&	\multirow{4}{*}{2D pair}	&	\multirow{4}{*}{U-net} 	&	\multirow{4}{*}{aff} &
								 45.4$\pm$8.5 & 43.0$\pm$2.0	& .65$\pm$.05	&	metrics for air	&  \multirow{4}{*}{Massa2020\cite{Massa2020}} \\ 
&\cellcolor{red!25} Brain &  & & & & & 3D T1 GRE Gd &  & & &44.6$\pm$7.4 & 43.4$\pm$1.2	& .63$\pm$.03  & air, bones, & \\ 
&\cellcolor{red!25}  &  & & & & & 2D T2 SE &  & & &   45.7$\pm$8.8 & 43.4$\pm$1.2	& .64$\pm$.03 & soft tissues;  & \\ 
&\cellcolor{red!25} &  & & & & & 2D T2 FLAIR &  & & & 51.2$\pm$4.5 & 44.9$\pm$1.2	& .61$\pm$.04  & DSC bones & \\\cline{2-16}

&\multirow{2}{*}{\cellcolor{red!25}}Brain	&	\multirow{2}{*}{28}	&		&	\multirow{2}{*}{6}	&		&	\multirow{2}{*}{1.5}	&	\multirow{2}{*}{T2}	&	2D pair	&	U-net 	& \multirow{2}{*}{rig}	&	65$\pm$4	&	28.8$\pm$0.6	& .972$\pm$.004	& same metrics for &\multirow{2}{*}{Li2020\cite{Li2020comp}} \\  
& \cellcolor{red!25} & & & & & & & 2D unp & GAN &  & 94$\pm$6& 26.3$\pm$0.6 & .955$\pm$0.007 & synthetic MRI &  \\ \hline 
\hline

\parbox[t]{2mm}{\multirow{6}{*}{\rotatebox[origin=c]{90}{\textbf{Head \& Neck}}}}&\cellcolor{gray!25}Nasophar	&	23	&		&	10	&		&	1.5&	T2	&	2D pair	&	U-net	&	def	&	131$\pm$24	&		&		&	\begin{tabular}[c]{@{}c@{}}MAE ME\\ tissue/bone\end{tabular}  & Wang2019\cite{Wang2019} \\ \cline{2-16}

&\cellcolor{gray!25}H$\&$N	&	28	&	4	&		&	8x	&	1.5	&	2D T1$\pm$Gd, T2	&	2D pair	&	GAN	&	aff	&	76$\pm$15	&	29.1$\pm$1.6	&	.92$\pm$.02	&	DSC MAE bone	&  Tie2020\cite{Tie2020} \\ \cline{2-16} 

&\cellcolor{gray!25}H$\&$N	&	\multicolumn{2}{c|}{60}			&	30	&		&	3	&	T1	&	2D unp	&	GAN	&	n.a.	&	19.6$\pm$0.7	&	62.4$\pm$0.5	&	.78$\pm$0.2	&		&  Kearney2020\cite{Kearney2020} \\ \cline{2-16}

&\cellcolor{gray!25}H$\&$N	&	7	&		&	8	&	LoO	&	1.5	&	3D T1, T2	&	2D pair	&		GAN	&	def	&	83$\pm$49	&		&		&	ME	&	Largent2020\cite{Largent2020}\\ \cline{2-16}

&\cellcolor{gray!25}H$\&$N	&	10	&		&		&	LoO	&	1.5	&	3D T1, T2	&	2D pair	&	GAN$^*$	&	def	&	42-62	&		&		&	RMSE, CC	&  Qian2020\cite{Qian2020} \\ \cline{2-16} 

&\cellcolor{gray!25}H$\&$N	&	32	&		&	8	&	5x	&	3	&	3D UTE	&	2D pair	&	U-net	&	def	&	104$\pm$21	&		&		&	DSC, spatial corr	&  Su2020\cite{Su2020} \\ \hline
\hline

\parbox[t]{2mm}{\multirow{9}{*}{\rotatebox[origin=c]{90}{\textbf{Pelvis}}}}&\cellcolor{cyan!25}Prostate & 22 &		&		&	LoO &	n.a.	&	T1	&	2.5Dp pair 	&		CNN+ 	&	rig	& 43$\pm$3 & 33.5$\pm$0.8 &  & & Xiang2018\cite{Xiang2018} \\  \cline{2-16} 

&\cellcolor{cyan!25}Pelvis  & 20 & & & LoO & n.a. & 3D T2 & 3Dp pair & GAN$^*$ & rig & 51$\pm$16 & 24.5$\pm$2.6 &  & NCC, HD body & Lei2019\cite{Lei2019mri} \\  \cline{2-16}

&\cellcolor{cyan!25}Prostate	&	20	&		&		&	5x	&	1.5	&	2D T1 TSE	&	\begin{tabular}[c]{@{}c@{}}2D pair\\3Dp pair\end{tabular}&		U-net	&	def	&	\begin{tabular}[c]{@{}c@{}}41$\pm$5\\38$\pm$5\end{tabular}	&		&		&	DSC bone	& Fu2019\cite{Fu2019}\\  \cline{2-16} 
    
&\cellcolor{cyan!25}Pelvis human	&	27	&		&		&	\multirow{2}{*}{3x}	&	3	&	3D T1 GRE 	&	3Dp	&		\multirow{2}{*}{U-net}	&	\multirow{2}{*}{def}	&	32$\pm$8	&	36.5$\pm$1.6	&		&	MAE/DSC bone &  \multirow{2}{*}{Florkow2019\cite{Florkow2020}} \\ 
&\cellcolor{cyan!25}Pelvis canine & 18 & & & & 1.5 & mDixon$^a$ & pair  & & & 36$\pm$4 & 36.1$\pm$1.7 &  &surf dist$<$0.5 mm  & \\  \cline{2-16}

&\cellcolor{cyan!25}Pelvis	&	15	&		&	4	&	5x	&	3	&	3D T2	&	2D pair 	&	\begin{tabular}[c]{@{}c@{}}CNN\\U-net\end{tabular}		&	def	&	\begin{tabular}[c]{@{}c@{}}38$\pm$6 \\43$\pm$9\end{tabular}	&	\begin{tabular}[c]{@{}c@{}}29.5$\pm$1.2	\\28.2$\pm$1.6\end{tabular}&	\begin{tabular}[c]{@{}c@{}}.96$\pm$.01\\.95$\pm$.01 \end{tabular}	&	\multirow{2}{*}{ME, PCC}	&  Bahrami2020\cite{Bahrami2020} \\  \cline{2-16} 

&\cellcolor{cyan!25}Pelvis	&	100	&		&	 &		&	3	&	2D T2 FSE	&	2D unp	&	GAN	&	No &		&		&		&	FID	&  Fetty2020\cite{Fetty2020} \\ \hline \hline

\parbox[t]{2mm}{\rotatebox[origin=c]{90}{\textbf{Thor}}}&\cellcolor{yellow!25} \begin{tabular}[c]{@{}c@{}}Breast\\ \ \end{tabular}&	14	&		&	2	&	LoO	&	n.a. &	n.a.  &	2D pair	&	U-net$^1$	&	def	&	&	 &	 & DSC .74-.76 	&  Jeon2019\cite{Jeon2019} \\ \hline

\end{tabular}
\end{adjustbox}
{\ \\ \scriptsize $^{v}$volunteers, not patients; $^1$to segment CT into 5-classes; $^a$multiple combinations of Dixon images was investigated but omitted here; $^b$dataset from \url{http://www.med.harvard.edu/AANLIB/}; $^t$ robustness to training size was investigated. Abbreviations: val=validation, x-fold=cross-fold, conf=configuration, arch=architecture, GRE=gradient echo, (T)SE=(turbo) spin-echo, mDixon = multi-contrast Dixon reconstruction, LoO=leave-one-out, (R)MSE=(root) meas squared error, ME=mean error, DSC=Dice similarity coefficient, (N)CC=normalized cross correlation.  }
\end{sidewaystable}


\setcounter{table}{2}
\renewcommand{\thetable}{\arabic{table}a}

\begin{sidewaystable}[]
\captionv{22}{MR-only sCT with dose}{a. Overview sCT methods for MR-only radiotherapy with image-based and dose evaluation.
\label{tab:MRonlyDose1}
}
\vspace{-20pt}
\begin{adjustbox}{width=\columnwidth,center}
\begin{tabular}{|cc|c|c|c|c|c|c|c|c|c|c|c|c|c|c|c|c|c|c|}
\hline
\multicolumn{2}{|c|}{\multirow{2}{*}{\textbf{Tumor}}} & \multicolumn{4}{c|}{\textbf{Patients}}            & \multicolumn{2}{c|}{\textbf{MRI}}       & \multicolumn{2}{c|}{\textbf{DL method}}         & \multirow{3}{*}{\textbf{Reg}} & \multicolumn{3}{c|}{\textbf{Image-similarity}}                 & \multirow{3}{*}{\textbf{Plan}} & \multicolumn{4}{c|}{\textbf{Dose}}  & \multirow{3}{*}{\textbf{Reference}}     \\ \cline{3-10} \cline{12-14} \cline{16-19}  

 \multicolumn{2}{|c|}{\multirow{2}{*}{\textbf{site}}}   & \multirow{2}{*}{\textbf{train}} & \multirow{2}{*}{\textbf{val}} & \multirow{2}{*}{\textbf{test}} & \textbf{x-} & \textbf{field} & \multirow{2}{*}{\textbf{sequence}} & \multirow{2}{*}{\textbf{conf}} & \multirow{2}{*}{\textbf{arch}} &  & \textbf{MAE} & \textbf{PSNR}  & \multirow{2}{*}{\textbf{others}} &               & \textbf{DD} & \textbf{GPR} & \multirow{2}{*}{\textbf{DVH}} & \multirow{2}{*}{\textbf{others}} &        \\ 
 
& & & & & \textbf{fold} & [T] & &  & & & [HU] & [dB]  & & &[\%]& [\%] & & &  \\ \hline

\parbox[t]{2mm}{\multirow{10}{*}{\rotatebox[origin=c]{90}{\textbf{Abdomen}}}}& \cellcolor{green!25}Liver	&	\multirow{2}{*}{21}	&		&		&	\multirow{2}{*}{LoO}	&	\multirow{2}{*}{3}	& 3D T1	&	3Dp  & \multirow{2}{*}{GAN}	&	\multirow{2}{*}{def}	&	\multirow{2}{*}{73$\pm$18}	&	\multirow{2}{*}{22.7$\pm$3.6}	&		\multirow{2}{*}{NCC} &
	\multirow{2}{*}{$p$}	& & \multirow{2}{*}{99.4$\pm$1.0$^3$} & \multirow{2}{*}{$<$1\%}  & range & 
\multirow{2}{*}{LiuY2019\cite{Liu2019}} \\ 
&\cellcolor{green!25} &  & &  &   &  & GRE & pair &  &  &  &		&	 & 
	&  &  &   &  $\gamma_{2}$ $\gamma_{1}$ &  \\ \cline{2-20}  

&\cellcolor{green!25}Abdomen	&	\multirow{2}{*}{12}	&		 	\multirow{2}{*}{}	& 	 &	\multirow{2}{*}{4x}	& 0.3 &  \multirow{2}{*}{GRE}  &	2D pair & \multirow{2}{*}{GAN$^*$}	&	\multirow{2}{*}{def}	& 90$\pm$19 & 27.4$\pm$1.6 &  &
	x	& $<\pm$0.6 & 98.7$\pm$1.5$^2$  & $<\pm$0.15  &  \multirow{2}{*}{$\gamma_{3}$}  & \multirow{2}{*}{Fu2020\cite{Fu2020}} \\ 
&\cellcolor{green!25} &  & &  &   & 1.5 &  & 2D unp   &  &             & 94$\pm$30   &27.2$\pm$2.2	& 	  & +B$_0$
 & 	$<\pm$0.6 & 98.5$\pm$1.6$^2$ &    &  &  \\ \cline{2-20}
 
& \cellcolor{green!25} Abdomen	&	\multirow{2}{*}{46}	&		&  	\multirow{2}{*}{31}	& \multirow{2}{*}{3x}	 &	\multirow{2}{*}{3}	& 3D T1   &	\multirow{2}{*}{2.5D pair} & \multirow{2}{*}{U-net}	&	syn	& \multirow{2}{*}{79$\pm$18} &  &  MAE ME &
	\multirow{2}{*}{x}	&  &  & \multirow{2}{*}{$<$2Gy}  &   & 
\multirow{2}{*}{Liu2020\cite{Liu2020}} \\ 
&\cellcolor{green!25} &  & &  &   &  & GRE &  &   & rig &    &	& organs	  & 
 & 	 & &    &  &  \\ \cline{2-20}
 
& \cellcolor{green!25}Abdomen	&	39	&		&  	19	&	 &	0.35	& GRE   &	2D pair & U-net	&	def	& 79$\pm$18 &  &  ME &
	x+B$_0$	& $<$0.1 & 98.7$\pm$1.1$^2$ & $<$2.5\%  & $\gamma_{3}$  $\gamma_{1}$  & 
Cusumano2020\cite{Cusumano2020} \\ \cline{2-20}

&\cellcolor{green!25}Abdomen	&	\multirow{2}{*}{54}	&	\multirow{2}{*}{18}	&  \multirow{2}{*}{12}		&	 \multirow{2}{*}{3x} &	1.5	& 3D T1   &	3Dp & \multirow{2}{*}{U-net}	&	\multirow{2}{*}{def}	& \multirow{2}{*}{62$\pm$13} & \multirow{2}{*}{30.0$\pm$1.8}	&  ME, DSC &
	x	& $<$0.1 & 99.7$\pm$0.3$^2$ & $<$2\%  & beam  & 
\multirow{2}{*}{Florkow2020\cite{Florkow2020dose}} \\ 
&\cellcolor{green!25}paed &  & &  &   & 3 & GRE, T2 TSE & pair &    &  &  &	& tissues	  & $p$
 & 	$<$0.5 & 96.2$\pm$4.0$^2$ & $<$3\%   & depth &  \\ \hline  \hline

\parbox[t]{2mm}{\multirow{20}{*}{\rotatebox[origin=c]{90}{\textbf{Brain}}}}&\cellcolor{red!25}Brain	&	\multirow{2}{*}{26}	&		&		&	\multirow{2}{*}{2x}	&	\multirow{2}{*}{1.5}	& 3D T1	&	\multirow{2}{*}{m2D$^+$ pair} & \multirow{2}{*}{CNN}	&	\multirow{2}{*}{rig}	&	\multirow{2}{*}{67$\pm$11}	&		&		ME tissues &
	\multirow{2}{*}{x}	& \multirow{2}{*}{-0.1$\pm$0.3} & \multirow{2}{*}{99.8$\pm$0.7$^2$} &   & beam $\gamma_{3}$ & 
\multirow{2}{*}{Dinkla2018\cite{Dinkla2018}} \\ 
&\cellcolor{red!25} &  & &  &   &  & GRE & &  &  &  &	&		DSC dist body & 
	&  &  &   & depth $\gamma_{1}$ &  \\ \cline{2-20}  
	
&Brain\cellcolor{red!25}	&	\multirow{2}{*}{40}	&		&	\multirow{2}{*}{10}	&		&	\multirow{2}{*}{1.5}	& 3D T1 &	\multirow{2}{*}{2D pair} & \multirow{2}{*}{CNN}	&	\multirow{2}{*}{def}	&	\multirow{2}{*}{75$\pm$23}	&	&		\multirow{2}{*}{DSC} &
	\multirow{2}{*}{x}	& \multirow{2}{*}{$<$0.2$\pm$0.5} & \multirow{2}{*}{99.2$^3$} &   &  & 
\multirow{2}{*}{LiuF2019\cite{LiuF2019}} \\ 
&\cellcolor{red!25}\cellcolor{red!25} &  & &  &   &  & GRE Gd &  &  &  &  &	&	&	
	&  &  &   &  &  \\ \cline{2-20}  
	
&Brain\cellcolor{red!25}	&	\multirow{2}{*}{54}	&	\multirow{2}{*}{9}	&	\multirow{2}{*}{14}	&	\multirow{2}{*}{5x}	&	\multirow{2}{*}{1.5}	& 2D T1  &	\multirow{2}{*}{2D pair} & \multirow{2}{*}{GAN}	&	\multirow{2}{*}{rig}	&	\multirow{2}{*}{47$\pm$11}	&	&		each &
	\multirow{2}{*}{x}	& \multirow{2}{*}{-0.7$\pm$0.5} & \multirow{2}{*}{99.2$\pm$0.8$^2$} & \multirow{2}{*}{$<$1\%}  & 2D/3D & 
\multirow{2}{*}{Kazemifar2019\cite{Kazemifar2019}} \\ 
&\cellcolor{red!25} &  & &  &   &  & SE Gd &  &  &  &  &	&	fold	& 
	&  &  &   & $\gamma_{3}$ $\gamma_{1}$ &  \\ \cline{2-20}	

&\cellcolor{red!25}Brain	&	\multirow{2}{*}{55}	&\multirow{2}{*}{28} &	\multirow{2}{*}{4}	&		&	\multirow{2}{*}{1.5}	& 3D T1	&	2D pair & \multirow{2}{*}{U-net}	&	\multirow{2}{*}{rig}	&	116$\pm$26	&		&		\multirow{2}{*}{ME} &
	x	&  & $>98^2$,98$\pm$2$^2$ &  & range & 
\multirow{2}{*}{Neppl2019\cite{Neppl2019}} \\ 
&\cellcolor{red!25} &  & &  &   &  & GRE & 3Dp pair &  &  & 137$\pm$32 &	&	&	
	 $p$ & & $>98^2$,97$\pm$3$^2$  &   &  $\gamma_{1}$ &  \\ \cline{2-20}  

&\cellcolor{red!25}Brain	&	\multirow{2}{*}{25}	&	\multirow{2}{*}{2}	&  \multirow{2}{*}{25}		&		&	\multirow{2}{*}{1.5}	& 3D T1  &	3Dp & \multirow{2}{*}{GAN}	&	\multirow{2}{*}{rig}	& \multirow{2}{*}{55$\pm$7} & 	&	 ME &
	\multirow{2}{*}{x}	& \multirow{2}{*}{$<$2} & \multirow{2}{*}{98.4$\pm$3.5$^2$} & \multirow{2}{*}{$<$1.65\%}   & range & 
\multirow{2}{*}{Shafai2019\cite{Shafai2019}} \\ 
&\cellcolor{red!25} &  & &  &   &  & GRE &pair &   &  &  &	& 	 DSC & 
 & 	&    &   & $\gamma_{3}$ $\gamma_{1}$ &  \\ \cline{2-20}  

& \cellcolor{red!25}Brain& \multirow{2}{*}{47} & & 13 & \multirow{2}{*}{5x} & \multirow{2}{*}{3} & \multirow{2}{*}{T1} & \multirow{2}{*}{2D pair} & \multirow{2}{*}{U-net} & \multirow{2}{*}{rig} &\multirow{2}{*}{81$\pm$15} & &  ME air,& \multirow{2}{*}{x} &\multirow{2}{*}{2.3$\pm$0.1} &  &   &  align  & \multirow{2}{*}{Gupta2019\cite{Gupta2019}} \\
&\cellcolor{red!25} &  & &  &   &  &  & &  &  &  &		& tissues & 
	&  &  &   & CBCT &  \\ \cline{2-20} 

&\cellcolor{red!25}Brain &	\multirow{2}{*}{12}	&	\multirow{2}{*}{2}	&  \multirow{2}{*}{1}		&	\multirow{2}{*}{LoO}	&	\multirow{2}{*}{3}	& 3D T1  &	\multirow{2}{*}{2D+ pair} & \multirow{2}{*}{U-net}	&	\multirow{2}{*}{rig}	& \multirow{2}{*}{54$\pm$7} & 	&	 ME, DSC &
	\multirow{2}{*}{$p$}	& \multirow{2}{*}{0.00$\pm$0.01} &  &   & \multirow{2}{*}{range} & 
\multirow{2}{*}{Spadea2019\cite{Spadea2019}} \\ 
&\cellcolor{red!25} &  & &  &   &  & GRE &  &   &  &  &	& 	tissues & 
 & 	&    &   &  &  \\ \cline{2-20}  %

& \cellcolor{red!25}Brain	&	\multirow{2}{*}{15}	&		 	\multirow{2}{*}{}	& \multirow{2}{*}{}	 &	\multirow{2}{*}{5x}	& n.a.&  T1, T2  &	\multirow{2}{*}{2Dp pair} & \multirow{2}{*}{GAN}	&	\multirow{2}{*}{def}	& \multirow{2}{*}{108$\pm$24} &  &  \multirow{2}{*}{tissues} &
	\multirow{2}{*}{x}	& \multirow{2}{*}{0.7} & \multirow{2}{*}{99.2$\pm$1.0$^2$}  & \multirow{2}{*}{$<$1\%}  & beam $\gamma_{3}$   & 
\multirow{2}{*}{Koike2019\cite{Koike2020}} \\ 
& \cellcolor{red!25} &  & &  &   &  & FLAIR$^c$ &   &   &  &    &	& 	  & 
 & 	 & &    & depth $\gamma_{1}$ &  \\ \cline{2-20}  
			
& \cellcolor{red!25}Brain	&	\multirow{2}{*}{30$^{t,m}$}	&	\multirow{2}{*}{10}	&  \multirow{2}{*}{20}		&	 \multirow{2}{*}{3x} &	1.5	& 3D T1   &	 \multirow{2}{*}{2D+$^*$ pair} & \multirow{2}{*}{GAN$^*$}	&	\multirow{2}{*}{rig}	& \multirow{2}{*}{61$\pm$14} & \multirow{2}{*}{26.7$\pm$1.9}	&  ME DSC &
	x	& -0.1$\pm$0.3 & 99.5$\pm$0.8$^2$ & $<$1\%  & beam  & 
\multirow{2}{*}{Maspero2020\cite{Maspero2020}} \\ 
& \cellcolor{red!25}paed &  & &  &   & 3 & GRE$\pm$Gd & &    &  &  &	& SSIM	  & $p$
 & 	 0.1$\pm$0.4 & 99.6$\pm$1.1$^2$ & $<$3\%   & depth $\gamma_3$ &  \\ \cline{2-20} 

& \cellcolor{red!25}Brain	&	\multirow{2}{*}{66}	&		 	\multirow{2}{*}{}	& \multirow{2}{*}{11}	 &	\multirow{2}{*}{5x}	& \multirow{2}{*}{1.5} &  2D T1  &	\multirow{2}{*}{2D unp} & \multirow{2}{*}{GAN}	&	\multirow{2}{*}{rig}	& \multirow{2}{*}{78$\pm$11} &  &   &
	\multirow{2}{*}{$p$}	& \multirow{2}{*}{0.3$\pm$0.3} & \multirow{2}{*}{99.2$\pm$1.0$^2$}  & \multirow{2}{*}{$<$3\%}  & beam $\gamma_{3}$   & \multirow{2}{*}{Kazemifar2020\cite{Kazemifar2020dose}} \\ 
&\cellcolor{red!25} &  & &  &   &  & SE Gd &  &   &  &    &	& 	  & 
 & 	 & &    & depth $\gamma_{1}$ &  \\ \cline{2-20}
 
& \cellcolor{red!25}Brain	&	\multirow{2}{*}{242$^{m,t}$}	&		 	\multirow{2}{*}{81}	& \multirow{2}{*}{79}	 &		& 3 &  3D T1  &	3Dp & CNN	&	\multirow{2}{*}{def}	& 81$\pm$22 & &\multirow{2}{*}{tissues} &
	\multirow{2}{*}{x}	& 0.13$\pm$0.13 & 99.6$\pm$0.3$^2$  & $<\pm$0.15  &  \multirow{2}{*}{$\gamma_{3}$}  &   \multirow{2}{*}{Andres2020\cite{Andres2020}} \\ 
&\cellcolor{red!25} &  & &  &   & 1.5 & GRE$\pm$Gd  &   pair &  U-net & & 90$\pm$21   &	& 	  & 
 & 	0.31$\pm$0.18 & 99.4$\pm$0.5$^2$ &   &  &  \\ \hline 


\end{tabular}
\end{adjustbox}
{\ \\ \scriptsize 
$^*$ comparison with other architecture has been provided
$^3 \gamma_{\tiny 3\%,3\textrm{mm}}$ = $\gamma_3$, $^2 \gamma_{\tiny 2\%,2\textrm{mm}}$ = $\gamma_2$, $^1 \gamma_{\tiny 1\%,1\textrm{mm}}$  = $\gamma_1$; 
$^+$ trained in 2D on multiple view and aggregated after inference
$^t$ robustness to training size was investigated
$^c$ multiple combinations  (also $\pm$ Dixon reconstruction, where present) of the sequences were investigated but omitted;
$^m$ data from multiple centers; x: photon plan; p: proton plan; paed: paedriatic.

}
\end{sidewaystable}


\setcounter{table}{2}
\renewcommand{\thetable}{\arabic{table}b}

\begin{sidewaystable}[]
\captionv{22}{MR-only sCT with dose}{Overview sCT methods for MR-only radiotherapy with image-based and dose evaluation.
\label{tab:MRonlyDose2}
}
\vspace{-20pt}
\begin{adjustbox}{width=\columnwidth,center}
\begin{tabular}{|cc|c|c|c|c|c|c|c|c|c|c|c|c|c|c|c|c|c|c|}
\hline
\multicolumn{2}{|c|}{\multirow{2}{*}{\textbf{Tumor}}} & \multicolumn{4}{c|}{\textbf{Patients}}            & \multicolumn{2}{c|}{\textbf{MRI}}       & \multicolumn{2}{c|}{\textbf{DL method}}         & \multirow{3}{*}{\textbf{Reg}} & \multicolumn{3}{c|}{\textbf{Image-similarity}}                 & \multirow{3}{*}{\textbf{Plan}} & \multicolumn{4}{c|}{\textbf{Dose}}  & \multirow{3}{*}{\textbf{Reference}}     \\ \cline{3-10} \cline{12-14} \cline{16-19}  

 \multicolumn{2}{|c|}{\multirow{2}{*}{\textbf{site}}}   & \multirow{2}{*}{\textbf{train}} & \multirow{2}{*}{\textbf{val}} & \multirow{2}{*}{\textbf{test}} & \textbf{x-} & \textbf{field} & \multirow{2}{*}{\textbf{sequence}} & \multirow{2}{*}{\textbf{conf}} & \multirow{2}{*}{\textbf{arch}} &  & \textbf{MAE} & \textbf{PSNR}  & \multirow{2}{*}{\textbf{others}} &               & \textbf{DD} & \textbf{GPR} & \multirow{2}{*}{\textbf{DVH}} & \multirow{2}{*}{\textbf{others}} &        \\ 
 
& & & & & \textbf{fold} & [T] & &  & & & [HU] & [dB]  & & &[\%]& [\%] & & &  \\ \hline

\parbox[t]{2mm}{\multirow{15}{*}{\rotatebox[origin=c]{90}{\textbf{Pelvis}}}} &\cellcolor{cyan!25}Prostate& 36 & & 15 &   & 3 & T2 TSE & 2D pair & U-net & def
 &	30$\pm$5 &	&		ME tissues &
  	x	& 0.16$\pm$0.09 & 99.4$^2$ & $<$0.2Gy  & $\gamma_{3}$ $\gamma_{1}$ & Chen2018\cite{Chen2018} \\ \cline{2-20}   

&\cellcolor{cyan!25}Prostate & 39 & & & 4x & 3 & 3D T2 & 2D pair & U-net & def &33$\pm$8 & &\begin{tabular}[c]{@{}c@{}}ME DSC\\ dist body\end{tabular} & x &-0.01$\pm$0.64 & 98.5$\pm$0.7$^2$ & $<3\%$  & $\gamma_{3}$ $\gamma_{1}$ & Arabi2018\cite{Arabi2018} \\ \cline{2-20} 

&\cellcolor{cyan!25}Prostate	&	\multirow{2}{*}{17}	&		&		&	\multirow{2}{*}{LoO}	&	\multirow{2}{*}{1.5}	& \multirow{2}{*}{T2}  &	3Dp  & \multirow{2}{*}{GAN$^*$}	&	\multirow{2}{*}{rig}	&	\multirow{2}{*}{51$\pm$17}	& \multirow{2}{*}{24.2$\pm$2.5}	&	 NCC, bone:	 &
	\multirow{2}{*}{$p$}	& \multirow{2}{*}{-0.07$\pm$0.07} & \multirow{2}{*}{98$\pm$6$^2$} & \multirow{2}{*}{$<$1\%}  & range, $\gamma_{3}$ & 
\multirow{2}{*}{LiuY2019b\cite{LiuY2019b}} \\ 
&\cellcolor{cyan!25} &  & &  &   &  &  & unp &  &  &  &	&	 dist, uniform & 
	&  &  &   & peak, $\gamma_{1}$ &  \\ \cline{2-20}  
	
&\cellcolor{cyan!25}Prostate	&	\multirow{2}{*}{25}	&		& 	\multirow{2}{*}{14}	&	\multirow{2}{*}{3x}	&	\multirow{2}{*}{3}	& 3D T2  &	\multirow{2}{*}{2D pair} & U-net$^*$	&	\multirow{2}{*}{def}	& 34$\pm$8 & 	&	 tissues&
	\multirow{2}{*}{x}	& $<$1\% & 99.2$\pm$1$^1$ & $<$1\%  &  & 
\multirow{2}{*}{Largent2019\cite{Largent2019}} \\ 
&\cellcolor{cyan!25} &  & &  &   &  & TSE &  & GAN$^*$  &  & 34$\pm$8 &	& 	 ME & 
 &$<$1\%	& 99.1$\pm$1$^1$    &   &  &  \\ \cline{2-20}  %

&\cellcolor{cyan!25}Pelvis	&	\multirow{2}{*}{11$^m$}	&		 	\multirow{2}{*}{}	& \multirow{2}{*}{8}	 &		& 3 &  T2  &	\multirow{2}{*}{2D pair} & \multirow{2}{*}{GAN$^*$}	&	\multirow{2}{*}{def}	& \multirow{2}{*}{49$\pm$6} &  & ME &
	\multirow{2}{*}{x}	& \multirow{2}{*}{0.7$\pm$0.4} & \multirow{2}{*}{99.2$\pm$1.0$^2$}  & \multirow{2}{*}{$<$1.5\%}  &    & 
\multirow{2}{*}{Boni2020\cite{Boni2020}} \\ 
&\cellcolor{cyan!25} &  & &  &   & 1.5 & TSE &   &   &  &    &	& 	organs  & 
 & 	 & &    &  &  \\ \cline{2-20}

&\cellcolor{cyan!25} Pelvis	&	\multirow{2}{*}{26}	&		 \multirow{2}{*}{15}		& \multirow{2}{*}{10+19$^m$}	 &		& 0.35 & 3D T2  &	\multirow{2}{*}{2.5D pair} & \multirow{2}{*}{GAN$^*$}	&	\multirow{2}{*}{def}	& \multirow{2}{*}{41$\pm$4} & \multirow{2}{*}{31.4$\pm$1} & ME MSE &
	\multirow{2}{*}{x}	& \multirow{2}{*}{$<\pm$1} & & \multirow{2}{*}{$<$1.5\%}  &   & 
\multirow{2}{*}{Fetty2020\cite{Fetty2020dose}} \\ 
&\cellcolor{cyan!25} &  & &  &   & 1.5/3 &   &    &  & &    &	& 	bone  & 
 & 	 &  &   &  &  \\ \cline{2-20}
 	
& \cellcolor{cyan!25}Pelvis & 39 & & 14 &   & 0.35  & GRE & 2D pair  & U-net     & def & 54$\pm$12  &	& tissues	  & x+B$_0$
 & 	$<$0.5 & 99.0$\pm$0.7$^2$ & $<$1\%   &  $\gamma_{3}$ $\gamma_{1}$ &  Cusumano2020\cite{Cusumano2020}\\ \cline{2-20}
 	
&	\cellcolor{cyan!25}Rectum	&	\multirow{2}{*}{46$^m$}	&		 		& \multirow{2}{*}{44}	 &		& \multirow{2}{*}{1.5} & \multirow{2}{*}{3D T2}  &	\multirow{2}{*}{2D pair} & \multirow{2}{*}{GAN}	&	\multirow{2}{*}{def}	& \multirow{2}{*}{35$\pm$7} & & ME &
	\multirow{2}{*}{x}	& \multirow{2}{*}{$<\pm$0.8} & \multirow{2}{*}{99.8$\pm$0.1$^2$}  & \multirow{2}{*}{$<$1\%}  &  \multirow{2}{*}{$\gamma_{3}$ $\gamma_{1}$}  & 
\multirow{2}{*}{Bird2020\cite{Bird2021}} \\ 
&\cellcolor{cyan!25} &  & &  &   &  &   &    &  & &    &	& 	bone  & 
 & 	 &  &   &  &  \\ \hline \hline
	
\parbox[t]{2mm}{\multirow{10}{*}{\rotatebox[origin=c]{90}{\textbf{Head \& Neck}}}} &\cellcolor{gray!25}H$\&$N	&	\multirow{2}{*}{34}	&		&		&	\multirow{2}{*}{3x}	&	\multirow{2}{*}{1.5}	& 3D T2	&	3Dp & \multirow{2}{*}{U-net}	&	\multirow{2}{*}{def}	&	\multirow{2}{*}{75$\pm$9}	&		&		ME &
	\multirow{2}{*}{x}	& \multirow{2}{*}{-0.07$\pm$0.22} & \multirow{2}{*}{95.6$\pm$2.9$^2$} &   & \multirow{2}{*}{$\gamma_{3}$} & 
\multirow{2}{*}{Dinkla2019\cite{Dinkla2019}} \\ 
&\cellcolor{gray!25} &  & &  &   &  & TSE & pair &  &  &  &	&		DSC bone & 
	&  &  &   &  &  \\ \cline{2-20}  

&\cellcolor{gray!25}H$\&$N	&	\multirow{2}{*}{15}	&		&  \multirow{2}{*}{12}		&	&	\multirow{2}{*}{3}	& T1   &	2Dp$^*$ & \multirow{2}{*}{GAN$^*$}	&	\multirow{2}{*}{def}	&  & 68$\pm$2	&   SSIM &
	\multirow{2}{*}{$p$}	& \multirow{2}{*}{$<$0.5} & \multirow{2}{*}{$<$98$^2$} & \multirow{2}{*}{$<$0.5}  &   & 
\multirow{2}{*}{Klages2019\cite{Klages2020}} \\ 
&\cellcolor{gray!25} &  & &  &   &  & GRE & pair &    &  &  &	& RMSE	  & 
 & 	&    &   &  &  \\ \cline{2-20}

&\cellcolor{gray!25}H$\&$N	&	\multirow{2}{*}{30}	&		&  \multirow{2}{*}{15}		&	&	\multirow{2}{*}{3}	& T1$\pm$Gd   &	\multirow{2}{*}{2D pair} & GAN$^*$	&	\multirow{2}{*}{rig}	& 70$\pm$12 & 29.4$\pm$1.3	&   SSIM &
	\multirow{2}{*}{$p$}	& -0.3$\pm$0.2 & 97.8$\pm$0.9$^2$ &   &   & 
\multirow{2}{*}{Qi2020\cite{Qi2020}} \\ 
&\cellcolor{gray!25} &  & &  &   &  & T2 TSE$^c$ & &  U-net      & & 71$\pm$12 & 29.2$\pm$1.3 &	 DSC, DRR	  & 
 & 	-0.2$\pm$0.2 & 97.6$\pm$1.3$^2$ &  &  &  \\ \cline{2-20}  

&\cellcolor{gray!25}H$\&$N	&	\multirow{2}{*}{135$^t$}	&	\multirow{2}{*}{10}	&  \multirow{2}{*}{28}		&	  &	\multirow{2}{*}{3}	& 3D T1   &	2D pair & \multirow{2}{*}{GAN$^*$}	&	\multirow{2}{*}{def}	& 70$\pm$9 & 	& ME, DSC &
	\multirow{2}{*}{x}	& -0.1$\pm$0.3 & 98.7$\pm$1.0$^2$ & $<$1.5\%  & beam  & 
\multirow{2}{*}{Peng2020\cite{Peng2020}} \\ 
&\cellcolor{gray!25} &  & &  &   &  & GRE & 2D unp &    &  & 101$\pm$8&	& tissues	  & 
 & 	0.1$\pm$0.4 & 98.5$\pm$1.1$^2$ & $<$1.5\%   & depth &  \\ \cline{2-20}

&\cellcolor{gray!25}H\&N	&	\multirow{2}{*}{27}	&		 & 	 &	\multirow{2}{*}{3x}	& \multirow{2}{*}{3} & 3D T1  &	2D+ & \multirow{2}{*}{GAN}	&	\multirow{2}{*}{def}	& \multirow{2}{*}{65$\pm$4} &  & ME &
	\multirow{2}{*}{p}	& \multirow{2}{*}{$<\pm$0.2} & \multirow{2}{*}{93.5$\pm$3.4$^2$}& \multirow{2}{*}{$<$1.5\%}  & NTCP  & 
\multirow{2}{*}{Thummerer2020\cite{Thummerer2020comparison}} \\ 
&\cellcolor{gray!25} &  & &  &   &  & GRE  &  pair  &  & &    &	& 	DSC  & 
 & 	 &  &   & RS $\gamma_3$  &  \\ \hline   \hline

\parbox[t]{2mm}{\multirow{2}{*}{\rotatebox[origin=c]{90}{\textbf{Thor}}}}&\cellcolor{yellow!25}Breast	&	\multirow{2}{*}{12$^t$}	&		&  \multirow{2}{*}{18}		&	\multirow{2}{*}{LtO}	&	\multirow{2}{*}{1.5}	& 3D GRE  &	2Dp$^+$  & \multirow{2}{*}{GAN$^*$}	&	\multirow{2}{*}{def}	& 94$\pm$11 & 	&	 \multirow{2}{*}{NCC} &
	\multirow{2}{*}{$p$}	& \multirow{2}{*}{$<$0.5} & \multirow{2}{*}{98.4$\pm$3.5$^2$} &   & DRR  & 
\multirow{2}{*}{Olberg2019\cite{Olberg2019}} \\ 
&\cellcolor{yellow!25} &  & &  &   &  & mDixon & pair &    &  & 103$\pm$15 &	& 	  & 
 & 	&    &   & dis bone &  \\ \hline 
\multicolumn{20}{c}{\textbf{Multiple sites with one network}} \\ \hline
 & \cellcolor{cyan!25}Prostate	&	32	&		&	27	&		&	3	&	3D T1  	&	\multirow{3}{*}{2D pair}	&	\multirow{3}{*}{GAN} &	\multirow{3}{*}{rig}	&	60$\pm$6 &	&		\multirow{3}{*}{ME} & 
\multirow{3}{*}{x}	& -0.3$\pm$0.4 & 99.4$\pm$0.6$^3$ &\multirow{3}{*}{$<$1\%} & \multirow{3}{*}{$\gamma_{2}$} & \multirow{3}{*}{Maspero2018\cite{Maspero2018}} \\ 
&\cellcolor{cyan!25}Rectum & & & 18 & & 1.5 & GRE  & & & & 56$\pm$5 & &  &    & -0.3$\pm$0.5 & 98.5$\pm$1.1$^3$ & & &\\ 
&\cellcolor{cyan!25}Cervix & & & 14 & & 1.5/3 & mDixon & & & & 59$\pm$6 &  &    &  & -0.1$\pm$0.3$^a$ & 99.6$\pm$1.9$^3$ & & &\\ \hline
\end{tabular}
\end{adjustbox}
{\ \\ \scriptsize 
$^*$ comparison with other architecture has been provided
$^3 \gamma_{\tiny 3\%,3\textrm{mm}}$ = $\gamma_3$, $^2 \gamma_{\tiny 2\%,2\textrm{mm}}$ = $\gamma_2$, $^1 \gamma_{\tiny 1\%,1\textrm{mm}}$  = $\gamma_1$; 
$^+$ trained in 2D on multiple view and aggregated after inference
$^t$ robustness to training size was investigated
$^c$ multiple combinations  (also $\pm$ Dixon reconstruction, where present) of the sequences were investigated but omitted;
$^m$ data from multiple centers; x: photon plan; p: proton plan.
}
\end{sidewaystable}

Considering the imaging protocol, we can observe that most of the MRIs were acquired at 1.5~T (51.9\%), followed by 3~T (42.6\%), and the remaining 6.5\% at 1~T or 0.35/0.3~T.
The most popular MRI sequences adopted depends on the anatomical site: T1 gradient recalled-echo (T1 GRE) for abdomen and brain; T2 turbo spin-echo (TSE) for pelvis and H\&N. Unfortunately, for more than ten studies, either sequence or magnetic field were not adequately reported. \\
Generally, a single MRI sequence is used as input. However, eight studies investigated using multiple input sequences or Dixon reconstructions~\cite{Xu2019multi,Tie2020,Kearney2020,Florkow2020,Koike2020,Maspero2018,Florkow2020dose,Olberg2019} based on the assumption that more input contrast may facilitate sCT generation.
A relevant aspect related to MRI is which kind on pre-processing is applied to the data before being fed to the network. Generally intensity normalisation techniques like z-score~\cite{Florkow2019impact}, percentile-~\cite{Maspero2018,Maspero2020} or range-based normalisation, histogram matching~\cite{Xiang2018,Fu2019,Fu2020,Tie2020} or linear rescaling were applied~\cite{Reinhold2019, Spadea2019}. However, techniques like bias field~\cite{Han2017,Xiang2018,Arabi2018,Andres2020,Fu2019,Fetty2020,Koike2020,Fu2020,Tie2020,Klages2020,Yang2020,Liu2019,Lei2019mri,Liu2020,Shafai2019,Largent2020}, intensity homogeneity~\cite{Arabi2018,Lei2019mri,Han2017,Xiang2018,Andres2020,Fu2019,Fu2020,Fetty2020,Koike2020,Tie2020,Yang2020,Liu2019,Liu2020,Shafai2019,Largent2020} were also applied to minimise inter-patient intensity variations.\\
Some studies compared the performance of sCT generation depending on the sequence acquired.  For example, Massa et al.~\cite{Massa2020} compared sCT from the most adopted MRI sequences in the brain, e.g. T1 GRE with (+Gd) and without Gadolinium (-Gd), T2 SE and T2 fluid-attenuated inversion recovery (FLAIR), obtaining the lowest MAE and highest PSNR for T1 GRE sequences with Gadolinium administration. Florkow et al.~\cite{Florkow2020} investigated how the performance of a 3D patch-based paired U-net was impacted by different combinations of T1 GRE images along with its Dixon reconstructions, finding that using multiple Dixon images is beneficial in the human and canine pelvis.
Qi et al.~\cite{Qi2020} studied the impact of combining T1 ($\pm$Gd) and T2 TSE, obtaining that their 2D paired GAN model trained on multiple sequences outperformed any model on a single sequence.\\
When focusing on the DL model configuration, we found that 2D models were the most popular ones, followed by 3D patch-based and 2.5D models. Only one study adopted a multi-2D (m2D) configuration~\cite{Dinkla2018}. Three studies also investigated whether the impact of combining sCTs from multiple 2D models after inference (2D+) shows that 2D+ is beneficial compared to single 2D view~\cite{Spadea2019,Klages2020,Maspero2020}. 
When comparing the performances of 2D against 3D models, Fu et al.~\cite{Fu2019} found that a modified 3D U-net outperformed a 2D U-net; while Neppl et al.~\cite{Neppl2019} one month later published that their 3D U-net under-performed a 2D U-net not only on image similarity metrics but also considering photon and proton dose differences. These contradicting results will be discussed later.
Paired models were the most adopted, with only ten studies investigating unpaired training~\cite{Xu2020,Jin2019,Li2020comp,Kearney2020,Fetty2020,Kazemifar2020dose,LiuY2019b,Fu2020,Peng2020,Yang2020}. 
Interestingly, Li et al.~\cite{Li2020comp} compared a 2D U-net trained in a paired manner against a cycle-GAN trained in an unpaired manner, finding that image similarity was higher with the U-net.
Similarly, two other studies compared 2D paired against unpaired GANs, achieving slightly better similarity and lower dose difference with paired training in the abdomen \cite{Fu2020} and H\&N \cite{Peng2020}. 
Mixed paired/unpaired training was proposed by Jin et al.~\cite{Jin2019} who found such a technique beneficial against either paired or unpaired training.
Yang et al.~\cite{Yang2020} found that structure-constrained loss functions and spectral normalisation ameliorated unpaired training performances in the pelvic and abdominal regions.\\
An interesting study on the impact of the directions of patch-based 2D slices, patch size and GAN architecture was conducted by Klages et al.~\cite{Klages2020} who reported that 2D+ is beneficial against a single view (2D) training, overlapping/non-overlapping patches is not a crucial point, and that upon good registration training of paired GANs outperforms unpaired training (cycle-GANs).\\
If we now turn to the architectures employed, we can observe that GAN covers the majority of the studies ($\sim$55\%), followed by U-net ($\sim$35\%) and other CNNs ($\sim$10\%).
A detailed examination of different 2D paired GANs against U-net with different loss functions by Largent et al.~\cite{Largent2019} showed that U-net and GANs could achieve similar image- and dose-base performances. 
Fetty et al.~\cite{Fetty2020dose} focused on comparing different generators of a 2D paired GAN against the performance of an ensemble of models, finding that the ensemble was overall better than single models being more robust to generalisation on data from different scanners/centres.
When considering CNNs architectures, it is worth mentioning using 2.5D dilated CNNs by Dinkla et al.~\cite{Dinkla2018} where the m2D training was claimed to increase the robustness of inference in a 2D+ manner, maintaining a big receptive field and a low number of weights.

An exciting aspect investigated by four studies is the impact of the training size~\cite{Maspero2020,Andres2020,Olberg2019,Peng2020,Yang2020}, which will be further reviewed in the discussion section.

Finally, when considering the metric performances, we found that 21 studies reported only image similarity metrics, and 30 also investigated the accuracy of sCT-based dose calculation on photon RT (19), proton RT (8), or both (3). Two studies performed treatment planning, considering the contribution of magnetic fields~\cite{Cusumano2020,Fu2020}, which is crucial for MR-guided RT.
Also, only four publications studied the robustness of sCT generation in a multiple centres~\cite{Andres2020,Boni2020,Maspero2020,Bird2021}.

Overall, DL-based sCT resulted in DD on average $<$1\% and $\gamma_{\tiny 2\%,2\textrm{mm}}$ GPR$>$95\%, except for one study~\cite{Thummerer2020comparison}. For each anatomical site, the metrics on image similarity and dose were not always calculated consistently. Such aspect will be detailed in the next section.

\subsection{CBCT-to-CT generation}

CBCT-to-CT conversion via DL is the most recent CT synthesis application, with the first paper published in 2018\cite{kida2018cone}.
Some of the works (5 out of 15) focused only on improving CBCT image quality for better IGRT\cite{kida2018cone, harms2019paired,chen2020synthetic,kida2020visual, yuan2020convolutional}. The remaining 10 proved the validity of the transformation with dosimetric studies for photons \cite{liang2019generating,li2019preliminary,Maspero2020,Liu2020,barateau2020comparison,eckl2020evaluation,liu2020cbct}, protons \cite{Thummerer2020comparison} and for both photons and protons\cite{Landry2019comparing,kurz2019cbct, zhang2020improving}.

Only three studies investigated unpaired training\cite{kurz2019cbct,liang2019generating,maspero2020single}; in eleven cases, paired training  was implemented by matching the CBCT and ground truth CT by rigid or deformable registration. In Eck et al. \cite{eckl2020evaluation}, however, CBCT and CT were not registered for the training phase, as the authors claimed the first fraction CBCT was geometrically close enough to 
the planning CT for the network. Deformable registration was then performed for image similarity analysing. In this work, the quality of contours propagated to sCT from CT was compared to manual contours drawn on the CT to assess each step of the IGART workflow: 
 image similarity, anatomical segmentation and dosimetric accuracy.
The network, a 2D cycle GAN implemented on a vendor's provided research software, was independently trained and tested on different sites, H\&N, thorax and pelvis, leading to best results for the pelvic region. 
 
Other authors studied training a single network for different anatomical regions. In Maspero et al. \cite{maspero2020single}, authors compared the performances of three cycle-GANs trained independently on three anatomical sites (H\&N, breast and lung) vs a single trained with all the anatomical sites together, finding similar results in terms of image similarity. \\

\setcounter{table}{3}
\renewcommand{\thetable}{\arabic{table}}
\begin{landscape}
\begin{table}
\captionv{22}{CBCT to sCT}{Overview sCT methods for adaptive radiotherapy with CBCT.
\label{tab:CBCT}}
\begin{adjustbox}{width=\columnwidth,center}
\begin{tabular}{|cc|c|c|c|c|c|c|c|c|c|c|c|c|c|c|c|c|c|c|}
\hline
\multicolumn{2}{|c|}{\multirow{2}{*}{\textbf{Tumor}}}      & \multicolumn{4}{c|}{\textbf{Patients}}               & \multicolumn{2}{c|}{\textbf{DL method}}     &\multirow{3}{*}{\textbf{Reg}}          & \multicolumn{4}{c|}{\textbf{Image-similarity}}& \multirow{3}{*}{\textbf{Plan}} & \multicolumn{5}{c|}{\textbf{Dose}}& \multirow{3}{*}{\textbf{Reference}}    \\ \cline{3-8} \cline{10-13} \cline{15-19}
 \multicolumn{2}{|c|}{\multirow{2}{*}{\textbf{site}}}   & \multirow{2}{*}{\textbf{train}} & \multirow{2}{*}{\textbf{val}} & \multirow{2}{*}{\textbf{test}} & \textbf{x-} & \multirow{2}{*}{\textbf{conf}} & \multirow{2}{*}{\textbf{arch}} &  & \textbf{MAE} & \textbf{PSNR} & \textbf{SSIM}  & \multirow{2}{*}{\textbf{others}} &               & \textbf{DD} & \textbf{DPR} & \textbf{GPR} & \multirow{2}{*}{\textbf{DVH}} & \multirow{2}{*}{\textbf{others}} &        \\ 
 & & & & & \textbf{fold} &  &  &  & [HU] & [dB] & & & &[\%]& [\%] & [\%] & & &  \\ \hline
\parbox[t]{2mm}{\rotatebox[origin=c]{90}{\textbf{Abd}}}&\cellcolor{green!25}Pancreas   & 30                     &                 &                    & LoO                 & \begin{tabular}[c]{@{}c@{}}3Dp  \\ pair \end{tabular}  & GAN$^*$   & def   & 56.9$\pm$13.8  & 28.8$\pm$2.5 & .71$\pm$.03  & \begin{tabular}[c]{@{}c@{}}NCC\\ SNU\end{tabular}       & x             &          &     & &   \textless1Gy &           & Liu 2020\cite{liu2020cbct}                   \\ \hline \hline
\parbox[t]{2mm}{\rotatebox[origin=c]{90}{\textbf{Thor}}}&\cellcolor{yellow!25}Thorax & 53 & & 15 && 2D pair & GAN & def &  94$\pm$32  &  & & ME DSC HD tis & x& & 76.7$\pm$17.3$^2$ & 93.8$\pm$5.9$^2$ & $<$2.6& $\gamma_3$   & Eckl2020\cite{eckl2020evaluation}           \\ \hline \hline
& \cellcolor{red!25}Brain      & 24   & &     & \multirow{2}{*}{LoO}    & 3Dp     & \multirow{2}{*}{GAN}           & \multirow{2}{*}{rig}   & 13$\pm$2  & 37.5$\pm$2.3 &   & NCC       &   \multirow{2}{*}{No}            &                 & &      &          &  & \multirow{2}{*}{Harms2019\cite{harms2019paired}}            \\ \cline{1-1}
\parbox[t]{2mm}{\multirow{10}{*}{\rotatebox[origin=c]{90}{\textbf{Pelvis}}}}&\cellcolor{cyan!25}Pelvis & 20 &  & & &  pair & &   &16$\pm$5 & 30.7$\pm$3.7 & & SNU &  
& & & & & &   \\ \cline{2-20}
&\cellcolor{cyan!25}Prostate   & \multirow{2}{*}{16}  & & \multirow{2}{*}{4}  & \multirow{2}{*}{5x}  & \multirow{2}{*}{2D pair}   & \multirow{2}{*}{U-net} & \multirow{2}{*}{def}  &   & \multirow{2}{*}{50.9} &\multirow{2}{*}{.967}  &  SNU   & \multirow{2}{*}{No}
   &   & &  &  &  & \multirow{2}{*}{Kida2018\cite{kida2018cone}} \\ 
&\cellcolor{cyan!25} &  & &  &   &  &   &    &  & & &  	RMSE  & 
 & 	 &  &  & &   & \\ \cline{2-20}  %
&\cellcolor{cyan!25}Prostate   & \multirow{2}{*}{27}  & \multirow{2}{*}{7}  & \multirow{2}{*}{8}  &     & \multirow{2}{*}{2D pair}   & \multirow{2}{*}{U-net$^*$}  & \multirow{2}{*}{def}   & \multirow{2}{*}{58}    &   & & \multirow{2}{*}{ME}&  x&  &      $>$98.4$^1$    &  99.5$^2$  &      &    $\gamma_1$ $\gamma_3$ DPR$_2$     & \multirow{2}{*}{Landry2019\cite{Landry2019comparing}}     \\ 
&\cellcolor{cyan!25} &  & &  &   &  &   &    & & & &  	 & $p$
 & 	 & 88.5$^3$ & $>$96.5$^2$ & &   DPR$_2$ RS & \\ \cline{2-20}  
&\cellcolor{cyan!25}Prostate   & 18                     &                 & 8                     & 4x                  & \begin{tabular}[c]{@{}c@{}}2D ens\\ unp\end{tabular}   & GAN$^w$           & rig   & 87$\pm$5   & &    & ME          & \begin{tabular}[c]{@{}c@{}}x\\$p$\end{tabular} & & \begin{tabular}[c]{@{}c@{}}99.9$\pm$0.3$^2$ \\ 80.5$\pm$5$^2$\end{tabular}   & \begin{tabular}[c]{@{}c@{}} \\ 95.9$\pm$2.0$^2$\end{tabular}       & \begin{tabular}[c]{@{}c@{}} \textless$\pm$1.5\%\\ $<$1\%\end{tabular}     & \begin{tabular}[c]{@{}c@{}}DPR$^1$ $\gamma_3$\\DPR$^3$ RS  \end{tabular} & Kurz2019\cite{kurz2019cbct}                 \\ \cline{2-20}
&\cellcolor{cyan!25}Prostate   & 16                     &                 & 4                     &                  & 2D  pair              & GAN$^*$           & rig   &   & &    & \begin{tabular}[c]{@{}c@{}}SSIM\\diffROI \end{tabular} &         No      &                 & &      &    &      & Kida2019\cite{kida2020visual}               \\ \cline{2-20}
& \cellcolor{cyan!25}Pelvis & 205  & & 15 && 2D pair & GAN    & def    & 42$\pm$5 &  &  &   ME DSC HD tis    &      x &  &  88.9$\pm$9.3$^2$ & 98.5$\pm$1.7$^2$ &  $<$1 & $\gamma_3$   & Eckl2020\cite{eckl2020evaluation}             \\ \hline \hline

&\cellcolor{gray!25}H\&N & 81   & 9            & 20          &                  & 2D  unp               & GAN$^*$           &  def   & 29.9$\pm$4.9   & 30.7$\pm$1.4 & .85$\pm$.03   & \begin{tabular}[c]{@{}c@{}}RMSE\\phantom \end{tabular}       &    x           &          &    &    \begin{tabular}[c]{@{}c@{}}98.4$\pm$1.7$^2$\\ 96.3$\pm$3.6$^1$\end{tabular}   &      &          & Liang2019 \cite{liang2019generating}\\ \cline{2-20}
&\cellcolor{gray!25}Nasophar& 50                     & 10                 & 10                    &       & 2D pair               & U-net         & rig   & 6-27& &    &\begin{tabular}[c]{@{}c@{}} ME \\ organs\end{tabular}      &   x      &        0.2$\pm$0.1& &95.5$\pm$1.6$^1$   & $<$1\%   &          & Li2019\cite{li2019preliminary}     \\ \cline{2-20}
&\cellcolor{gray!25}H\&N & 30                     & 7                  & 7                     &                  & 2D pair                & U-net$^*$         & rig   & 18.98& 33.26      & 0.8911& \begin{tabular}[c]{@{}c@{}}RMSE\\ tissues\end{tabular}        &      No         &            &     & &      &          & Chen2019\cite{chen2020synthetic} \\ \cline{2-20}
&\cellcolor{gray!25}H\&N & 50$^t$               &                 & 10                    &                  & 2.5D pair            & U-net         & rig   & 49.28  &14.25 & .85  & SNR         &         No      &                 & &      &   &       & Yuan2020\cite{yuan2020convolutional}\\ \cline{2-20}
&\cellcolor{gray!25}H\&N & 22                     &                 & 11                    & 3x                  & 2D$^+$ pair & U-net         & def   & 36$\pm$6     & &    & \begin{tabular}[c]{@{}c@{}}ME DSC\\ SNU\end{tabular}   & $p$           &    -0.1$\pm$0.3      & & 98.1$\pm$1.2$^2$    &      & \begin{tabular}[c]{@{}c@{}}RS \\ $\gamma_3$\end{tabular} &Thummerer2020\cite{thummerer2020CBCT}\\ \cline{2-20}
&\cellcolor{gray!25}H\&N & 30                     &                 & 14                    &                  & 2D pair                 & GAN           & def   & 82.4$\pm$10.6    & &    & \begin{tabular}[c]{@{}c@{}}ME\\tissues\end{tabular}          & x                              & 91.0$\pm$5.3$^2$    & &     & \begin{tabular}[c]{@{}c@{}}$<$1Gy\\$<$1\%\end{tabular} &         & Barateau2020\cite{barateau2020comparison}\\ \cline{2-20}
&\cellcolor{gray!25}H\&N & 25 & & 15 &&2D pair & GAN & def & 77.2$\pm$16.6  &  &  & ME DSC HD tis & x &&  91.5$\pm$4.3$^2$ & 95.0$\pm$2.4$^2$ & $<$2.4 & $\gamma_3$   & Eckl2020\cite{eckl2020evaluation}           \\ \hline
\multicolumn{15}{c|}{\textbf{Multiple sites with one network}} \\ \hline
\multicolumn{2}{|c|}{\begin{tabular}[c]{@{}c@{}}\cellcolor{gray!25}H$\&$N\\  \cellcolor{green!25}Lung\\ \cellcolor{yellow!25}Breast\end{tabular}}     & \begin{tabular}[c]{@{}c@{}}15\\  15 \\ 15\end{tabular}  & \begin{tabular}[c]{@{}c@{}}8 \\ 8 \\ 8\end{tabular} & \begin{tabular}[c]{@{}c@{}}10 \\ 10 \\ 10\end{tabular} &                  & 2D unp$^*$                & GAN$^*$           & rig   & \begin{tabular}[c]{@{}c@{}}53$\pm$12\\ 83$\pm$10\\ 66$\pm$18\end{tabular}            & \begin{tabular}[c]{@{}c@{}}30.5$\pm$2.2\\ 28.5$\pm$1.6\\ 29.0$\pm$2.1\end{tabular} & \begin{tabular}[c]{@{}c@{}}.81$\pm$.04\\ .78$\pm$.04\\ .76$\pm$.02\end{tabular} & ME          & x             & \begin{tabular}[c]{@{}c@{}}0.1$\pm$0.5 \\ 0.2$\pm$0.9 \\ 0.1$\pm$0.4 \end{tabular}       &   &     \begin{tabular}[c]{@{}c@{}}97.8$\pm$1$^2$\\ 94.9$\pm$3$^2$\\ 92$\pm$8$^2$\end{tabular}                    & \textless 2\%     & $\gamma_3$      & Maspero2020\cite{maspero2020single}         \\ \hline
\multicolumn{2}{|c|}{\begin{tabular}[c]{@{}c@{}}\cellcolor{cyan!25}Pelvis\\ \cellcolor{gray!25}H\&N\end{tabular}   }            &  \begin{tabular}[c]{@{}c@{}}135 \\ \ \end{tabular}                  & 15                 & \begin{tabular}[c]{@{}c@{}}15\\ 10\end{tabular}         & 10x                 & 2.5D pair              & GAN$^*$           &  def   & \begin{tabular}[c]{@{}c@{}}24$\pm$5\\ 24$\pm$4\end{tabular}                 & \begin{tabular}[c]{@{}c@{}}20.1$\pm$3.4\\ 22.8$\pm$3.4\end{tabular}             &    &     &     \begin{tabular}[c]{@{}c@{}}x\\ $p$\end{tabular} & \begin{tabular}[c]{@{}c@{}}\textless1\% \\ \ \end{tabular}    & &   &   & \begin{tabular}[c]{@{}c@{}}\\ RS\end{tabular} & Zhang2020\cite{zhang2020improving}      \\ \hline 
\end{tabular}
\end{adjustbox}
{\ \\ \scriptsize 
$^*$ comparison with other architecture has been provided;
$^3$ dose pass rate (DPR) 1\% or $\gamma_{\tiny 1\%,1\textrm{mm}}$  = $\gamma_1$; $^2$ DPR 2\% or $\gamma_{\tiny 2\%,2\textrm{mm}}$ = $\gamma_2$; DPR 3\% or $\gamma_{\tiny 3\%,3\textrm{mm}}$ = $\gamma_3$;   
$^+$ trained in 2D on multiple view and aggregated after inference;
$^w$ different nets were trained and the different outputs were weighted to obtain final sCT; $^t$ robustness to training size was investigated; x: photon plan; p: proton plan.
}
\end{table}
\end{landscape}
Zhang et al. \cite{zhang2020improving} trained a 2.5D conditional GAN\cite{zhu2017unpaired} with feature matching on a large cohort of 135 pelvic patients. Then, they tested the network on additional 15 pelvic patients acquired with a different CT scanner and ten H\&N patients. The network predicted sCT with similar MAE for both testing groups, demonstrating the potentialities to transfer pre-trained models to different anatomical regions. They also compared different GAN flavours and U-net finding the latter statistically worse than any GAN configuration. \\
Three works tested unpaired training with cycle-GANs \cite{liang2019generating,kurz2019cbct,maspero2020single}. In particular, Liang et al. \cite{liang2019generating} compared unsupervised training among cycle-GAN, DCGAN\cite{radford2015unsupervised} and PGGAN\cite{karras2017progressive} on the same dataset, finding the first to perform better both in terms of image similarity and dose agreement.

Considering the anatomical regions investigated, most of the studies dealt with H\&N and pelvic regions. Liu et al. \cite{liu2020cbct} investigated CBCT-to-CT in the framework of breath-hold stereotactic pancreatic radiotherapy, where they trained a 3D patch cycle-GAN introducing an attention gate (AG)\cite{oktay2018attention} to deal with moving organs. They found that the cycle-GAN with AG performed better than U-net and cycle-GAN without AG. Moreover, the DL approach led to a statistically significant improvement in sCT vs CBCT, although some residual discrepancies were still present for this particular anatomical site.


\subsection{PET attenuation correction}
DL methods for deriving sCT for PET AC have been published since 2017 \cite{leynes2017direct}. Two possible image translations are available in this category: i) MR-to-CT for MR attenuation correction (MRAC), where 14 papers were found; ii) uncorrected PET-to-CT, with three published articles. 

In the first case, most methods have been tested with paired data in H\&N (9 papers) and the pelvic region (4 papers) except Baydoun et al.~\cite{Baydoun2020} who investigated the thorax district. The number of patients used for training ranged between 10 and 60. Most of the MR images employed in these studies have been acquired directly through 3T PET/MRI hybrid scanners, where specific MR sequences, such as UTE (ultra-short echo time) and ZTE (zero time echo) are used to enhance short $T_{2}$ tissues, such as in the cortical bone and Dixon reconstruction is employed to derive fat and water images. \\
Leynes et al.\cite{leynes2017direct} compared the Dixon-based sCT vs sCT predicted by U-net receiving both Dixon and ZTE. Results showed that DL prediction reduced the RMSE in corrected PET SUV by a factor of 4 for bone lesions and 1.5 for soft tissue lesions. Following this first work, other authors showed the improvement of DL-based AC over the traditional atlas-based MRAC proposed by the vendors \cite{gong2018attenuation, Jang2018deep, torrado2019dixon, ladefoged2019deep, blanc2019attenuation, Baydoun2020, gong2020b}, also comparing several network configurations \cite{pozaruk2020augmented, gong2020a}.\\ 
\begin{sidewaystable}[]
\captionv{22}{attenuation correction}{Overview methods on sCT for PET AC. 
\label{tab:PET}
}
\vspace{-20pt}
\begin{adjustbox}{width=\columnwidth,center}
\begin{tabular}{|c|c|c|c|c|c|c|c|c|c|c|c|c|c|c|c|}
\hline
\multirow{3}{*}{\textbf{Region}} & \multicolumn{4}{c|}{\textbf{Patients}}    & \multicolumn{2}{c|}{\textbf{MRI}}               & \multicolumn{2}{c|}{\textbf{DL method}}          & \multirow{3}{*}{\textbf{Reg}} & \multicolumn{2}{c|}{\textbf{Image-similarity}}  & \multicolumn{2}{c|}{\textbf{PET-related}}   &   \multirow{3}{*}{\textbf{Others}}                          & \multirow{2}{*}{\textbf{Reference}}    \\ \cline{2-9} \cline{11-14}
     & \multirow{2}{*}{\textbf{train}} & \multirow{2}{*}{\textbf{val}} & \multirow{2}{*}{\textbf{test}} & \textbf{x-} & \textbf{field}& \multirow{2}{*}{\textbf{contrast}} & \multirow{2}{*}{\textbf{conf}} & \multirow{2}{*}{\textbf{arch}} &  & \textbf{MAE} & \textbf{DSC}   & \textbf{tracer} &\textbf{PETerr}     &   &   \\ 
 & & & & \textbf{fold} & [T] &  &  & & &  [HU]  & & & [\%]&  &  \\ \hline

\cellcolor{cyan!25}Pelvis& 10             &           & 16               &                  & 3$^{H}$   & \begin{tabular}[c]{@{}c@{}}Dixon\\$\pm$ZTE\end{tabular}    & \begin{tabular}[c]{@{}c@{}}3Dp\\pair\end{tabular}            & U-net         & def              &             &              & {\footnotesize\begin{tabular}[c]{@{}c@{}}$^{18}$F-FDG\\$^{68}$Ga-PSMA\end{tabular}} && \begin{tabular}[c]{@{}c@{}}RMSE \\ SUV diff\end{tabular}    & Leynes2017\cite{leynes2017direct}           \\ \hline

\cellcolor{cyan!25}Pelvis& 15             &           & 4                & 4         & 3$^{H}$    & \begin{tabular}[c]{@{}c@{}}T1 GRE$^p$ \\ Dixon\end{tabular}    & 2D pair            & U-net         & def              & &  &{\footnotesize$^{18}$F-FDG} & \begin{tabular}[c]{@{}c@{}} 1.8$\pm$2.4\\ 1.7$\pm$2.0$^f$ \\ 1.8$\pm$2.4$^s$\\ 3.8$\pm$3.9$^b$\end{tabular} & $\mu$-map diff       & Torrado2019\cite{torrado2019dixon}          \\ \hline

\cellcolor{cyan!25}Pelvis& 12             &           & 6                &                  & 3$^{H}$    & \begin{tabular}[c]{@{}c@{}}T1 GRE$^c$ \\ T2 TSE \end{tabular}        & 3Dp pair            & CNN$^1$           & def               &                 & \begin{tabular}[c]{@{}c@{}}.99$\pm$.00$^s$\\ .48$\pm$.21$^a$\\ .94$\pm$.01$^f$\\ .88$\pm$0.03$^w$\\ .98$\pm$0.01$^s$    \end{tabular}    & {\footnotesize$^{18}$F-FDG}    & & RMSE     & Bradshaw2018\cite{bradshaw2018feasibility}  \\ \hline

\cellcolor{cyan!25}Prostate      & 18             &           & 10               &        & 3$^{H}$    & Dixon             & 2D pair           & GAN$^*$           & def              &                 & & {\footnotesize$^{68}$Ga-PSMA}  & \begin{tabular}[c]{@{}c@{}}\\.75$\pm$.64$^{max}$\\.52$\pm$.62$^{mea}$ \end{tabular}            &  \begin{tabular}[c]{@{}c@{}}SSIM\\$\mu$-map diff  \end{tabular}    & Pozaruk2020\cite{pozaruk2020augmented}      \\ \hline \hline

\cellcolor{gray!25}Head  & 30             &           & \begin{tabular}[c]{@{}c@{}}10 \\ 5$^{pet}$ \end{tabular}               &                  & 1.5  & \begin{tabular}[c]{@{}c@{}}T1 GRE \\ Gd \end{tabular}      & 2D pair           & CNN$^1$         & def              &                 & \begin{tabular}[c]{@{}c@{}}.971$\pm$.005$^a$\\ .936$\pm$.011$^s$\\ .803$\pm$.021$^b$\end{tabular}      &n.a.    & -0.7$\pm$1.1$^{pet}$               & & Liu2018\cite{liu2018deep}                   \\ \hline

\cellcolor{gray!25}Head  & 30$^p$+6             &           &     8          &          & 1.5$^p$+3$^{H}$  &  UTE             & 2D pair        & U-net$^1$         & def              &                 & \begin{tabular}[c]{@{}c@{}}.76$\pm$.03$^a$\\ .96$\pm$.01$^s$\\ .88$\pm$.01$^b$\end{tabular}     & {\footnotesize$^{18}$F-FDG}&   $<$1 &       & Jang2018\cite{Jang2018deep}          \\ \hline

\cellcolor{gray!25}H\&N  & 32          &           & 8              & 5             & \multirow{2}{*}{3$^{H}$}    & Dixon         & \multirow{2}{*}{2D pair}         & \multirow{2}{*}{U-net}         & \multirow{2}{*}{rig}              &     13.8$\pm$1.4            & .76$\pm$.04$^b$ & \multirow{2}{*}{{\footnotesize$^{18}$F-FDG}} & \multirow{2}{*}{$<$3} &    & \multirow{2}{*}{Gong2018\cite{gong2018attenuation}}          \\ 
\cellcolor{gray!25} & 12 &  & 2 & 7 & & $\pm$ZTE & & & & 12.6$\pm$1.5 &.80$\pm$.04$^b$ &  && &\\ \hline

\cellcolor{gray!25}Head  &\multirow{2}{*}{60}          &           & \multirow{2}{*}{19}               & \multirow{2}{*}{4}         & \multirow{2}{*}{3$^{H}$}    & mDixon             & 3Dp            & \multirow{2}{*}{U-net}         & \multirow{2}{*}{rig}              &   &  \multirow{2}{*}{.90$\pm$.07$^j$}  &\multirow{2}{*}{\footnotesize$^{18}$F-FET} & & biol tumor& \multirow{2}{*}{Ladefoged2019\cite{ladefoged2019deep}}   \\
\cellcolor{gray!25}paed & & & & & & +UTE & pair & & & & &&&vol, SUV&  
\\ \hline

\cellcolor{gray!25}Head  & 40             &           &               & 2         & 3    & T1 GRE        & \begin{tabular}[c]{@{}c@{}}3Dp\\pair\end{tabular}            & GAN           & def              & \begin{tabular}[c]{@{}c@{}}101$\pm$40 \\302$\pm$79$^b$\\ 407$\pm$228$^a$\\ 10$\pm$5$^s$\end{tabular} & .80$\pm$.07$^b$              & {\footnotesize$^{18}$F-FDG}&\begin{tabular}[c]{@{}c@{}}3.2$\pm$3.4\\1.2$\pm$13.8$^b$\\ 3.2$\pm$13.6$^s$\\ 3.2$\pm$13.6$^a$\end{tabular}           & \begin{tabular}[c]{@{}c@{}}rel vol dif\\ surf dist ME\\ RMSE PSNR\\ SSIM SUV\end{tabular} & Arabi2019\cite{arabi2019novel}              \\ \hline

\cellcolor{gray!25}Head  & 44             & 11           & 11               &                  & 1.5  & T1 GRE                & 2.5D pair          & U-net         & rig              &                 & &{\footnotesize\begin{tabular}[c]{@{}c@{}}$^{11}$C-WAY\\$^{11}$C-DASB\end{tabular}} & \begin{tabular}[c]{@{}c@{}} -0.49$\pm$1.7 \\-1.52$\pm$.73\end{tabular}              & \begin{tabular}[c]{@{}c@{}}synt $\mu$-map,\\ kin anal\end{tabular}                   & Spuhler2019\cite{spuhler2019synthesis}      \\ \hline

\cellcolor{gray!25}Head  & 23             &           & 47               &                  & 3$^{H}$    & ZTE               & \begin{tabular}[c]{@{}c@{}}3Dp \\pair\end{tabular}           & U-net         & def              &                 & .81$\pm$.03$^b$ &{\footnotesize$^{18}$F-FDG} & -0.2$\pm$5.6               & Jac  & Blanc-Durand2019 \cite{blanc2019attenuation} \\ \hline

\cellcolor{gray!25}Head  & 32             &           &               & 4         & 3$^{H}$    & Dixon$^c$             & \begin{tabular}[c]{@{}c@{}}3Dp  \\pair\end{tabular}          & GAN$^*$   & def             &   15.8$\pm$2.4\%           & .74$\pm$.05$^b$ & {\footnotesize$^{18}$F-FDG}& -1.0$\pm$13 &  SUV &     Gong2020a \cite{gong2020a}                   \\ \hline

\cellcolor{gray!25}Head  & 35             &           &               & 5         & 3    & \begin{tabular}[c]{@{}c@{}}mDixon \\UTE$^c$ \end{tabular}       & 2.5D pair         & U-net         & rig              &  10.94$\pm$.01\%               & .87$\pm$.03$^b$      & {\footnotesize\begin{tabular}[c]{@{}c@{}}$^{11}$C-PiB\\$^{18}$F-MK$_{6240}$\end{tabular}} &   $<$2 &       & Gong2020b\cite{gong2020b}          \\ \hline \hline

\cellcolor{yellow!25}Thorax& 14             &           &               & LoO                 & 3 $^{H}$   & Dixon$^c$             & 2D pair            & GAN$^*$           & def              & 67.45$\pm$9.89       &        & {\footnotesize$^{18}$F-NaF}        &                       & \begin{tabular}[c]{@{}c@{}}PSNR SSIM\\ RMSE\end{tabular}              & Baydoun2020\cite{Baydoun2020}          \\ \hline

\multicolumn{15}{c|}{\textbf{Other than MR-based sCT}} \\ \hline

\cellcolor{purple!50}Body & 100            &          &         28      &          &    \multicolumn{2}{c|}{ PET, no att corrected}      & 2D pair          & U-net         & Y$^i$              &  111$\pm$16               & .94$\pm$.01$^b$      & {\footnotesize$^{18}$F-FDG} &   -0.6$\pm$2.0 &    abs err   & Liu2018\cite{Liu2018pet}          \\ \hline

\cellcolor{purple!50}Body  & 80             &           &    39           &          & \multicolumn{2}{c|}{ PET, no att corrected}  & 3Dp pair   & GAN   & Y$^i$               &  109$\pm$19               & .87$\pm$.03$^b$      & {\footnotesize$^{18}$F-FDG}&  $<$1.0 &  NCC PSNR ME     & Dong2019\cite{Dong2019synthetic}          \\ \hline

\cellcolor{purple!50}Body  & 100             &           &    25        &         &   \multicolumn{2}{c|}{ PET, no att corrected}        & 2.5D pair         & GAN         & Y$^i$              &             &       & {\footnotesize$^{18}$F-FDG} & -0.8$\pm$8.6   &   SUV ME    & Armanious2020\cite{Armanious2020}          \\ \hline

\end{tabular}
\end{adjustbox}
{\ \\ \scriptsize 
$^*$ comparison with other architecture has been provided;
$^p$ data from another MRI sequence used as pre-training;
$^{pr}$ patients acquired with different scanner;
$^{H}$ MRI data from hybrid PET/MRI scanner;
$^{max}$ in SUV max;
$^{mea}$ in SUV mean;
$^a$ in air or bowel gas; $^b$ in the bony structures; $^s$ in the soft tissue; $^f$ in the fatty tissue; 
$^w$ in water;
$^1$ trained to segment the CT/sCT into classes;
$^j$ expressed in terms of Jaccard index and not DSC;
$^c$ multiple combinations  (also$\pm$Dixon reconstruction, where present) of the sequences were investigated but omitted;
$^i$ intrinsically registered: PET-CT data; paed: paedriatic.
}
\end{sidewaystable}
Torrado et al. \cite{torrado2019dixon} pre-trained their U-net on 19 healthy brains acquired  with $T_{1}$ GRE MRI and, subsequently, they trained the network using Dixon images of colorectal and prostate cancer patients. They showed that pre-training led to faster training with a slightly smaller residual error than U-net weights' random initialisation.\\
Pozaruk et al. \cite{pozaruk2020augmented} proposed data augmentation over 18 prostate cancer patients by perturbing the deformation field used to match the MR/CT pair for feeding the network. They compared the performance of GAN with augmentation vs 1) Dixon based and 2) Dixon + bone segmentation from the vendor, 3) U-net with and 4) without augmentation. They found significant differences between the 3 DL methods and classic MRAC routines. GAN with augmentation performed slightly better than the U-net with/without augmentation, although the differences were not statistically relevant.\\ 
Gong et al. \cite{gong2020a} used unregistered MR/CT pair for a 3D patch cycle GAN, comparing the results vs atlas-based MRAC and CNN with registered pair. Both DL methods performed better than atlas MRAC in DSC, MAE and $\textrm{PET}_{err}$. No significant difference was found between CNN and cycle-GAN. They concluded that cycle-GAN has the potentiality to skip the limit of using a perfectly aligned dataset for training. However, it requires more input data to improve output.\\
Baydoun et al. \cite{Baydoun2020} tried different network configurations (VGG16\cite{simonyan2014very}, VGG19\cite{simonyan2014very}, and ResNet\cite{he2016deep}) as a benchmark with a 2D conditional GAN receiving either two Dixon input (water and fat) or four (water, fat, in-phase and opposed-phase). The GAN always performed better than VGG19 and ResNet, with more accurate results obtained with four inputs.   

In the effort to reduce the time for image acquisition and patient discomfort, some authors proposed to obtain the sCT directly from diagnostic images, $T_{1}$- or $T_{2}$-weighted, both using images from standalone MRI scanners \cite{liu2018deep, Arabi2018,spuhler2019synthesis} or hybrid machines \cite{bradshaw2018feasibility}.
In particular, Bradshaw et al. \cite{bradshaw2018feasibility} trained a combination of three CNNs with $T_{1}$ GRE and $T_{2}$ TSE MRI (single sequence or both) to derive an sCT stratified in classes (air, water, fat and bone) which was compared with the scanner default MRAC output.  The RMSE on PET reconstruction computed on SUV and  was significantly lower with the deep learning method and $T_{1}$/$T_{2}$ input. However, recently, Gong et al. \cite{gong2020b} tested on a brain patient cohort a CNN with either $T_{1}$ or Dixon and multiple echo UTE (mUTE) as input finding that using mUTE outperformed $T_{1}$.
Liu et al. \cite{liu2018deep} trained a CNN  to predict CT tissue classes from diagnostic 1.5~T $T_{1}$ GRE of 30 patients. They tested on ten independent patients of the same cohort, whose results are reported in table \ref{tab:PET} in terms of DSC.  Then, they predicted sCT for five patients acquired prospectively with a 3T MRI/PET scanner ($T_{1}$ GRE), and they computed the $\textrm{PET}_{err}$, resulting $<$1\%. They concluded that DL approaches are flexible and promising to be applied to heterogeneous datasets acquired with different scanners and settings.

DL methods have also been proposed to estimate sCT from uncorrected PET. Thanks to the more considerable number of single PET exams, these methods have been tested on the full-body acquisitions and larger patient populations (up to 100 for training and 39 for testing). Although the global MAE is higher than site-specific MR-to-CT studies (about 110HU vs ~10-15 HU), $\textrm{PET}_{err}$ is below 1\% on average, demonstrating the validity of the approach for the scope of PET AC.

\section{Discussion}

This review encompassed DL-based approaches to generate sCT from other radiotherapy imaging modalities, focusing on published journal articles. The research topic was earlier introduced at conferences in 2016 \cite{nie2016estimating}. Since 2016, we have observed increasing interest in using DL for sCT generation.
DL methods' success is probably related to the growth of available computational resources in the last decade that allowed training large volume datasets~\cite{Lecun2015} achieving fast image translation, i.e., in the order of a few seconds~\cite{vanDyk2020}. Fast image-to-image translation facilitates applying DL in clinical cases and demonstrates its feasibility for clinical scenarios.  
In this review, we considered three clinical purposes for deriving sCT from other image modality, which are discussed in the following:
\begin{enumerate}[label=\textbf{\Roman*}]
\item \textbf{MR-only RT.} 
The generation of sCT for MR-only RT with DL is the most populated category. Its 51 papers demonstrate the potential of using DL for sCT generation from MRI. Several training techniques and configurations have been proposed. For anatomical regions, as pelvis and brain/H\&N, high image similarity and dosimetric accuracy, i.e., dose differences $<1\%$, can be achieved for photon RT and proton therapy. In region strongly affected by motion~\cite{Stemkens2018,Paganelli2018}, e.g. abdomen and thorax, the first feasibility studies seem to be promising~\cite{LiuY2019b,Olberg2019,Fu2020,Cusumano2020,Florkow2020dose}.
However, no study proposed the generation of DL-based 4D sCT yet, as from classical methods \cite{Freedman2019}. 
An exciting application is the DL-based sCT generation for the paediatric population~\cite{Maspero2020,Florkow2020dose}, which is considered more radiation-sensitive than an adult population \cite{Goodman2019} and could enormously benefit from MR-only, especially when patients' simulations are repeated~\cite{Karlsson2009}. \\
The geometric accuracy of sCT needs to be thoroughly tested to enable the clinical adoption of sCT for treatment planning purposes, primarily when MRI or sCT are used to substitute CT for position verification purposes. So far, the number of studies that investigated such an aspect from DL-based sCT is still scarce. Only Gupta et al.~\cite{Gupta2019}, for the brain, and Olberg et al.~\cite{Olberg2019}, for breast cancer, have investigated this aspect assessing the accuracy of alignment based on CBCT and digitally reconstructed radiography, respectively. Future studies are required to strengthen the clinical use of sCT, especially considering that geometric accuracy has been already extensively investigated for sCT generated with classical methods for 3~T and below~\cite{Walker2016,Gustafsson2017,MasperoRectum2018}.\\
DL-based sCT generation in the context of MR-guided radiotherapy\cite{Lagendijk2008,Lagendijk2014,Fallone2014rotating,Mutic2014viewray,Keall2014,Jaffray2014facility} may reduce the treatment time,
facilitating daily image guidance and plan adaptation based on sole MRI~\cite{Winkel2019,Hall2019}. For this application, the accuracy of dose calculation in the magnetic field's presence must be assessed before clinical implementation. So far, the studies investigating this aspect are still few, e.g. for abdominal~\cite{Fu2020} and pelvic tumours~\cite{Cusumano2020} and only considered low magnetic fields.
Recently, Groot Koerkamp et al.~\cite{GrootKoerkamp2021} published the first dosimetric evaluation of DL-based sCT for high magnetic field MR-guided RT achieving dose differences $<1\%$ for breast cases.
The results are promising, but we advocate for further studies on additional anatomical sites and magnetic field strengths. 
\item \textbf{CBCT-to-CT for image-guided (adaptive) radiotherapy.} 
In-room CBCT imaging is widespread in photon and proton RT for daily patient set-up~\cite{Boda2011}. However, CBCT is not commonly exploited for daily plan adaptation and dose recalculation due to the artefacts associated with scatter and reconstruction algorithms that affect the quality of the electron density predicted by CBCT \cite{elstrom2011evaluation}. Traditional methods to cope with this issue have been based on image registration \cite{peroni2012automatic, veiga2015cone}, scatter correction \cite{park2015proton}, look-up-table to rescale HU intensities \cite{kurz2016feasibility} and histogram matching \cite{arai2017feasibility}. DL's introduction for converting CBCT to sCT has substantially improved image quality leading to faster results than image registration and analytical corrections \cite{thummerer2020CBCT}. Speed is crucial for the translation of the method into the clinical routine. However, one of the problems arising in CBCT-to-CT conversion for clinical application is the different field of view (FOV) between CBCT and CT. Usually, the training is performed by registering, cropping and resampling the volume to the CBCT size, which is smaller than the planning CT.\\
Nonetheless, for replanning purposes, the limited FOV may hinder calculating the plan to the sCT. Some authors have proposed to assign water equivalent density within the CT body contour for the missing information \cite{barateau2020comparison}. In other cases, the sCT patch has been stitched to the planning CT to cover the entire dose volume\cite{maspero2020single}.  Ideally, appropriate FOV coverage should be employed when re-calculating the plan for online adaptive RT. 
Besides the dosimetric aspect, improved image quality may increase accuracy during image guidance for patient set-up and OAR segmentation. These are  necessary steps for online adaptive radiotherapy, especially for anatomical sites prone to large movements, as speculated by Liu et el. \cite{liu2020cbct} in the framework of pancreatic treatments.\\
CBCT-to-CT resulted in accurate dose calculations both for photon and proton radiotherapy. For proton RT, the set-up accuracy and dose calculation are even more relevant to avoid range shift errors that could jeopardise the benefit of treatment \cite{paganetti2012range}. Because there is an intrinsic error in converting HU to relative proton stopping power \cite{goma2018revisiting}, it has been shown that deep learning methods can translate CBCT directly to  stopping power\cite{harms2020cone}. This approach has not been covered in this review, but it is an exciting approach that will probably lead to further investigations.

Interestingly, increasing the quality of CBCT can be tackled as an image-to-image translation problem and as an inverse problem, i.e. from a reconstruction perspective. Specifically, by having the raw data measurements (projections), DL could improve tomography. In this sense, many investigations have been proposed but considered out of the scope of this review. For the interested reader, we suggest the following resources~\cite{Hansen2018,Wang2018,Wang2019book,Li2019reco,Maier2019}. Currently, it is unclear whether formulating
(CB)CT quality enhancement as a synthesis or reconstruction problem would be beneficial.
First attempts showed that training convolutional networks for reconstruction enhanced their generalisation capability to other anatomy~\cite{Lonning2019}; however, research on such aspects is still ongoing.

\item \textbf{PET attenuation correction.} The
sCT in this category is obtained either from MRI or from uncorrected PET. In the first case, the work's motivation is to overcome the current limitations in generating attenuation maps ($\mu$-maps) from MR images in MRI/PET hybrid acquisitions that miscalculated the bone contribution \cite{izquierdo2014comparison}. In the second case, the limits to overcome are different:  i) to avoid extra-radiation dose when the sole PET exam is required, ii) to avoid misregistration errors when standalone CT and PET machines are used, iii) to be independent of the MR contrast in MRI/PET acquisitions. 
Besides the network configuration, MRI used for the input, or the number of patients included in the studies, DL-based sCT have consistently outperformed current MRAC methods available on commercial software. The results of this review support the idea that DL-based sCT will substitute current AC methods, being also able to overcome most of the limitations mentioned above. These aspects seem to contradict the stable number of papers in this category in the last three years. Nonetheless, we have to consider that the recent trend has been to derive the $\mu$-map from uncorrected PET via DL directly. Because this review considered only image-to-CT translation, these works were not included, but they can be found in a recent review by Lee~\cite{lee2020review}.  However, it is worth mentioning a recent study from Shiri et al. \cite{shiri2020deep}, where the largest patient cohort ever (1150 patients split in 900 for training, 100 for validation and 150 for test) was used for the scope.  Direct $\mu$-map prediction via DL is an auspicious opportunity that may direct future research efforts in this context.
\end{enumerate}

\vspace{-5pt}
\textbf{Deep learning considerations and trends}\\
\addcontentsline{toc}{subsection}{Deep learning considerations and trends}
The number of patients used for training the networks is quite variable, ranging from a minimum of 7 (in I)~\cite{Qian2020} to a maximum of 205 (in II)~\cite{eckl2020evaluation} and 242~\cite{Andres2020} (in I).
In most cases, the patient number is limited to the availability of training pairs. Data augmentation is performed as linear and non-linear transformation \cite{shorten2019survey} to increase the training accuracy, as demonstrated in Pozaruk et al. \cite{pozaruk2020augmented}. However, few publications investigated the impact of increasing the training size \cite{Maspero2020,Peng2020,Olberg2019,yuan2020convolutional,Andres2020}, finding that image similarity increases when training up to fifty patients. This investigation can indicate the minimum amount of patients necessary to include in the training to achieve the state of the art performances. 
The optimal patient number may also depend on the anatomical site and its inter-fraction and intra-fraction variability. Besides, attention should be dedicated to balancing the training set, as performed in \cite{Maspero2020,Andres2020}. Otherwise, the network may overfit, as previously demonstrated for segmentation tasks \cite{li2019overfitting}. 

GANs were the most popular architecture, but we cannot conclude that it is the best network scheme for sCT. Indeed, some studies compared U-net or other CNN vs GAN finding GAN performing statistically better \cite{Baydoun2020,zhang2020improving}; others found similar results \cite{pozaruk2020augmented,gong2020a} or even worse performances ~\cite{gong2020b,Li2020comp}. 
We can speculate that, as demonstrated by \cite{Largent2019}, a vital role is played by the loss function, which, despite being the effective driver for network learning, has been investigated less than the network architecture, as highlighted for image restoration~\cite{Zhao2018}.
Another important aspect is the growing trend, except category III, in unpaired training (5 and 7 papers in 2019 and 2020, respectively).
The quality of the registration when training in a paired manner 
influences the quality of deep learning-based sCT~\cite{Florkow2019impact}.
In this sense, unpaired training offers an option to alleviate the need for
well-matched training pairs. When comparing paired vs unpaired training, we observed that paired training leads to slightly better performances. However, the differences were not always statistically significant \cite{Peng2020,Yang2020,Li2020comp}.
As proposed by Yang et al.~\cite{Yang2020}, unsupervised training decreases the semantic information from one domain to another \cite{Yang2020}. Such an issue may be solved by introducing a structure-consistency loss, which extracts structural features from the image defining the loss in the feature space. Yang et al.'s results showed improvements in this sense relative to other unsupervised methods. They also showed that pre-registering unpaired MR-CT further improves unsupervised training results, which can be an option when input and target images are available, but perfect alignment is not achievable.
In some cases, unpaired training even demonstrated to be superior to paired training~\cite{Wolterink2017}.
A trend lately emerged is the use of architecture initially thought for unpaired training, e.g. cycle-GAN to be used for paired training~\cite{Lei2019mri, harms2019paired}.\\
Focusing on the body sites, we observed that most of the investigations were conducted in the brain, H\&N and pelvic regions. Fewer studies are available for the thorax and the abdomen, representing a more challenging patient  population due to the organ motion \cite{rehman2020deep}. 

In MR-only RT, we found contradicting results regarding the best performing spatial configuration for the papers that directly compared 2D vs 3D training~\cite{Fu2019,Neppl2019}. It is undoubtedly clear that 2D+ increases the sCT quality compared to a single 2D views, as demonstrated in Spadea et al.~\cite{Spadea2019} and Maspero et al.~\cite{Maspero2020}; however, when comparing 2D against 3D training, patch size is a vital aspect~\cite{Klages2020}. 3D deep networks require a more significant number of training parameters than 2D networks \cite{Singh20203d}. For sCT generation, the approaches adopted have chosen to use patch size much smaller than the whole volume, probably hindering the contextual information considered. Generally, downsampling approaches have been proposed to increase the network' perceptive field, e.g. for segmentation tasks~\cite{Kamnitsas2016}, but they have not been applied to sCT generation. We believe this will be an exciting area of research.

For what concerns the latest development from the deep learning perspective, in 2018, Oktay et al.~\cite{oktay2018attention} proposed a new mechanism, called attention gate (AG), to focus on  target structures that can vary in shape and size. Liu et al. \cite{liu2020cbct} incorporated the AG in the generator of a cycle-GAN to learn organ variation from CBCT-CT pairs in the context of pancreas adaptive RT, showing that its contribution significantly improved the predictions compared to a network without AG. Other papers also adopted attention~\cite{Yang2020,Kearney2020}.
Embedding has also been proposed to increase the network's expressivity of the network and applied by Xiang et al.~\cite{Xiang2018} (I).
As AG's mechanism is a way to focus the $attention$ on specific portions of the image, it can potentially open the path for new research topics. In 2019, Schlemper and colleagues \cite{schlemper2019attention} evaluated the AG for different tasks in medical image processing: classification, object detection, segmentation. So, in the online IGART, we can envision that such a mechanism could lead to multi-task applications, such as deriving sCT while delineating the structure of interests. 

\textbf{Benefits and challenges for clinical implementations}\\
\addcontentsline{toc}{subsection}{Benefits and challenges of clinical implementations}
Deep learning-based sCT generations may reduce the need for additional or non-standard MRI sequences, e.g. UTE or ZTE. Avoiding additional sequences will shorten the total acquisition time, speed up the workflow, increasing patient throughput. As already mentioned, speed is particularly interesting for MR-guided RT and for adaptive RT in II, which is considered crucial for online correction. For what concern categories II and III, the generation of DL-based sCT possibly enables dose decreasing during imaging by reducing the need for CT in case of anatomical changes (in II) or by possibly diminishing the amount of radioactive material injected (in III). 

Finally, it is worth commenting on the current status of the clinical adoption of DL-based sCT. We could not find that any of the methods considered are now clinically implemented and used.
We speculate that this is probably related to the fact that the field is still relatively young, with the first publications only from 2017 and that time for clinical implementations
generally last years, if not decades~\cite{Keeling2020,Bertholet2020}.
Additionally, as already mentioned, for categories I/II, the impact of sCT for position verification still needs to be thoroughly investigated. 
The implementation may also be more comfortable for category III if the methods would be directly integrated into scanners.
In general, the involvement of vendors may streamline the clinical adoption of DL-based sCT. In this sense,
we can report that vendors are currently active in evaluating their methods in research settings, e.g. for brain~\cite{Andres2020}, pelvis~\cite{Bird2021} in I, and for H\&N, thorax and pelvis in II~\cite{eckl2020evaluation}. In the last month, Palmer et al.~\cite{Palmer2021synt} also reported using a pre-released version of a DL-based sCT generation approach for H\&N in MR-only RT.
Another essential aspect that needs to be satisfied is the compliance to the currently adopted regulations~\cite{MDR2017}, where vendors can offer vital support~\cite{Fiorino2020technology,Beckers2021}. 


A key aspect of clinical implementation is the precise definition of a DL-based solution's requirements before being accepted. 
If we consider the reported metrics, we cannot find uniform criteria for reporting. Multiple metrics have been defined, and it is not clear which region of interests they should be computed. For example, the image-based similarity was reported on the body contour or in tissues generally defined by different thresholds; for task-specific metrics, the methods employed are even more heterogeneous. For example, in I and II, gamma pass rates can be performed in 2D, 3D and different dose thresholds level have been employed, e.g. 10\%, 30\%, 50\% or 90\% of the prescribed or maximum dose. In III, the $\textrm{PET}_{err}$ can be computed either on SUV, max SUV or larger VOI, making it difficult to compare different network configurations' performances.
We think that this lack of standardisation in reporting the results is also detrimental to clinical adoption.
A first attempt at revising the metrics currently adopted has been performed by Liesbeth et al.~\cite{Liesbeth2020}. However, this is still insufficient, considering the differences in how such metrics can be calculated and reported. In this sense, we advocate for consensus-based requirements that may facilitate reporting in future clinical trials \cite{Liu2020reporting}.
Also, no public datasets arranged in the form of grand challenges (\url{https://grand-challenge.org/}) are available to enable a fair and open evaluation of different approaches~\cite{Dowling2019}.\\
To date, four scientific studies have already investigated the performance of DL-based sCT in a multi-centre setting~\cite{Maspero2020,Boni2020,Fetty2020dose,Bird2021}. These studies have been reported only for MR-only RT. Future work should focus on assessing the performance of DL-based sCT generation for II and III. On the contrary, investigations on sCT generation with classical methods using multi-centre data are more diffuse for all the three categories~\cite{Teuho2016,Wyatt2017,Persson2017,Greer2019,Bird2019,Loi2020}.
Of particular relevance when considering the generalisation of a DL model
for sCT generation may be the application of transfer learning~\cite{Pan2010,Cheplygina2019}. Mainly, transfer learning may be exploited to facilitate fine-tuning a model pre-trained on a specific MRI contrast or CBCT image protocols; or generalise among multiple anatomies. No paper was found up to December 2020 investigating this aspect, but it could be an exciting research area. More recently, Li et al.~\cite{Li2021} showed that transfer learning facilitated training a DL model on different MRI contrasts for sCT generation.

The quality of sCT cannot be judged by a user, except when its quality is inferior. Therefore, software-based quality assurance (QA) procedures should be put in place. In general, having at disposal phantoms to verify the quality of the sCT may enable regular QA procedures, as for QA of CT~\cite{Mutic2003}. This would be relatively straightforward for II; however, in MR-based sCT, phantoms' manufacturing is quite challenging due to the need for contrast in MRI and CT. Recently, the first phantoms have been proposed for such task~\cite{Gallas2015,Niebuhr2019,Singhrao2020,Colvill2020} showing the potential of additive manufacturing.

Alternatively, it would be relevant if a CNN could automatically 
generate a metric to assess the quality of sCTs, as, for example, already presented for automatic segmentation~\cite{Chen2020cnn}. 
In this sense, Bragman et al.~\cite{Bragman2018uncertainty} introduce uncertainty for such a task by adopting a multi-task network and a Bayesian probabilistic framework. More recently, two other works proposed to use uncertainty either from the combination of independently trained networks~\cite{Maspero2020} or via dropout-based variational inference~\cite{Hemsley2020}. So far, the field of uncertainty estimation with deep learning~\cite{Abdar2020review} has been superficially touched for sCT generation. It would be interesting to see future work focusing on developing criteria for automatically identifying failure cases using uncertainty prediction. Patients with inaccurate synthetic CTs will be flagged for CT rescan or manual adjustment of the sCT if deemed feasible.

\textbf{Beyond sCT for radiotherapy}\\
\addcontentsline{toc}{subsection}{Beyond sCT for radiotherapy}
We found other possible applications of DL-based image generation during the database search, which are beyond the categories mentioned so far or the radiotherapy application.
For example, Kawahara et al.~\cite{Kawahara2020} proposed to generate synthetic dual-energy CT from CT  to assess the body material composition using 2D paired GANs.
Also, commercial solutions start to be evaluated for the generation of DL-based sCT from MRI for lesion detection of suspected sacroiliitis~\cite{Jans2020mri} or to facilitate surgical planning of the spine~\cite{Staartjes2021}.
An exciting application is also the generation of sCT to facilitate multi-modal image registration, as proposed by Mckenzie et al. \cite{Mckenzie2020}. 

All the techniques of category I could be directly applied to MR-guided high-intensity focused ultrasound, where otherwise an additional CT would be required to plan the treatment properly\cite{Siedek2019}.

Additionally, the methods here reviewed to generate sCT can be applied to translating other image modalities. Interesting examples in the RT realm are provided by Jiang et al.~\cite{Jiang2019cross}, who investigated using MRI-to-CT translation to increase the segmentation's robustness. 
Kieselmann et al.~\cite{Kieselmann2020} generated synthetic MRI from CT to train segmentation networks that exploit the wealth of delineation on another modality.
A detailed review of other image-to-image translation applications in radiotherapy has been recently compiled by Wang et al.~\cite{Wang2020review}.

\section{Conclusion}
Deep learning-based methods for sCT generation have been reviewed in the context of I) MR to replace CT in radiotherapy treatment planning, II) CBCT-based adaptive radiotherapy, and III) in generating attenuation maps for PET. \\
For each category, we presented a detailed comparison in terms of imaging protocols,  DL architectures, and performances according to the most popular metrics reported.
We found that DL-based sCT generation is an active and growing area of research. For several anatomical sites, e.g. H\&N/brain and pelvis, sCT seems feasible, with deep learning achieving dose difference to CT-based planning $<1\%$ in the radiotherapy context and better performance for PET attenuation correction to the standard MRAC methods. \\
We can conclude that the deep learning-based generation of sCT has a bright future, with an extensive amount of research work being done on the topic. Further steps to spread DL-based sCT techniques into the clinic will be necessary to evaluate their generalisation among multiple centres and propose
comprehensive commissioning and QA methods, 
to ensure treatment efficacy and patient safety.

\section{Acknowledgements}

Matteo Maspero is grateful to prof.dr.ir. Cornelis (Nico) A.T. van den Berg, head of the Computational Imaging Group for MR diagnostics \& therapy, Center for Image Sciences, UMC Utrecht, the Netherlands for the general support provided during this manuscript's compilation.

\section{Conflict of interest}
None of the authors has conflict of interests to disclose.

\section*{Appendix}
\begin{figure}[h]
  \centering
  \includegraphics[width=\textwidth]{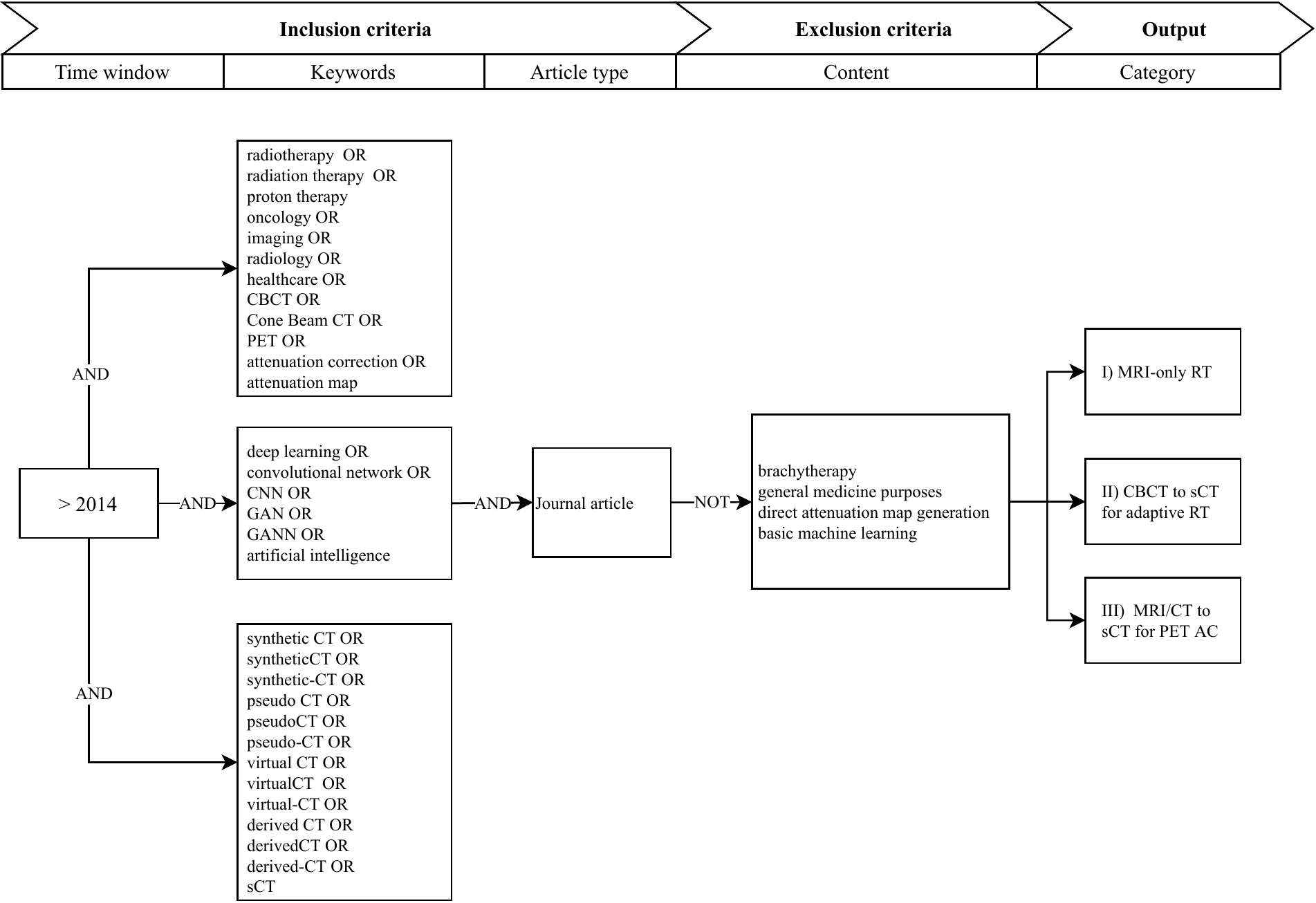}
  \caption{\textbf{Schematic of the search inclusion/exclusion criteria} adopted for this review selecting the time window, keywords, type of article, content and the three categories defined. 
  }\vspace{10pt}
   \label{fig:database}
\end{figure}

\addcontentsline{toc}{section}{\numberline{}Appendix}
\label{sec:appA}
The query used in selected databases - PubMed, Scopus and Web of Science - in the fields (Title/Abstract/Keywords) was the following (Figure~\ref{fig:database}):

(("radiotherapy") OR ("radiation therapy") OR ("proton therapy") OR ("oncology") OR ("imaging") OR
("radiology") OR ("healthcare") OR ("CBCT") OR ("cone-beam CT") OR ("PET") OR ("attenuation correction") OR ("attenuation map")) AND (("synthetic CT") OR ("syntheticCT") OR ("synthetic-CT") OR ("pseudo CT") OR ("pseudoCT") OR ("pseudo-CT") OR ("virtual CT") OR ("virtualCT") OR ("virtual-CT") OR ("derived CT") OR ("derivedCT") OR ("derived-CT") OR (sCT)) AND (("deep learning") OR ("convolutional network") OR ("CNN") OR ("GAN") OR ("GANN") OR (artificial
intelligence));

\section{Acronyms and abbreviations}
{\small
\textbf{2Dp}: 2D patches; \textbf{3Dp}: 3D patches; \textbf{AC}: attenuation correction; \textbf{aff}: affine; \textbf{AT}: attention gate; \textbf{back}: backwards pass; \textbf{CBCT}: cone-beam computed tomography; \textbf{CC}: cross-correlation; \textbf{CNNs}: Convolutional neural networks; \textbf{cor}: coronal; \textbf{CT}: computed tomography; \textbf{D}: discriminator; \textbf{DD}: dose difference; \textbf{def}: deformable; \textbf{DL}: deep learning; \textbf{DPR}: dose pass rate; \textbf{DSC}: Dice similarity coefficient; \textbf{DVH}: dose-volume histogram; \textbf{ens}: ensemble; \textbf{FID}: Frechet inception distance; \textbf{FLAIR}: fluid-attenuated inversion recovery; \textbf{forw}: forward pass; \textbf{FOV}: field of view; \textbf{G}: generator; \textbf{GANs}: generative adversarial networks; \textbf{Gd}: Gadolinium; \textbf{GPR}: gamma pass rate; \textbf{GRE}: gradient recalled-echo; \textbf{H\&N}: head \& neck; \textbf{IGART}: image-guided adaptive radiation therapy; \textbf{m2D}: multi-2D; \textbf{MAE}: mean absolute error; \textbf{MR}: magnetic resonance; \textbf{MRAC}: magnetic resonance attenuation correction; \textbf{MSE}: mean squared error; \textbf{mUTE}: multiple echo UTE; \textbf{NCC}: normalised cross-correlation; \textbf{OARS}: organs-at-risk; \textbf{p}: proton; \textbf{paed}: paediatric; \textbf{PET$_{|err|}$}: absolute error PET reconstruction; \textbf{PET$_{err}$}: relative error PET reconstruction; \textbf{PET}: positron emission tomography; \textbf{PSNR}: peak signal-to-noise ratio; \textbf{QA}: quality assurance; \textbf{rig}: rigid; \textbf{RMSE}: root mean squared error; \textbf{ROI}: region-of-interest; \textbf{RS}: range shift; \textbf{RT}: radiotherapy; \textbf{sag}: sagittal; \textbf{sCT}: synthetic computed tomography; \textbf{SSIM}: structural similarity index measure; \textbf{SUV}: standard uptake values; \textbf{tra}: transverse; \textbf{TSE}: turbo spin-echo; \textbf{UTE}: ultra-short echo time; \textbf{VOI}: volume-of-interest; \textbf{x}: photon; \textbf{ZTE}: zero time echo; \bm{$\mu-map$}: attenuation maps.
}

\section*{References}
\addcontentsline{toc}{section}{\numberline{}References}
\vspace{20pt}
\vspace*{-25mm}




\bibliography{./biblio}      

\begin{thebibliography}{100}

\bibitem{Husband2016}
J.~Husband, R.~H. Reznek, and J.~E. Husband,
\newblock {\em Imaging in oncology},
\newblock CRC Press, 2016.

\bibitem{Beaton2019}
L.~Beaton, S.~Bandula, M.~N. Gaze, and R.~A. Sharma,
\newblock {How rapid advances in imaging are defining the future of precision
  radiation oncology},
\newblock Br J Cancer {\bf 120}, 779--790 (2019).

\bibitem{Verellen2007}
D.~Verellen, M.~De~Ridder, N.~Linthout, K.~Tournel, G.~Soete, and G.~Storme,
\newblock {Innovations in image-guided radiotherapy},
\newblock Nat Rev Canc {\bf 7}, 949--960 (2007).

\bibitem{jaffray2012image}
D.~A. Jaffray,
\newblock Image-guided radiotherapy: from current concept to future
  perspectives,
\newblock Nat Rev Clin Oncol {\bf 9}, 688 (2012).

\bibitem{seco2015imaging}
J.~Seco and M.~F. Spadea,
\newblock Imaging in particle therapy: state of the art and future perspective,
\newblock Acta Oncol {\bf 54}, 1254--1258 (2015).

\bibitem{IAEA2017}
IAEA,
\newblock {\em {Radiotherapy in Cancer Care: Facing the Global Challenge}},
\newblock Non-serial Publications, INTERNATIONAL ATOMIC ENERGY AGENCY, Vienna,
  2017.

\bibitem{Seco2006}
J.~Seco and P.~M. Evans,
\newblock Assessing the effect of electron density in photon dose calculations,
\newblock Medical Physics {\bf 33(2)}, 540--552 (2006).

\bibitem{Unterrainer2020pet}
M.~Unterrainer et~al.,
\newblock Recent advances of PET imaging in clinical Radiat Oncol,
\newblock Radiat Oncol {\bf 15}, 1:15 (2020).

\bibitem{Dirix2014}
P.~Dirix, K.~Haustermans, and V.~Vandecaveye,
\newblock {The value of magnetic resonance imaging for radiotherapy planning},
\newblock  {\bf 24}, 151--159 (2014).

\bibitem{Schmidt2015}
M.~A. Schmidt and G.~S. Payne,
\newblock {Radiotherapy planning using MRI},
\newblock Phys Med Biol {\bf 60}, R323 (2015).

\bibitem{Devic2012}
S.~Devic,
\newblock {MRI simulation for radiotherapy treatment planning.},
\newblock Med Phys {\bf 39}, 6701 (2012).

\bibitem{Nyholm2009}
T.~Nyholm, M.~Nyberg, M.~G. Karlsson, and M.~Karlsson,
\newblock {Systematisation of spatial uncertainties for comparison between a MR
  and a CT-based radiotherapy workflow for prostate treatments},
\newblock Radiat Oncol {\bf 4}, 1--9 (2009).

\bibitem{Ulin2010}
K.~Ulin, M.~M. Urie, and J.~M. Cherlow,
\newblock {Results of a multi-institutional benchmark test for cranial CT/MR
  image registration},
\newblock Int J Radiat Oncol Biol Phys {\bf 77}, 1584--1589 (2010).

\bibitem{Fraas1987}
B.~A. Fraass, D.~L. McShan, R.~F. Diaz, R.~K. Ten~Haken, A.~Aisen, S.~Gebarski,
  G.~Glazer, and A.~S. Lichter,
\newblock {Integration of magnetic resonance imaging into radiation therapy
  treatment planning: i. technical considerations},
\newblock Int J Radiat Oncol Biol Phys {\bf 13}, 1897--908 (1987).

\bibitem{Lee2003}
Y.~K. Lee, M.~Bollet, G.~Charles-Edwards, M.~A. Flower, M.~O. Leach, H.~McNair,
  E.~Moore, C.~Rowbottom, and S.~Webb,
\newblock {Radiotherapy treatment planning of prostate cancer using magnetic
  resonance imaging alone},
\newblock Radiother Oncol {\bf 66}, 203--216 (2003).

\bibitem{Nyholm_counter}
T.~Nyholm and J.~Jonsson,
\newblock {Counterpoint: Opportunities and Challenges of a Magnetic Resonance
  Imaging-Only Radiotherapy Work Flow},
\newblock Semin Radiat Oncol {\bf 24}, 175--80 (2014).

\bibitem{Kapanen2013}
M.~Kapanen, J.~Collan, A.~Beule, T.~Seppälä, K.~Saarilahti, and M.~Tenhunen,
\newblock {Commissioning of MRI-only based treatment planning procedure for
  external beam radiotherapy of prostate},
\newblock Magn Reson Med {\bf 70}, 127--35 (2013).

\bibitem{Owrangi2018}
A.~M. Owrangi, P.~B. Greer, and C.~K. Glide-Hurst,
\newblock {{MRI}-only treatment planning: benefits and challenges},
\newblock Phys Med Biol {\bf 63}, 05TR01 (2018).

\bibitem{Karlsson2009}
M.~Karlsson, M.~G. Karlsson, T.~Nyholm, C.~Amies, and B.~Zackrisson,
\newblock {Dedicated Magnetic Resonance Imaging in the Radiotherapy Clinic},
\newblock Int. J. Radiat. Oncol. Biol. Phys. {\bf 74}, 644--51 (2009).

\bibitem{Lagendijk2014}
J.~J. Lagendijk, B.~W. Raaymakers, C.~A. Van~den Berg, M.~A. Moerland, M.~E.
  Philippens, and M.~Van~Vulpen,
\newblock {MR guidance in radiotherapy},
\newblock Phys Med Biol {\bf 59}, R349 (2014).

\bibitem{Jonsson2010}
J.~H. Jonsson, M.~G. Karlsson, M.~Karlsson, and T.~Nyholm,
\newblock {Treatment planning using MRI data: an analysis of the dose
  calculation accuracy for different treatment regions},
\newblock Radiat Oncol {\bf 5}, 62 (2010).

\bibitem{Edmund2017}
J.~M. Edmund and T.~Nyholm,
\newblock {A review of substitute {CT} generation for {MRI}-only radiation
  therapy},
\newblock Radiat~Oncol {\bf 12} (2017).

\bibitem{Johnstone2018}
E.~Johnstone, J.~J. Wyatt, A.~M. Henry, S.~C. Short, D.~Sebag-Montefiore,
  L.~Murray, C.~G. Kelly, H.~M. McCallum, and R.~Speight,
\newblock {Systematic Review of Synthetic Computed Tomography Generation
  Methodologies for Use in Magnetic Resonance Imaging-Only Radiation Therapy},
\newblock Int J Radiat Oncol Biol Phys {\bf 100}, 199--217 (2018).

\bibitem{Wafa2018}
B.~Wafa and A.~Moussaoui,
\newblock {A review on methods to estimate a {CT} from {MRI} data in the
  context of {MRI}-alone {RT}},
\newblock Med Tech J {\bf 2}, 150--178 (2018).

\bibitem{Kerkmeijer2018}
L.~Kerkmeijer, M.~Maspero, G.~Meijer, J.~van der Voort~van Zyp, H.~de~Boer, and
  C.~van~den Berg,
\newblock {Magnetic Resonance Imaging only Workflow for Radiotherapy Simulation
  and Planning in Prostate Cancer},
\newblock Clinic Oncol {\bf 30}, 692--701 (2018).

\bibitem{Bird2019}
D.~Bird, A.~M. Henry, D.~Sebag-Montefiore, D.~L. Buckley, B.~Al-Qaisieh, and
  R.~Speight,
\newblock {A Systematic Review of the Clinical Implementation of Pelvic
  Magnetic Resonance Imaging-Only Planning for External Beam Radiation
  Therapy},
\newblock Int J Radiat Oncol Biol Phys {\bf 105}, 479--492 (2019).

\bibitem{Thorwarth2021}
D.~Thorwarth and D.~A. Low,
\newblock Technical Challenges of Real-Time Adaptive {MR}-Guided Radiotherapy,
\newblock Front Oncol {\bf 11} (2021).

\bibitem{Hoffmann2020}
A.~Hoffmann, B.~Oborn, M.~Moteabbed, S.~Yan, T.~Bortfeld, A.~Knopf, H.~Fuchs,
  D.~Georg, J.~Seco, M.~F. Spadea, O.~Jäkel, C.~Kurz, and K.~Parodi,
\newblock {{MR}-guided proton therapy: a review and a preview},
\newblock Radiat Oncol {\bf 15} (2020).

\bibitem{taasti2020developments}
V.~T. Taasti, P.~Klages, K.~Parodi, and L.~P. Muren,
\newblock Developments in deep learning based corrections of cone beam computed
  tomography to enable dose calculations for adaptive radiotherapy,
\newblock Physics and Imaging in Radiat Oncol {\bf 15}, 77--79 (2020).

\bibitem{ZhuCTBCTnoise}
L.~Zhu, J.~Wang, and L.~Xing,
\newblock Noise suppression in scatter correction for cone-beam CT,
\newblock Med Phys {\bf 36}, 741--752 (2009b).

\bibitem{ZhuCTBCTscatter}
L.~Zhu, Y.~Xie, J.~Wang, and L.~Xing,
\newblock Scatter correction for cone-beam CT in radiation therapy,
\newblock Med Phys {\bf 36}, 2258--2268 (2009c).

\bibitem{mehranian2016vision}
A.~Mehranian, H.~Arabi, and H.~Zaidi,
\newblock Vision 20/20: magnetic resonance imaging-guided attenuation
  correction in PET/MRI: challenges, solutions, and opportunities,
\newblock Med Phys {\bf 43}, 1130--1155 (2016).

\bibitem{mecheter2020mr}
I.~Mecheter, L.~Alic, M.~Abbod, A.~Amira, and J.~Ji,
\newblock MR Image-Based Attenuation Correction of Brain PET Imaging: Review of
  Literature on Machine Learning Approaches for Segmentation,
\newblock Journal of Digital Imaging , 1--18 (2020).

\bibitem{Catana2020}
C.~Catana,
\newblock Attenuation correction for human PET/MRI studies,
\newblock Phys Med Biol {\bf 65}, TR02 (2020).

\bibitem{lecun2015deep}
Y.~LeCun, Y.~Bengio, and G.~Hinton,
\newblock Deep learning,
\newblock Nature {\bf 521}, 436--444 (2015).

\bibitem{Goodfellow2016}
I.~Goodfellow, Y.~Bengio, A.~Courville, and Y.~Bengio,
\newblock {\em {Deep learning}},
\newblock Number~2 in Adaptive Computation and Machine Learning, MIT press
  Cambridge, 2016.

\bibitem{Meyer2018}
P.~Meyer, V.~Noblet, C.~Mazzara, and A.~Lallement,
\newblock {Survey on deep learning for radiotherapy},
\newblock Comp Biol Med {\bf 98}, 126--146 (2018).

\bibitem{Sahiner2018}
B.~Sahiner, A.~Pezeshk, L.~M. Hadjiiski, X.~Wang, K.~Drukker, K.~H. Cha, R.~M.
  Summers, and M.~L. Giger,
\newblock {Deep learning in medical imaging and radiation therapy},
\newblock Med Phys {\bf 46}, e1--e36 (2018).

\bibitem{Boon2018}
I.~Boon, T.~A. Yong, and C.~Boon,
\newblock {Assessing the Role of Artificial Intelligence ({AI}) in Clinical
  Oncology: Utility of Machine Learning in Radiotherapy Target Volume
  Delineation},
\newblock Medicines {\bf 5}, 131 (2018).

\bibitem{Wang2019rev}
C.~Wang, X.~Zhu, J.~C. Hong, and D.~Zheng,
\newblock {Artificial Intelligence in Radiotherapy Treatment Planning: Present
  and Future},
\newblock Tech Canc Res Treat {\bf 18}, 153303381987392 (2019).

\bibitem{Boldrini2019}
L.~Boldrini, J.-E. Bibault, C.~Masciocchi, Y.~Shen, and M.-I. Bittner,
\newblock {Deep Learning: A Review for the Radiation Oncologist},
\newblock Front Oncol {\bf 9} (2019).

\bibitem{Jarrett2019}
D.~Jarrett, E.~Stride, K.~Vallis, and M.~J. Gooding,
\newblock {Applications and limitations of machine learning in Radiat Oncol},
\newblock Brit J Radiol {\bf 92}, 20190001 (2019).

\bibitem{Kiser2019}
K.~J. Kiser, C.~D. Fuller, and V.~K. Reed,
\newblock {Artificial intelligence in Radiat Oncol treatment planning: a brief
  overview},
\newblock J Med Art Intel {\bf 2}, 9--9 (2019).

\bibitem{Krizhevsky2012}
A.~Krizhevsky, I.~Sutskever, and G.~E. Hinton,
\newblock {Imagenet classification with deep convolutional neural networks},
\newblock Adv Neur Inf Proc Syst {\bf 25}, 1097--1105 (2012).

\bibitem{Litjens2017}
G.~Litjens, T.~Kooi, B.~E. Bejnordi, A.~A.~A. Setio, F.~Ciompi, M.~Ghafoorian,
  J.~A. Van Der~Laak, B.~Van~Ginneken, and C.~I. S{\'a}nchez,
\newblock {A survey on deep learning in medical image analysis},
\newblock Med Image Anal {\bf 42}, 60--88 (2017).

\bibitem{nie2016estimating}
D.~Nie, X.~Cao, Y.~Gao, L.~Wang, and D.~Shen,
\newblock Estimating CT image from MRI data using 3D fully convolutional
  networks,
\newblock pages 170--178 (2016).

\bibitem{lee2020review}
J.~S. Lee,
\newblock A review of deep learning-based approaches for attenuation correction
  in positron emission tomography,
\newblock IEEE Transactions on Radiation and Plasma Medical Sciences  (2020).

\bibitem{Yu2020}
B.~Yu, Y.~Wang, L.~Wang, D.~Shen, and L.~Zhou,
\newblock {\em {Medical Image Synthesis via Deep Learning}}, pages 23--44,
\newblock Springer International Publishing, Cham, 2020.

\bibitem{Wang2020review}
T.~Wang, Y.~Lei, Y.~Fu, J.~F. Wynne, W.~J. Curran, T.~Liu, and X.~Yang,
\newblock {A review on medical imaging synthesis using deep learning and its
  clinical applications},
\newblock J App Clin Med Phys  (2020).

\bibitem{Lecun2015}
Y.~LeCun, Y.~Bengio, and G.~Hinton,
\newblock {Deep learning},
\newblock Nature {\bf 521}, 436--444 (2015).

\bibitem{ronneberger2015u}
O.~Ronneberger, P.~Fischer, and T.~Brox,
\newblock U-net: Convolutional networks for biomedical image segmentation,
\newblock in {\em International Conference on Medical image computing and
  computer-assisted intervention}, pages 234--241, Springer, 2015.

\bibitem{goodfellow2014generative}
I.~Goodfellow, J.~Pouget-Abadie, M.~Mirza, B.~Xu, D.~Warde-Farley, S.~Ozair,
  A.~Courville, and Y.~Bengio,
\newblock Generative adversarial nets,
\newblock Advances in neural information processing systems {\bf 27},
  2672--2680 (2014).

\bibitem{Isola2017}
P.~Isola, J.-Y. Zhu, T.~Zhou, and A.~A. Efros,
\newblock {Image-to-image translation with conditional adversarial networks},
\newblock in {\em Proc IEEE CVPR}, pages 1125--1134, 2017.

\bibitem{wu2017survey}
X.~Wu, K.~Xu, and P.~Hall,
\newblock A survey of image synthesis and editing with generative adversarial
  networks,
\newblock Tsinghua Science and Technology {\bf 22}, 660--674 (2017).

\bibitem{creswell2018generative}
A.~Creswell, T.~White, V.~Dumoulin, K.~Arulkumaran, B.~Sengupta, and A.~A.
  Bharath,
\newblock Generative adversarial networks: An overview,
\newblock IEEE Signal Processing Magazine {\bf 35}, 53--65 (2018).

\bibitem{yi2019generative}
X.~Yi, E.~Walia, and P.~Babyn,
\newblock Generative adversarial network in medical imaging: A review,
\newblock Medical image analysis {\bf 58}, 101552 (2019).

\bibitem{zhu2017unpaired}
J.-Y. Zhu, T.~Park, P.~Isola, and A.~A. Efros,
\newblock Unpaired image-to-image translation using cycle-consistent
  adversarial networks,
\newblock in {\em Proceedings of the IEEE international conference on computer
  vision}, pages 2223--2232, 2017.

\bibitem{Paganelli2018pati}
C.~Paganelli, G.~Meschini, S.~Molinelli, M.~Riboldi, and G.~Baroni,
\newblock {Patient-specific validation of deformable image registration in
  radiation therapy: Overview and caveats},
\newblock Med Phys {\bf 45}, e908--e922 (2018).

\bibitem{Wang2004}
Z.~Wang, A.~Bovik, H.~Sheikh, and E.~Simoncelli,
\newblock Image Quality Assessment: From Error Visibility to Structural
  Similarity,
\newblock {IEEE} Trans Imag Proc {\bf 13}, 600--612 (2004).

\bibitem{Dice1945}
L.~R. Dice,
\newblock Measures of the amount of ecologic association between species,
\newblock Ecology {\bf 26}, 297--302 (1945).

\bibitem{Huttenlocher1993}
D.~P. Huttenlocher, G.~A. Klanderman, and W.~J. Rucklidge,
\newblock Comparing images using the Hausdorff distance,
\newblock IEEE Transactions on pattern analysis and machine intelligence {\bf
  15}, 850--863 (1993).

\bibitem{Reinke2021}
A.~Reinke et~al.,
\newblock {Common limitations of image processing metrics: A picture story},
\newblock arXiv preprint arXiv:2104.05642  (2021).

\bibitem{Low2010}
D.~A. Low,
\newblock {Gamma dose distribution evaluation tool},
\newblock  {\bf 250}, 012071 (2010).

\bibitem{clasie2012numerical}
B.~M. Clasie, G.~C. Sharp, J.~Seco, J.~B. Flanz, and H.~M. Kooy,
\newblock Numerical solutions of the $\gamma$-index in two and three
  dimensions,
\newblock Physics in Medicine \& Biology {\bf 57}, 6981 (2012).

\bibitem{Hussein2017}
M.~Hussein, C.~Clark, and A.~Nisbet,
\newblock {Challenges in calculation of the gamma index in radiotherapy
  {\textendash} Towards good practice},
\newblock Phys Med {\bf 36}, 1--11 (2017).

\bibitem{Drzymala1991}
R.~Drzymala, R.~Mohan, L.~Brewster, J.~Chu, M.~Goitein, W.~Harms, and M.~Urie,
\newblock Dose-volume histograms,
\newblock Int J Radiat Oncol Biol Phys {\bf 21}, 71--78 (1991).

\bibitem{paganetti2012range}
H.~Paganetti,
\newblock Range uncertainties in proton therapy and the role of Monte Carlo
  simulations,
\newblock Phys Med Biol {\bf 57}, R99 (2012).

\bibitem{pileggi2018proton}
G.~Pileggi, C.~Speier, G.~C. Sharp, D.~Izquierdo~Garcia, C.~Catana, J.~Pursley,
  F.~Amato, J.~Seco, and M.~F. Spadea,
\newblock Proton range shift analysis on brain pseudo-CT generated from T1 and
  T2 MR,
\newblock Acta Oncologica {\bf 57}, 1521--1531 (2018).

\bibitem{Andres2020}
E.~A. Andres et~al.,
\newblock {Dosimetry-driven quality measure of brain pseudo Computed Tomography
  generated from deep learning for MRI-only radiotherapy treatment planning},
\newblock Int J Radiat Oncol Biol Phys {\bf 108}, 813--823 (2020).

\bibitem{eckl2020evaluation}
M.~Eckl, L.~Hoppen, G.~R. Sarria, J.~Boda-Heggemann, A.~Simeonova-Chergou,
  V.~Steil, F.~A. Giordano, and J.~Fleckenstein,
\newblock Evaluation of a cycle-generative adversarial network-based cone-beam
  CT to synthetic CT conversion algorithm for adaptive radiation therapy,
\newblock Physica Medica {\bf 80}, 308--316 (2020).

\bibitem{Peng2020}
Y.~Peng et~al.,
\newblock {Magnetic resonance-based synthetic computed tomography images
  generated using generative adversarial networks for nasopharyngeal carcinoma
  radiotherapy treatment planning},
\newblock Radiother Oncol {\bf 150}, 217--224 (2020).

\bibitem{Qian2020}
P.~Qian, K.~Xu, T.~Wang, Q.~Zheng, H.~Yang, A.~Baydoun, J.~Zhu, B.~Traughber,
  and R.~F. Muzic,
\newblock {Estimating CT from MR Abdominal Images Using Novel Generative
  Adversarial Networks},
\newblock J Grid Comp {\bf 18}, 1--16 (2020).

\bibitem{Xu2019multi}
K.~Xu, J.~Cao, K.~Xia, H.~Yang, J.~Zhu, C.~Wu, Y.~Jiang, and P.~Qian,
\newblock {Multichannel residual conditional GAN-leveraged abdominal pseudo-CT
  generation via Dixon MR images},
\newblock IEEE Access {\bf 7}, 163823--163830 (2019).

\bibitem{ladefoged2019deep}
C.~N. Ladefoged, L.~Marner, A.~Hindsholm, I.~Law, L.~H{\o}jgaard, and F.~L.
  Andersen,
\newblock Deep learning based attenuation correction of PET/MRI in pediatric
  brain tumor patients: Evaluation in a clinical setting,
\newblock Frontiers in neuroscience {\bf 12}, 1005 (2019).

\bibitem{Maspero2020}
M.~Maspero, L.~G. Bentvelzen, M.~H. Savenije, F.~Guerreiro, E.~Seravalli, G.~O.
  Janssens, C.~A. van~den Berg, and M.~E. Philippens,
\newblock {Deep learning-based synthetic CT generation for paediatric brain
  MR-only photon and proton radiotherapy},
\newblock Radiother Oncol {\bf 153}, 197--204 (2020).

\bibitem{Florkow2020dose}
M.~C. Florkow et~al.,
\newblock {Deep learning-enabled MRI-only photon and proton therapy treatment
  planning for paediatric abdominal tumours},
\newblock Radiother Oncol {\bf 153}, 220--227 (2020).

\bibitem{Jeon2019}
W.~Jeon, H.~J. An, J.-i. Kim, J.~M. Park, H.~Kim, K.~H. Shin, and E.~K. Chie,
\newblock {Preliminary Application of Synthetic Computed Tomography Image
  Generation from Magnetic Resonance Image Using Deep-Learning in Breast Cancer
  Patients},
\newblock J Radiat Prot Res {\bf 44}, 149--155 (2019).

\bibitem{bradshaw2018feasibility}
T.~J. Bradshaw, G.~Zhao, H.~Jang, F.~Liu, and A.~B. McMillan,
\newblock Feasibility of deep learning--based PET/MR attenuation correction in
  the pelvis using only diagnostic MR images,
\newblock Tomography {\bf 4}, 138 (2018).

\bibitem{Fu2020}
J.~Fu, K.~Singhrao, M.~Cao, V.~Yu, A.~P. Santhanam, Y.~Yang, M.~Guo, A.~C.
  Raldow, D.~Ruan, and J.~H. Lewis,
\newblock {Generation of abdominal synthetic CTs from 035 T MR images using
  generative adversarial networks for MR-only liver radiotherapy},
\newblock Biom Phys Eng Express {\bf 6}, 015033 (2020).

\bibitem{Li2020comp}
Y.~Li, W.~Li, J.~Xiong, J.~Xia, and Y.~Xie,
\newblock {Comparison of Supervised and Unsupervised Deep Learning Methods for
  Medical Image Synthesis between Computed Tomography and Magnetic Resonance
  Images},
\newblock BioMed Research International {\bf 2020} (2020).

\bibitem{Xu2020}
L.~Xu, X.~Zeng, H.~Zhang, W.~Li, J.~Lei, and Z.~Huang,
\newblock {BPGAN: Bidirectional CT-to-MRI prediction using multi-generative
  multi-adversarial nets with spectral normalization and localization},
\newblock Neural Networks {\bf 128}, 82--98 (2020).

\bibitem{Fu2019}
J.~Fu, Y.~Yang, K.~Singhrao, D.~Ruan, F.-I. Chu, D.~A. Low, and J.~H. Lewis,
\newblock {Deep learning approaches using 2D and 3D convolutional neural
  networks for generating male pelvic synthetic computed tomography from
  magnetic resonance imaging},
\newblock Med Phys {\bf 46}, 3788--3798 (2019).

\bibitem{Neppl2019}
S.~Neppl et~al.,
\newblock {Evaluation of proton and photon dose distributions recalculated on
  2D and 3D Unet-generated pseudoCTs from T1-weighted MR head scans},
\newblock Acta Oncol {\bf 58}, 1429--1434 (2019).

\bibitem{Fetty2020}
L.~Fetty, M.~Bylund, P.~Kuess, G.~Heilemann, T.~Nyholm, D.~Georg, and
  T.~L{\"o}fstedt,
\newblock {Latent space manipulation for high-resolution medical image
  synthesis via the StyleGAN},
\newblock Zeits Med Phy {\bf 30} (2020).

\bibitem{Xiang2018}
L.~Xiang, Q.~Wang, D.~Nie, L.~Zhang, X.~Jin, Y.~Qiao, and D.~Shen,
\newblock {Deep embedding convolutional neural network for synthesizing CT
  image from T1-Weighted MR image},
\newblock Med Imag Anal {\bf 47}, 31--44 (2018).

\bibitem{Cusumano2020}
D.~Cusumano et~al.,
\newblock {A deep learning approach to generate synthetic CT in low field
  MR-guided adaptive radiotherapy for abdominal and pelvic cases},
\newblock Radiother Oncol {\bf 153}, 205--212 (2020).

\bibitem{harms2019paired}
J.~Harms, Y.~Lei, T.~Wang, R.~Zhang, J.~Zhou, X.~Tang, W.~J. Curran, T.~Liu,
  and X.~Yang,
\newblock Paired cycle-GAN-based image correction for quantitative cone-beam
  computed tomography,
\newblock Med Phys {\bf 46}, 3998--4009 (2019).

\bibitem{maspero2020single}
M.~Maspero, A.~C. Houweling, M.~H. Savenije, T.~C. van Heijst, J.~J. Verhoeff,
  A.~N. Kotte, and C.~A. van~den Berg,
\newblock A single neural network for cone-beam computed tomography-based
  radiotherapy of head-and-neck, lung and breast cancer,
\newblock Phys Imag Radiat Oncol {\bf 14}, 24--31 (2020).

\bibitem{zhang2020improving}
Y.~Zhang, N.~Yue, M.-Y. Su, B.~Liu, Y.~Ding, Y.~Zhou, H.~Wang, Y.~Kuang, and
  K.~Nie,
\newblock Improving CBCT Quality to CT Level using Deep-Learning with
  Generative Adversarial Network,
\newblock Med Phys  (2020).

\bibitem{Maspero2018}
M.~Maspero, M.~H. Savenije, A.~M. Dinkla, P.~R. Seevinck, M.~P. Intven, I.~M.
  Jurgenliemk-Schulz, L.~G. Kerkmeijer, and C.~A. van~den Berg,
\newblock {Dose evaluation of fast synthetic-CT generation using a generative
  adversarial network for general pelvis MR-only radiotherapy},
\newblock Phys Med Biol {\bf 63}, 185001 (2018).

\bibitem{Han2017}
X.~Han,
\newblock {MR-based synthetic CT generation using a deep convolutional neural
  network method},
\newblock Med Phys {\bf 44}, 1408--1419 (2017).

\bibitem{Emami2018}
H.~Emami, M.~Dong, S.~P. Nejad-Davarani, and C.~K. Glide-Hurst,
\newblock {Generating synthetic CTs from magnetic resonance images using
  generative adversarial networks},
\newblock Med Phys {\bf 45}, 3627--3636 (2018).

\bibitem{Jin2019}
C.-B. Jin, H.~Kim, M.~Liu, W.~Jung, S.~Joo, E.~Park, Y.~S. Ahn, I.~H. Han,
  J.~I. Lee, and X.~Cui,
\newblock {Deep CT to MR synthesis using paired and unpaired data},
\newblock Sensors {\bf 19}, 2361 (2019).

\bibitem{Lei2019mri}
Y.~Lei, J.~Harms, T.~Wang, Y.~Liu, H.-K. Shu, A.~B. Jani, W.~J. Curran, H.~Mao,
  T.~Liu, and X.~Yang,
\newblock {MRI-only based synthetic CT generation using dense cycle consistent
  generative adversarial networks},
\newblock Med Phys {\bf 46}, 3565--3581 (2019).

\bibitem{Yang2020}
H.~Yang, J.~Sun, A.~Carass, C.~Zhao, J.~Lee, J.~L. Prince, and Z.~Xu,
\newblock {Unsupervised MR-to-CT Synthesis Using Structure-Constrained
  CycleGAN},
\newblock IEEE Trans Med Imaging {\bf 39}, 4249--4261 (2020).

\bibitem{Massa2020}
H.~Massa, J.~Johnson, and A.~McMillan,
\newblock {Comparison of deep learning synthesis of synthetic CTs using
  clinical MRI inputs},
\newblock Phys Med Biol {\bf 65}, NT03 (2020).

\bibitem{Wang2019}
Y.~Wang, C.~Liu, X.~Zhang, and W.~Deng,
\newblock {Synthetic CT generation based on T2 weighted MRI of nasopharyngeal
  carcinoma (NPC) using a deep convolutional neural network (DCNN)},
\newblock Front Oncol {\bf 9} (2019).

\bibitem{Tie2020}
X.~Tie, S.-K. Lam, Y.~Zhang, K.-H. Lee, K.-H. Au, and J.~Cai,
\newblock {Pseudo-CT generation from multi-parametric MRI using a novel
  multi-channel multi-path conditional generative adversarial network for
  nasopharyngeal carcinoma patients},
\newblock Med Phys {\bf 47}, 1750--1762 (2020).

\bibitem{Kearney2020}
V.~Kearney, B.~P. Ziemer, A.~Perry, T.~Wang, J.~W. Chan, L.~Ma, O.~Morin, S.~S.
  Yom, and T.~D. Solberg,
\newblock {Attention-Aware Discrimination for MR-to-CT Image Translation Using
  Cycle-Consistent Generative Adversarial Networks},
\newblock Radiol: Art Intel {\bf 2}, e190027 (2020).

\bibitem{Largent2020}
A.~Largent et~al.,
\newblock {Head-and-Neck MRI-only radiotherapy treatment planning: From
  acquisition in treatment position to pseudo-CT generation},
\newblock Cancer/Radioth{\'e}rapie {\bf 24}, 288--297 (2020).

\bibitem{Su2020}
P.~Su, S.~Guo, S.~Roys, F.~Maier, H.~Bhat, E.~Melhem, D.~Gandhi, R.~Gullapalli,
  and J.~Zhuo,
\newblock {Transcranial MR Imaging--Guided Focused Ultrasound Interventions
  Using Deep Learning Synthesized CT},
\newblock Am J Neurorad {\bf 41}, 1841--1848 (2020).

\bibitem{Florkow2020}
M.~C. Florkow et~al.,
\newblock {Deep learning--based MR-to-CT synthesis: The influence of varying
  gradient echo--based MR images as input channels}.

\bibitem{Bahrami2020}
A.~Bahrami, A.~Karimian, E.~Fatemizadeh, H.~Arabi, and H.~Zaidi,
\newblock {A new deep convolutional neural network design with efficient
  learning capability: Application to CT image synthesis from MRI},
\newblock Med Phys {\bf 47}, 5158--5171 (2020).

\bibitem{Liu2019}
Y.~Liu et~al.,
\newblock {MRI-based treatment planning for proton radiotherapy: dosimetric
  validation of a deep learning-based liver synthetic CT generation method},
\newblock Phys Med Biol {\bf 64}, 145015 (2019).

\bibitem{Liu2020}
L.~Liu, A.~Johansson, Y.~Cao, J.~Dow, T.~S. Lawrence, and J.~M. Balter,
\newblock {Abdominal synthetic CT generation from MR Dixon images using a U-net
  trained with 'semi-synthetic' CT data},
\newblock Phys Med Biol {\bf 65}, 125001 (2020).

\bibitem{Dinkla2018}
A.~M. Dinkla, J.~M. Wolterink, M.~Maspero, M.~H. Savenije, J.~J. Verhoeff,
  E.~Seravalli, I.~I{\v{s}}gum, P.~R. Seevinck, and C.~A. van~den Berg,
\newblock {MR-only brain radiation therapy: dosimetric evaluation of synthetic
  CTs generated by a dilated convolutional neural network},
\newblock Int J Radiat Oncol Biol Phys {\bf 102}, 801--812 (2018).

\bibitem{LiuF2019}
F.~Liu, P.~Yadav, A.~M. Baschnagel, and A.~B. McMillan,
\newblock {MR-based treatment planning in radiation therapy using a deep
  learning approach},
\newblock J App Clin Med Phys {\bf 20}, 105--114 (2019).

\bibitem{Kazemifar2019}
S.~Kazemifar, S.~McGuire, R.~Timmerman, Z.~Wardak, D.~Nguyen, Y.~Park,
  S.~Jiang, and A.~Owrangi,
\newblock {MRI-only brain radiotherapy: Assessing the dosimetric accuracy of
  synthetic CT images generated using a deep learning approach},
\newblock Radiother Oncol {\bf 136}, 56--63 (2019).

\bibitem{Shafai2019}
G.~Shafai-Erfani et~al.,
\newblock {MRI-based proton treatment planning for base of skull tumors},
\newblock Int J Part Ther {\bf 6}, 12--25 (2019).

\bibitem{Gupta2019}
D.~Gupta, M.~Kim, K.~A. Vineberg, and J.~M. Balter,
\newblock {Generation of synthetic CT images from MRI for treatment planning
  and patient positioning using a 3-channel U-Net trained on sagittal images},
\newblock Front Oncol {\bf 9}, 964 (2019).

\bibitem{Spadea2019}
M.~F. Spadea, G.~Pileggi, P.~Zaffino, P.~Salome, C.~Catana,
  D.~Izquierdo-Garcia, F.~Amato, and J.~Seco,
\newblock {Deep convolution neural network (DCNN) multiplane approach to
  synthetic CT generation from MR images—application in brain proton
  therapy},
\newblock Int J Radiat Oncol Biol Phys {\bf 105}, 495--503 (2019).

\bibitem{Koike2020}
Y.~Koike, Y.~Akino, I.~Sumida, H.~Shiomi, H.~Mizuno, M.~Yagi, F.~Isohashi,
  Y.~Seo, O.~Suzuki, and K.~Ogawa,
\newblock {Feasibility of synthetic computed tomography generated with an
  adversarial network for multi-sequence magnetic resonance-based brain
  radiotherapy},
\newblock J Radiat Res {\bf 61}, 92--103 (2020).

\bibitem{Kazemifar2020dose}
S.~Kazemifar, A.~M. Barrag{\'a}n~Montero, K.~Souris, S.~T. Rivas, R.~Timmerman,
  Y.~K. Park, S.~Jiang, X.~Geets, E.~Sterpin, and A.~Owrangi,
\newblock {Dosimetric evaluation of synthetic CT generated with GANs for
  MRI-only proton therapy treatment planning of brain tumors},
\newblock J App Clin Med Phys  (2020).

\bibitem{Chen2018}
S.~Chen, A.~Qin, D.~Zhou, and D.~Yan,
\newblock {U-net-generated synthetic CT images for magnetic resonance
  imaging-only prostate intensity-modulated radiation therapy treatment
  planning},
\newblock Med Phys {\bf 45}, 5659--5665 (2018).

\bibitem{Arabi2018}
H.~Arabi, J.~A. Dowling, N.~Burgos, X.~Han, P.~B. Greer, N.~Koutsouvelis, and
  H.~Zaidi,
\newblock {Comparative study of algorithms for synthetic CT generation from
  MRI: consequences for MRI-guided radiation planning in the pelvic region},
\newblock Med Phys {\bf 45}, 5218--5233 (2018).

\bibitem{LiuY2019b}
Y.~Liu et~al.,
\newblock {Evaluation of a deep learning-based pelvic synthetic CT generation
  technique for MRI-based prostate proton treatment planning},
\newblock Phys Med Biol {\bf 64}, 205022 (2019).

\bibitem{Largent2019}
A.~Largent et~al.,
\newblock {Comparison of deep learning-based and patch-based methods for
  pseudo-CT generation in MRI-based prostate dose planning},
\newblock Int J Radiat Oncol Biol Phys {\bf 105}, 1137--1150 (2019).

\bibitem{Boni2020}
K.~N.~B. Boni, J.~Klein, L.~Vanquin, A.~Wagner, T.~Lacornerie, D.~Pasquier, and
  N.~Reynaert,
\newblock {MR to CT synthesis with multicenter data in the pelvic area using a
  conditional generative adversarial network},
\newblock Phys Med Biol {\bf 65}, 075002 (2020).

\bibitem{Fetty2020dose}
L.~Fetty, T.~L{\"o}fstedt, G.~Heilemann, H.~Furtado, N.~Nesvacil, T.~Nyholm,
  D.~Georg, and P.~Kuess,
\newblock {Investigating conditional GAN performance with different generator
  architectures, an ensemble model, and different MR scanners for MR-sCT
  conversion},
\newblock Phys Med Biol {\bf 65}, 5004 (2020).

\bibitem{Bird2021}
D.~Bird et~al.,
\newblock {Multicentre, deep learning, synthetic-CT generation for ano-rectal
  MR-only radiotherapy treatment planning},
\newblock Radiother Oncol {\bf 156}, 23--28 (2021).

\bibitem{Dinkla2019}
A.~M. Dinkla, M.~C. Florkow, M.~Maspero, M.~H. Savenije, F.~Zijlstra, P.~A.
  Doornaert, M.~van Stralen, M.~E. Philippens, C.~A. van~den Berg, and P.~R.
  Seevinck,
\newblock {Dosimetric evaluation of synthetic CT for head and neck radiotherapy
  generated by a patch-based three-dimensional convolutional neural network},
\newblock Med Phys {\bf 46}, 4095--4104 (2019).

\bibitem{Klages2020}
P.~Klages, I.~Benslimane, S.~Riyahi, J.~Jiang, M.~Hunt, J.~O. Deasy,
  H.~Veeraraghavan, and N.~Tyagi,
\newblock {Patch-based generative adversarial neural network models for head
  and neck MR-only planning},
\newblock Med Phys {\bf 47}, 626--642 (2020).

\bibitem{Qi2020}
M.~Qi et~al.,
\newblock {Multi-sequence MR image-based synthetic CT generation using a
  generative adversarial network for head and neck MRI-only radiotherapy},
\newblock Med Phys {\bf 47}, 1880--1894 (2020).

\bibitem{Thummerer2020comparison}
A.~Thummerer, B.~A. de~Jong, P.~Zaffino, A.~Meijers, G.~G. Marmitt, J.~Seco,
  R.~J. Steenbakkers, J.~A. Langendijk, S.~Both, and M.~F. Spadea,
\newblock {Comparison of the suitability of CBCT-and MR-based synthetic CTs for
  daily adaptive proton therapy in head and neck patients},
\newblock Phys Med Biol {\bf 65}, 235036 (2020).

\bibitem{Olberg2019}
S.~Olberg et~al.,
\newblock {Synthetic CT reconstruction using a deep spatial pyramid
  convolutional framework for MR-only breast radiotherapy},
\newblock Med Phys {\bf 46}, 4135--4147 (2019).

\bibitem{Florkow2019impact}
M.~C. Florkow, F.~Zijlstra, L.~G. Kerkmeijer, M.~Maspero, C.~A. van~den Berg,
  M.~van Stralen, and P.~R. Seevinck,
\newblock {The impact of MRI-CT registration errors on deep learning-based
  synthetic CT generation},
\newblock in {\em Medical Imaging 2019: Image Processing}, volume 10949, page
  1094938, International Society for Optics and Photonics, 2019.

\bibitem{Reinhold2019}
J.~C. Reinhold, B.~E. Dewey, A.~Carass, and J.~L. Prince,
\newblock Evaluating the impact of intensity normalization on {MR} image
  synthesis,
\newblock in {\em Medical Imaging 2019: Image Processing}, edited by E.~D.
  Angelini and B.~A. Landman, {SPIE}, 2019.

\bibitem{kida2018cone}
S.~Kida, T.~Nakamoto, M.~Nakano, K.~Nawa, A.~Haga, J.~Kotoku, H.~Yamashita, and
  K.~Nakagawa,
\newblock Cone beam computed tomography image quality improvement using a deep
  convolutional neural network,
\newblock Cureus {\bf 10} (2018).

\bibitem{chen2020synthetic}
L.~Chen, X.~Liang, C.~Shen, S.~Jiang, and J.~Wang,
\newblock Synthetic CT generation from CBCT images via deep learning,
\newblock Med Phys {\bf 47}, 1115--1125 (2020).

\bibitem{kida2020visual}
S.~Kida, S.~Kaji, K.~Nawa, T.~Imae, T.~Nakamoto, S.~Ozaki, T.~Ohta, Y.~Nozawa,
  and K.~Nakagawa,
\newblock Visual enhancement of Cone-beam CT by use of CycleGAN,
\newblock Med Phys {\bf 47}, 998--1010 (2020).

\bibitem{yuan2020convolutional}
N.~Yuan, B.~Dyer, S.~Rao, Q.~Chen, S.~Benedict, L.~Shang, Y.~Kang, J.~Qi, and
  Y.~Rong,
\newblock Convolutional neural network enhancement of fast-scan low-dose
  cone-beam CT images for head and neck radiotherapy,
\newblock Phys Med Biol {\bf 65}, 035003 (2020).

\bibitem{liang2019generating}
X.~Liang, L.~Chen, D.~Nguyen, Z.~Zhou, X.~Gu, M.~Yang, J.~Wang, and S.~Jiang,
\newblock Generating synthesized computed tomography (CT) from cone-beam
  computed tomography (CBCT) using CycleGAN for adaptive radiation therapy,
\newblock Phys Med Biol {\bf 64}, 125002 (2019).

\bibitem{li2019preliminary}
Y.~Li, J.~Zhu, Z.~Liu, J.~Teng, Q.~Xie, L.~Zhang, X.~Liu, J.~Shi, and L.~Chen,
\newblock A preliminary study of using a deep convolution neural network to
  generate synthesized CT images based on CBCT for adaptive radiotherapy of
  nasopharyngeal carcinoma,
\newblock Phys Med Biol {\bf 64}, 145010 (2019).

\bibitem{barateau2020comparison}
A.~Barateau et~al.,
\newblock Comparison of CBCT-based dose calculation methods in head and neck
  cancer radiotherapy: from Hounsfield unit to density calibration curve to
  deep learning,
\newblock Med Phys {\bf 47}, 4683--4693 (2020).

\bibitem{liu2020cbct}
Y.~Liu, Y.~Lei, T.~Wang, Y.~Fu, X.~Tang, W.~J. Curran, T.~Liu, P.~Patel, and
  X.~Yang,
\newblock CBCT-based synthetic CT generation using deep-attention cycleGAN for
  pancreatic adaptive radiotherapy,
\newblock Med Phys  (2020).

\bibitem{Landry2019comparing}
G.~Landry, D.~Hansen, F.~Kamp, M.~Li, B.~Hoyle, J.~Weller, K.~Parodi, C.~Belka,
  and C.~Kurz,
\newblock Comparing Unet training with three different datasets to correct CBCT
  images for prostate radiotherapy dose calculations [J],
\newblock Phys Med Biol {\bf 64} (2019).

\bibitem{kurz2019cbct}
C.~Kurz, M.~Maspero, M.~H. Savenije, G.~Landry, F.~Kamp, M.~Pinto, M.~Li,
  K.~Parodi, C.~Belka, and C.~A. Van~den Berg,
\newblock CBCT correction using a cycle-consistent generative adversarial
  network and unpaired training to enable photon and proton dose calculation,
\newblock Phys Med Biol {\bf 64}, 225004 (2019).

\bibitem{thummerer2020CBCT}
A.~Thummerer, P.~Zaffino, A.~Meijers, G.~G. Marmitt, J.~Seco, R.~J.
  Steenbakkers, J.~A. Langendijk, S.~Both, M.~F. Spadea, and A.~C. Knopf,
\newblock Comparison of CBCT based synthetic CT methods suitable for proton
  dose calculations in adaptive proton therapy,
\newblock Phys Med Biol {\bf 65}, 095002 (2020).

\bibitem{radford2015unsupervised}
A.~Radford, L.~Metz, and S.~Chintala,
\newblock Unsupervised representation learning with deep convolutional
  generative adversarial networks,
\newblock arXiv preprint arXiv:1511.06434  (2015).

\bibitem{karras2017progressive}
T.~Karras, T.~Aila, S.~Laine, and J.~Lehtinen,
\newblock Progressive growing of gans for improved quality, stability, and
  variation,
\newblock arXiv preprint arXiv:1710.10196  (2017).

\bibitem{oktay2018attention}
O.~Oktay et~al.,
\newblock Attention u-net: Learning where to look for the pancreas,
\newblock arXiv preprint arXiv:1804.03999  (2018).

\bibitem{leynes2017direct}
A.~P. Leynes, J.~Yang, F.~Wiesinger, S.~S. Kaushik, D.~D. Shanbhag, Y.~Seo,
  T.~A. Hope, and P.~E. Larson,
\newblock {Direct pseudoCT generation for pelvis PET/MRI attenuation correction
  using deep convolutional neural networks with multi-parametric MRI: zero
  echo-time and dixon deep pseudoCT (ZeDD-CT)},
\newblock J Nuc Med , jnumed--117 (2017).

\bibitem{Baydoun2020}
A.~Baydoun et~al.,
\newblock {Dixon-based thorax synthetic CT generation using Generative
  Adversarial Network},
\newblock Intelligence-Based Medicine {\bf 3}, 100010 (2020).

\bibitem{gong2018attenuation}
K.~Gong, J.~Yang, K.~Kim, G.~El~Fakhri, Y.~Seo, and Q.~Li,
\newblock Attenuation correction for brain PET imaging using deep neural
  network based on Dixon and ZTE MR images,
\newblock Phys Med Biol {\bf 63}, 125011 (2018).

\bibitem{Jang2018deep}
H.~Jang, F.~Liu, G.~Zhao, T.~Bradshaw, and A.~B. McMillan,
\newblock {Deep learning based MRAC using rapid ultrashort echo time imaging},
\newblock Med Phys {\bf 45}, 3697--3704 (2018).

\bibitem{torrado2019dixon}
A.~Torrado-Carvajal, J.~Vera-Olmos, D.~Izquierdo-Garcia, O.~A. Catalano, M.~A.
  Morales, J.~Margolin, A.~Soricelli, M.~Salvatore, N.~Malpica, and C.~Catana,
\newblock Dixon-VIBE deep learning (DIVIDE) pseudo-CT synthesis for pelvis
  PET/MR attenuation correction,
\newblock Journal of nuclear medicine {\bf 60}, 429--435 (2019).

\bibitem{blanc2019attenuation}
P.~Blanc-Durand, M.~Khalife, B.~Sgard, S.~Kaushik, M.~Soret, A.~Tiss,
  G.~El~Fakhri, M.-O. Habert, F.~Wiesinger, and A.~Kas,
\newblock Attenuation correction using 3D deep convolutional neural network for
  brain 18F-FDG PET/MR: Comparison with Atlas, ZTE and CT based attenuation
  correction,
\newblock PloS one {\bf 14}, e0223141 (2019).

\bibitem{gong2020b}
K.~Gong, P.~K. Han, K.~A. Johnson, G.~El~Fakhri, C.~Ma, and Q.~Li,
\newblock Attenuation correction using deep Learning and integrated
  UTE/multi-echo Dixon sequence: evaluation in amyloid and tau PET imaging,
\newblock Eur J Nucl Med Mol Imaging , 1--11 (2020).

\bibitem{pozaruk2020augmented}
A.~Pozaruk, K.~Pawar, S.~Li, A.~Carey, J.~Cheng, V.~P. Sudarshan, M.~Cholewa,
  J.~Grummet, Z.~Chen, and G.~Egan,
\newblock Augmented deep learning model for improved quantitative accuracy of
  MR-based PET attenuation correction in PSMA PET-MRI prostate imaging,
\newblock Eur J Nucl Med Mol Imaging  (2020).

\bibitem{gong2020a}
K.~Gong, J.~Yang, P.~E. Larson, S.~C. Behr, T.~A. Hope, Y.~Seo, and Q.~Li,
\newblock MR-based attenuation correction for brain PET using 3D
  cycle-consistent adversarial network,
\newblock IEEE Transactions on Radiation and Plasma Medical Sciences  (2020).

\bibitem{liu2018deep}
F.~Liu, H.~Jang, R.~Kijowski, T.~Bradshaw, and A.~B. McMillan,
\newblock Deep learning MR imaging--based attenuation correction for PET/MR
  imaging,
\newblock Radiology {\bf 286}, 676--684 (2018).

\bibitem{arabi2019novel}
H.~Arabi, G.~Zeng, G.~Zheng, and H.~Zaidi,
\newblock Novel adversarial semantic structure deep learning for MRI-guided
  attenuation correction in brain PET/MRI,
\newblock Eur J Nucl Med Mol Imaging {\bf 46}, 2746--2759 (2019).

\bibitem{spuhler2019synthesis}
K.~D. Spuhler, J.~Gardus, Y.~Gao, C.~DeLorenzo, R.~Parsey, and C.~Huang,
\newblock Synthesis of patient-specific transmission data for PET attenuation
  correction for PET/MRI neuroimaging using a convolutional neural network,
\newblock J Nucl Med {\bf 60}, 555--560 (2019).

\bibitem{Liu2018pet}
F.~Liu, H.~Jang, R.~Kijowski, G.~Zhao, T.~Bradshaw, and A.~B. McMillan,
\newblock {A deep learning approach for 18 F-FDG PET attenuation correction},
\newblock EJNMMI physics {\bf 5}, 1--15 (2018).

\bibitem{Dong2019synthetic}
X.~Dong, T.~Wang, Y.~Lei, K.~Higgins, T.~Liu, W.~J. Curran, H.~Mao, J.~A. Nye,
  and X.~Yang,
\newblock {Synthetic CT generation from non-attenuation corrected PET images
  for whole-body PET imaging},
\newblock Phys Med Biol {\bf 64}, 215016 (2019).

\bibitem{Armanious2020}
K.~Armanious, T.~Hepp, T.~K{\"u}stner, H.~Dittmann, K.~Nikolaou,
  C.~La~Foug{\`e}re, B.~Yang, and S.~Gatidis,
\newblock {Independent attenuation correction of whole body [18 F] FDG-PET
  using a deep learning approach with Generative Adversarial Networks},
\newblock EJNMMI research {\bf 10}, 1--9 (2020).

\bibitem{simonyan2014very}
K.~Simonyan and A.~Zisserman,
\newblock Very deep convolutional networks for large-scale image recognition,
\newblock arXiv preprint arXiv:1409.1556  (2014).

\bibitem{he2016deep}
K.~He, X.~Zhang, S.~Ren, and J.~Sun,
\newblock Deep residual learning for image recognition,
\newblock in {\em Proceedings of the IEEE conference on computer vision and
  pattern recognition}, pages 770--778, 2016.

\bibitem{vanDyk2020}
J.~J. Van~Dyk, editor,
\newblock {\em {The Modern Technology of Radiation Oncology}}, volume~4,
\newblock Medical Physics Publisher, 2020.

\bibitem{Stemkens2018}
B.~Stemkens, E.~S. Paulson, and R.~H. Tijssen,
\newblock {Nuts and bolts of 4D-MRI for radiotherapy},
\newblock Phys Med Biol {\bf 63}, 21TR01 (2018).

\bibitem{Paganelli2018}
C.~Paganelli et~al.,
\newblock {MRI-guidance for motion management in external beam radiotherapy:
  current status and future challenges},
\newblock Phys Med Biol {\bf 63}, 22TR03 (2018).

\bibitem{Freedman2019}
J.~N. Freedman, H.~E. Bainbridge, S.~Nill, D.~J. Collins, M.~Kachelrie{\ss},
  M.~O. Leach, F.~McDonald, U.~Oelfke, and A.~Wetscherek,
\newblock {Synthetic 4D-CT of the thorax for treatment plan adaptation on
  MR-guided radiotherapy systems},
\newblock Phys Med Biol {\bf 64}, 115005 (2019).

\bibitem{Goodman2019}
T.~R. Goodman, A.~Mustafa, and E.~Rowe,
\newblock {Pediatric CT radiation exposure: where we were, and where we are
  now},
\newblock Pediatric Radiol {\bf 49}, 469--478 (2019).

\bibitem{Walker2016}
A.~Walker, P.~Metcalfe, G.~Liney, V.~Batumalai, K.~Dundas, C.~Glide-Hurst,
  G.~P. Delaney, M.~Boxer, M.~L. Yap, J.~Dowling, D.~Rivest-Henault, E.~Pogson,
  and L.~Holloway,
\newblock {MRI} geometric distortion: Impact on tangential whole-breast {IMRT},
\newblock J App Clin Med Phys {\bf 17}, 7--19 (2016).

\bibitem{Gustafsson2017}
C.~Gustafsson, F.~Nordstr\"{o}m, E.~Persson, J.~Brynolfsson, and L.~E. Olsson,
\newblock {Assessment of dosimetric impact of system specific geometric
  distortion in an {MRI} only based radiotherapy workflow for prostate},
\newblock Phys Med Biol {\bf 62}, 2976--2989 (2017).

\bibitem{MasperoRectum2018}
M.~Maspero, M.~D. Tyyger, R.~H. Tijssen, P.~R. Seevinck, M.~P. Intven, and
  C.~A. {van den Berg},
\newblock {Feasibility of magnetic resonance imaging-only rectum radiotherapy
  with a commercial synthetic computed tomography generation solution},
\newblock Phys Imag Radiat Oncol {\bf 7}, 58--64 (2018).

\bibitem{Lagendijk2008}
J.~J. Lagendijk, B.~W. Raaymakers, A.~J. Raaijmakers, J.~Overweg, K.~J. Brown,
  E.~M. Kerkhof, R.~W. van~der Put, B.~H{\aa}rdemark, M.~van Vulpen, and U.~A.
  van~der Heide,
\newblock {MRI}/linac integration,
\newblock Radiother Oncol {\bf 86}, 25--29 (2008).

\bibitem{Fallone2014rotating}
B.~G. Fallone,
\newblock {The rotating biplanar linac--magnetic resonance imaging system},
\newblock  {\bf 24}, 200--202 (2014).

\bibitem{Mutic2014viewray}
S.~Mutic and J.~F. Dempsey,
\newblock {The ViewRay system: magnetic resonance--guided and controlled
  radiotherapy},
\newblock  {\bf 24}, 196--199 (2014).

\bibitem{Keall2014}
P.~J. Keall et~al.,
\newblock {The Australian magnetic resonance imaging--linac program},
\newblock  {\bf 24}, 203--206 (2014).

\bibitem{Jaffray2014facility}
D.~A. Jaffray, M.~C. Carlone, M.~F. Milosevic, S.~L. Breen, T.~Stanescu,
  A.~Rink, H.~Alasti, A.~Simeonov, M.~C. Sweitzer, and J.~D. Winter,
\newblock {A facility for magnetic resonance--guided radiation therapy},
\newblock  {\bf 24}, 193--195 (2014).

\bibitem{Winkel2019}
D.~Winkel, G.~H. Bol, P.~S. Kroon, B.~van Asselen, S.~S. Hackett, A.~M.
  Werensteijn-Honingh, M.~P. Intven, W.~S. Eppinga, R.~H. Tijssen, L.~G.
  Kerkmeijer, H.~C. de~Boer, S.~Mook, G.~J. Meijer, J.~Hes,
  M.~Willemsen-Bosman, E.~N. de~Groot-van Breugel, I.~M.
  J\"{u}rgenliemk-Schulz, and B.~W. Raaymakers,
\newblock {Adaptive radiotherapy: The Elekta Unity {MR}-linac concept},
\newblock Clin Transl Radiat Oncol {\bf 18}, 54--59 (2019).

\bibitem{Hall2019}
W.~A. Hall et~al.,
\newblock {The transformation of radiation oncology using real-time magnetic
  resonance guidance: A review},
\newblock Eur J Cancer {\bf 122}, 42--52 (2019).

\bibitem{GrootKoerkamp2021}
M.~L.~G. Koerkamp, Y.~J.~M. de~Hond, M.~Maspero, C.~Kontaxis, S.~Mandija, J.~E.
  Vasmel, R.~K. Charaghvandi, M.~E.~P. Philippens, B.~van Asselen, H.~J. G.~D.
  van~den Bongard, S.~S. Hackett, and A.~C. Houweling,
\newblock {Synthetic {CT} for single-fraction neoadjuvant partial breast
  irradiation on an {MRI}-linac},
\newblock Phys Med Biol {\bf xx}, xxxxx (2021).

\bibitem{Boda2011}
J.~Boda-Heggemann, F.~Lohr, F.~Wenz, M.~Flentje, and M.~Guckenberger,
\newblock {kV cone-beam CT-based IGRT},
\newblock Strahlen Onkol {\bf 187}, 284--291 (2011).

\bibitem{elstrom2011evaluation}
U.~V. Elstr{\o}m, L.~P. Muren, J.~B. Petersen, and C.~Grau,
\newblock Evaluation of image quality for different kV cone-beam CT acquisition
  and reconstruction methods in the head and neck region,
\newblock Acta Oncologica {\bf 50}, 908--917 (2011).

\bibitem{peroni2012automatic}
M.~Peroni, D.~Ciardo, M.~F. Spadea, M.~Riboldi, S.~Comi, D.~Alterio, G.~Baroni,
  and R.~Orecchia,
\newblock Automatic segmentation and online virtualCT in head-and-neck adaptive
  radiation therapy,
\newblock Int J Radiat Oncol Biol Phys {\bf 84}, e427--e433 (2012).

\bibitem{veiga2015cone}
C.~Veiga, J.~Alshaikhi, R.~Amos, A.~M. Louren{\c{c}}o, M.~Modat, S.~Ourselin,
  G.~Royle, and J.~R. McClelland,
\newblock Cone-beam computed tomography and deformable registration-based
  “dose of the day” calculations for adaptive proton therapy,
\newblock Int J Part Ther {\bf 2}, 404--414 (2015).

\bibitem{park2015proton}
Y.-K. Park, G.~C. Sharp, J.~Phillips, and B.~A. Winey,
\newblock Proton dose calculation on scatter-corrected CBCT image: Feasibility
  study for adaptive proton therapy,
\newblock Med Phys {\bf 42}, 4449--4459 (2015).

\bibitem{kurz2016feasibility}
C.~Kurz, R.~Nijhuis, M.~Reiner, U.~Ganswindt, C.~Thieke, C.~Belka, K.~Parodi,
  and G.~Landry,
\newblock Feasibility of automated proton therapy plan adaptation for head and
  neck tumors using cone beam CT images,
\newblock Radiat Oncol {\bf 11}, 1--9 (2016).

\bibitem{arai2017feasibility}
K.~Arai et~al.,
\newblock Feasibility of CBCT-based proton dose calculation using a
  histogram-matching algorithm in proton beam therapy,
\newblock Physica Medica {\bf 33}, 68--76 (2017).

\bibitem{goma2018revisiting}
C.~Gom{\`a}, I.~P. Almeida, and F.~Verhaegen,
\newblock Revisiting the single-energy CT calibration for proton therapy
  treatment planning: a critical look at the stoichiometric method,
\newblock Phys Med Biol {\bf 63}, 235011 (2018).

\bibitem{harms2020cone}
J.~Harms, Y.~Lei, T.~Wang, M.~McDonald, B.~Ghavidel, W.~Stokes, W.~J. Curran,
  J.~Zhou, T.~Liu, and X.~Yang,
\newblock Cone-beam CT-derived relative stopping power map generation via deep
  learning for proton radiotherapy,
\newblock Med Phys {\bf 47}, 4416--4427 (2020).

\bibitem{Hansen2018}
D.~C. Hansen, G.~Landry, F.~Kamp, M.~Li, C.~Belka, K.~Parodi, and C.~Kurz,
\newblock {ScatterNet}: A convolutional neural network for cone-beam {CT}
  intensity correction,
\newblock Med Phys {\bf 45}, 4916--4926 (2018).

\bibitem{Wang2018}
G.~Wang, J.~C. Ye, K.~Mueller, and J.~A. Fessler,
\newblock {Image Reconstruction is a New Frontier of Machine Learning},
\newblock {IEEE} Trans Med Imag {\bf 37}, 1289--1296 (2018).

\bibitem{Wang2019book}
G.~Wang, Y.~Zhang, X.~Ye, and X.~Mou,
\newblock {\em {Machine Learning for Tomographic Imaging}},
\newblock {IOP} Publishing, 2019.

\bibitem{Li2019reco}
Y.~Li, K.~Li, C.~Zhang, J.~Montoya, and G.-H. Chen,
\newblock Learning to Reconstruct Computed Tomography Images Directly From
  Sinogram Data Under A Variety of Data Acquisition Conditions,
\newblock {IEEE} Trans Med Imag {\bf 38}, 2469--2481 (2019).

\bibitem{Maier2019}
A.~K. Maier, C.~Syben, B.~Stimpel, T.~W\"{u}rfl, M.~Hoffmann, F.~Schebesch,
  W.~Fu, L.~Mill, L.~Kling, and S.~Christiansen,
\newblock Learning with known operators reduces maximum error bounds,
\newblock Nat Machine Intell {\bf 1}, 373--380 (2019).

\bibitem{Lonning2019}
K.~L{\o}nning, P.~Putzky, J.-J. Sonke, L.~Reneman, M.~W. Caan, and M.~Welling,
\newblock Recurrent inference machines for reconstructing heterogeneous {MRI}
  data,
\newblock Med Image Anal {\bf 53}, 64--78 (2019).

\bibitem{izquierdo2014comparison}
D.~Izquierdo-Garcia, S.~J. Sawiak, K.~Knesaurek, J.~Narula, V.~Fuster,
  J.~Machac, and Z.~A. Fayad,
\newblock Comparison of MR-based attenuation correction and CT-based
  attenuation correction of whole-body PET/MR imaging,
\newblock European journal of nuclear medicine and molecular imaging {\bf 41},
  1574--1584 (2014).

\bibitem{shiri2020deep}
I.~Shiri, H.~Arabi, P.~Geramifar, G.~Hajianfar, P.~Ghafarian, A.~Rahmim, M.~R.
  Ay, and H.~Zaidi,
\newblock Deep-JASC: joint attenuation and scatter correction in whole-body 18
  F-FDG PET using a deep residual network,
\newblock European Journal of Nuclear Medicine and Molecular Imaging  (2020).

\bibitem{shorten2019survey}
C.~Shorten and T.~M. Khoshgoftaar,
\newblock A survey on image data augmentation for deep learning,
\newblock Journal of Big Data {\bf 6}, 1--48 (2019).

\bibitem{li2019overfitting}
Z.~Li, K.~Kamnitsas, and B.~Glocker,
\newblock Overfitting of neural nets under class imbalance: Analysis and
  improvements for segmentation,
\newblock in {\em International Conference on Medical Image Computing and
  Computer-Assisted Intervention}, pages 402--410, Springer, 2019.

\bibitem{Zhao2018}
Z.~Hang, G.~Orazio, F.~Iuri, and K.~Jan,
\newblock {Loss Functions for Neural Networks for Image Processing},
\newblock CoRR {\bf abs/1511.08861} (2015).

\bibitem{Wolterink2017}
J.~M. Wolterink, A.~M. Dinkla, M.~H. Savenije, P.~R. Seevinck, C.~A. van~den
  Berg, and I.~I{\v{s}}gum,
\newblock {Deep MR to CT synthesis using unpaired data},
\newblock in {\em Int Work SASHIMI}, pages 14--23, Springer, 2017.

\bibitem{rehman2020deep}
A.~Rehman and F.~G. Khan,
\newblock A deep learning based review on abdominal images,
\newblock Multimedia Tools and Applications , 1--32 (2020).

\bibitem{Singh20203d}
S.~P. Singh, L.~Wang, S.~Gupta, H.~Goli, P.~Padmanabhan, and B.~Guly{\'a}s,
\newblock {3D deep learning on medical images: a review},
\newblock Sensors {\bf 20}, 5097 (2020).

\bibitem{Kamnitsas2016}
K.~Kamnitsas, E.~Ferrante, S.~Parisot, C.~Ledig, A.~V. Nori, A.~Criminisi,
  D.~Rueckert, and B.~Glocker,
\newblock {DeepMedic for brain tumor segmentation},
\newblock in {\em International workshop on Brainlesion: Glioma, multiple
  sclerosis, stroke and traumatic brain injuries}, pages 138--149, Springer,
  2016.

\bibitem{schlemper2019attention}
J.~Schlemper, O.~Oktay, M.~Schaap, M.~Heinrich, B.~Kainz, B.~Glocker, and
  D.~Rueckert,
\newblock Attention gated networks: Learning to leverage salient regions in
  medical images,
\newblock Med Image Anal {\bf 53}, 197--207 (2019).

\bibitem{Keeling2020}
P.~Keeling, J.~Clark, and S.~Finucane,
\newblock Challenges in the clinical implementation of precision medicine
  companion diagnostics,
\newblock Expert review of molecular diagnostics {\bf 20}, 593--599 (2020).

\bibitem{Bertholet2020}
J.~Bertholet et~al.,
\newblock {Patterns of practice for adaptive and real-time radiation therapy
  (POP-ART RT) part II: Offline and online plan adaption for interfractional
  changes},
\newblock Radiother Oncol {\bf 153}, 88--96 (2020).

\bibitem{Palmer2021synt}
E.~Palm{\'e}r, A.~Karlsson, F.~Nordstr{\"o}m, K.~Petruson, C.~Siversson,
  M.~Ljungberg, and M.~Sohlin,
\newblock {Synthetic computed tomography data allows for accurate absorbed dose
  calculations in a magnetic resonance imaging only workflow for head and neck
  radiotherapy},
\newblock Phys Imag Radiat Oncol {\bf 17}, 36--42.

\bibitem{MDR2017}
{Council of European Union},
\newblock Regulation (EU) 2017/745 of the European Parliament and of the
  Council of 5 April 2017 on medical devices, amending Directive 2001/83/EC,
  Regulation (EC) No 178/2002 and Regulation (EC) No 1223/2009 and repealing
  Council Directives 90/385/EEC and 93/42/EEC, 2017,
\newblock \newline\url{http://data.europa.eu/eli/reg/2017/745/oj}.

\bibitem{Fiorino2020technology}
C.~Fiorino, M.~Guckenberger, M.~Schwarz, U.~A. van~der Heide, and B.~Heijmen,
\newblock {Technology-driven research for radiotherapy innovation},
\newblock Mol Oncol {\bf 14}, 1500--1513 (2020).

\bibitem{Beckers2021}
R.~Beckers, Z.~Kwade, and F.~Zanca,
\newblock {The EU medical device regulation: Implications for artificial
  intelligence-based medical device software in medical physics},
\newblock Phys Med {\bf 83}, 1--8 (2021).

\bibitem{Liesbeth2020}
V.~Liesbeth et~al.,
\newblock Overview of artificial intelligence-based applications in
  radiotherapy: recommendations for implementation and quality assurance,
\newblock Radiother Oncol  (2020).

\bibitem{Liu2020reporting}
X.~Liu, S.~C. Rivera, D.~Moher, M.~J. Calvert, and A.~K. Denniston,
\newblock {Reporting guidelines for clinical trial reports for interventions
  involving artificial intelligence: the CONSORT-AI extension},
\newblock Brit Med J {\bf 370} (2020).

\bibitem{Dowling2019}
J.~A. Dowling and J.~Korhonen,
\newblock {MR}-Only Methodology,
\newblock in {\em {MRI} for Radiotherapy}, pages 131--151, Springer
  International Publishing, 2019.

\bibitem{Teuho2016}
J.~Teuho, J.~Johansson, J.~Linden, A.~E. Hansen, S.~Holm, S.~H. Keller,
  G.~Delso, P.~Veit-Haibach, K.~Magota, V.~Saunavaara, T.~Tolvanen, M.~Teras,
  and H.~Iida,
\newblock {Effect of Attenuation Correction on Regional Quantification Between
  {PET}/{MR} and {PET}/{CT}: A Multicenter Study Using a 3-Dimensional Brain
  Phantom},
\newblock J Nuc Med {\bf 57}, 818--824 (2016).

\bibitem{Wyatt2017}
J.~J. Wyatt, J.~A. Dowling, C.~G. Kelly, J.~McKenna, E.~Johnstone, R.~Speight,
  A.~Henry, P.~B. Greer, and H.~M. McCallum,
\newblock Investigating the generalisation of an atlas-based synthetic-{CT}
  algorithm to another centre and {MR} scanner for prostate {MR}-only
  radiotherapy,
\newblock Phys Med Biol {\bf 62}, N548--N560 (2017).

\bibitem{Persson2017}
E.~Persson, C.~Gustafsson, F.~Nordstr\"{o}m, M.~Sohlin, A.~Gunnlaugsson,
  K.~Petruson, N.~Rintel\"{a}, K.~Hed, L.~Blomqvist, B.~Zackrisson, T.~Nyholm,
  L.~E. Olsson, C.~Siversson, and J.~Jonsson,
\newblock {{MR}-{OPERA}: A Multicenter/Multivendor Validation of Magnetic
  Resonance Imaging{\textendash}Only Prostate Treatment Planning Using
  Synthetic Computed Tomography Images},
\newblock Int J Radiat Oncol Biol Phys {\bf 99}, 692--700 (2017).

\bibitem{Greer2019}
P.~Greer, J.~Martin, M.~Sidhom, P.~Hunter, P.~Pichler, J.~H. Choi, L.~Best,
  J.~Smart, T.~Young, M.~Jameson, T.~Afinidad, C.~Wratten, J.~Denham,
  L.~Holloway, S.~Sridharan, R.~Rai, G.~Liney, P.~Raniga, and J.~Dowling,
\newblock {A Multi-center Prospective Study for Implementation of an {MRI}-Only
  Prostate Treatment Planning Workflow},
\newblock Front Oncol {\bf 9} (2019).

\bibitem{Loi2020}
G.~Loi, M.~Fusella, C.~Vecchi, S.~Menna, F.~Rosica, E.~Gino, N.~Maffei,
  E.~Menghi, A.~Savini, A.~Roggio, L.~Radici, E.~Cagni, F.~Lucio, L.~Strigari,
  S.~Strolin, C.~Garibaldi, C.~Roman{\`{o}}, M.~Piovesan, P.~Franco, and
  C.~Fiandra,
\newblock {Computed Tomography to Cone Beam Computed Tomography Deformable
  Image Registration for Contour Propagation Using Head and Neck, Patient-Based
  Computational Phantoms: A~Multicenter Study},
\newblock Pract Radiat Oncol {\bf 10}, 125--132 (2020).

\bibitem{Pan2010}
S.~J. Pan and Q.~Yang,
\newblock {A Survey on Transfer Learning},
\newblock {IEEE} Transactions on Knowledge and Data Engineering {\bf 22},
  1345--1359 (2010).

\bibitem{Cheplygina2019}
V.~Cheplygina, M.~de~Bruijne, and J.~P. Pluim,
\newblock Not-so-supervised: A survey of semi-supervised, multi-instance, and
  transfer learning in medical image analysis,
\newblock Med Image Anal {\bf 54}, 280--296 (2019).

\bibitem{Li2021}
W.~Li, S.~Kazemifar, T.~Bai, D.~Nguyen, Y.~Weng, Y.~Li, J.~Xia, J.~Xiong,
  Y.~Xie, A.~Owrangi, and S.~Jiang,
\newblock Synthesizing {CT} images from {MR} images with deep learning: model
  generalization for different datasets through transfer learning,
\newblock  {\bf 7}, 025020 (2021).

\bibitem{Mutic2003}
S.~Mutic, J.~R. Palta, E.~K. Butker, I.~J. Das, M.~S. Huq, L.-N.~D. Loo, B.~J.
  Salter, C.~H. McCollough, and J.~Van~Dyk,
\newblock {Quality assurance for computed-tomography simulators and the
  computed-tomography-simulation process: report of the AAPM Radiation Therapy
  Committee Task Group No. 66},
\newblock Med Phys {\bf 30}, 2762--2792 (2003).

\bibitem{Gallas2015}
R.~R. Gallas, N.~H{\"u}nemohr, A.~Runz, N.~I. Niebuhr, O.~J{\"a}kel, and
  S.~Greilich,
\newblock {An anthropomorphic multimodality (CT/MRI) head phantom prototype for
  end-to-end tests in ion radiotherapy},
\newblock Zeitsch Mediz Phys {\bf 25}, 391--399 (2015).

\bibitem{Niebuhr2019}
N.~Niebuhr, W.~Johnen, G.~Echner, A.~Runz, M.~Bach, M.~Stoll, K.~Giske,
  S.~Greilich, and A.~Pfaffenberger,
\newblock {The ADAM-pelvis phantom—an anthropomorphic, deformable and
  multimodal phantom for MRgRT},
\newblock Phys Med Biol {\bf 64}, 04NT05 (2019).

\bibitem{Singhrao2020}
K.~Singhrao, J.~Fu, H.~H. Wu, P.~Hu, A.~U. Kishan, R.~K. Chin, and J.~H. Lewis,
\newblock {A novel anthropomorphic multimodality phantom for MRI-based
  radiotherapy quality assurance testing},
\newblock Med physics {\bf 47}, 1443--1451 (2020).

\bibitem{Colvill2020}
E.~Colvill et~al.,
\newblock {Anthropomorphic phantom for deformable lung and liver CT and MR
  imaging for radiotherapy},
\newblock Phys Med Biol {\bf 65}, 07NT02 (2020).

\bibitem{Chen2020cnn}
X.~Chen, K.~Men, B.~Chen, Y.~Tang, T.~Zhang, S.~Wang, Y.~Li, and J.~Dai,
\newblock {CNN-based quality assurance for automatic segmentation of breast
  cancer in radiotherapy},
\newblock Front Oncol {\bf 10} (2020).

\bibitem{Bragman2018uncertainty}
F.~J. Bragman, R.~Tanno, Z.~Eaton-Rosen, W.~Li, D.~J. Hawkes, S.~Ourselin,
  D.~C. Alexander, J.~R. McClelland, and M.~J. Cardoso,
\newblock {Uncertainty in multitask learning: joint representations for
  probabilistic MR-only radiotherapy planning},
\newblock in {\em International Conference on Medical Image Computing and
  Computer-Assisted Intervention}, pages 3--11, Springer, 2018.

\bibitem{Hemsley2020}
M.~Hemsley, B.~Chugh, M.~Ruschin, Y.~Lee, C.-L. Tseng, G.~Stanisz, and A.~Lau,
\newblock {Deep Generative Model for Synthetic-CT Generation with Uncertainty
  Predictions},
\newblock in {\em International Conference on Medical Image Computing and
  Computer-Assisted Intervention}, pages 834--844, Springer, 2020.

\bibitem{Abdar2020review}
M.~Abdar et~al.,
\newblock {A review of uncertainty quantification in deep learning: Techniques,
  applications and challenges},
\newblock arXiv preprint arXiv:2011.06225  (2020).

\bibitem{Kawahara2020}
D.~Kawahara, A.~Saito, S.~Ozawa, and Y.~Nagata,
\newblock {Image synthesis with deep convolutional generative adversarial
  networks for material decomposition in dual-energy CT from a kilovoltage CT},
\newblock Comp Biol Med {\bf 128}, 104111 (2020).

\bibitem{Jans2020mri}
L.~B. Jans, M.~Chen, D.~Elewaut, F.~Van~den Bosch, P.~Carron, P.~Jacques,
  R.~Wittoek, J.~L. Jaremko, and N.~Herregods,
\newblock {MRI-based synthetic CT in the detection of structural lesions in
  patients with suspected sacroiliitis: comparison with MRI},
\newblock Radiol , 201537 (2020).

\bibitem{Staartjes2021}
V.~E. Staartjes, P.~R. Seevinck, W.~P. Vandertop, M.~van Stralen, and M.~L.
  Schr{\"o}der,
\newblock {Magnetic resonance imaging--based synthetic computed tomography of
  the lumbar spine for surgical planning: a clinical proof-of-concept},
\newblock Neurosurgical Focus {\bf 50}, E13 (2021).

\bibitem{Mckenzie2020}
E.~M. McKenzie, A.~Santhanam, D.~Ruan, D.~O'Connor, M.~Cao, and K.~Sheng,
\newblock {Multimodality image registration in the head-and-neck using a deep
  learning-derived synthetic CT as a bridge},
\newblock Med Phys {\bf 47}, 1094--1104 (2020).

\bibitem{Siedek2019}
F.~Siedek, S.~Y. Yeo, E.~Heijman, O.~Grinstein, G.~Bratke, C.~Heneweer,
  M.~Puesken, T.~Persigehl, D.~Maintz, and H.~Gr\"{u}ll,
\newblock {Magnetic Resonance-Guided High-Intensity Focused Ultrasound
  ({MR}-{HIFU}): Technical Background and Overview of Current Clinical
  Applications (Part 1)},
\newblock {R\"{o}Fo} - Fortschritte auf dem Gebiet der R\"{o}ntgenstrahlen und
  der bildgebenden Verfahren {\bf 191}, 522--530 (2019).

\bibitem{Jiang2019cross}
J.~Jiang, Y.-C. Hu, N.~Tyagi, P.~Zhang, A.~Rimner, J.~O. Deasy, and
  H.~Veeraraghavan,
\newblock {Cross-modality (CT-MRI) prior augmented deep learning for robust
  lung tumor segmentation from small MR datasets},
\newblock Med Phys {\bf 46}, 4392--4404 (2019).

\bibitem{Kieselmann2020}
J.~P. Kieselmann, C.~D. Fuller, O.~J. Gurney-Champion, and U.~Oelfke,
\newblock {Cross-modality deep learning: Contouring of MRI data from annotated
  CT data only},
\newblock Med Phys  (2020).

\end{thebibliography}




\bibliographystyle{./medphy.bst}    

\end{document}